\documentstyle[12pt,fleqn]{article}

\font\twlgot =eufm10 scaled \magstep1
\font\egtgot =eufm8
\font\sevgot =eufm7

\font\twlmsb =msbm10 scaled \magstep1
\font\egtmsb =msbm8
\font\sevmsb =msbm7

\newfam\gotfam
\def\pgot{\fam\gotfam\twlgot}
\textfont\gotfam\twlgot
\scriptfont\gotfam\egtgot
\scriptscriptfont\gotfam\sevgot
\def\got{\protect\pgot}

\newfam\msbfam
\textfont\msbfam\twlmsb
\scriptfont\msbfam\egtmsb
\scriptscriptfont\msbfam\sevmsb
\def\Bbb{\protect\pBbb}
\def\pBbb{\relax\ifmmode\expandafter\Bb\else\typeout{You cann't use
Bbb in text mode}\fi}
\def\Bb #1{{\fam\msbfam\relax#1}}

\let\Large=\large

\def\op#1{\mathop{{\it\fam0} #1}\limits}
\newcommand{\glos}[1]{\bigskip{\bf #1}\bigskip}
\newcommand{\id}{{\rm Id\,}}

\newcommand{\pr}{{\rm pr}}
\newcommand{\df}{{\rm def}}

\newcommand{\di}{{\rm dim\,}}
\newcommand{\fdi}{{\rm fdim\,}}

\newcommand{\Id}{{\rm Id}}
\newcommand{\Ker}{{\rm Ker\,}}
\newcommand{\im}{{\rm Im\, }}

\newcommand{\nm}[1]{\mid {#1}\mid}

\newcommand{\bite}{\begin{itemize}}
\newcommand{\eite}{\end{itemize}}
\newcommand{\benu}{\begin{enumerate}}
\newcommand{\eenu}{\end{enumerate}}
\newcommand{\bde}{\begin{description}}
\newcommand{\ede}{\end{description}}
\newcommand{\bquo}{\begin{quote}}
\newcommand{\equo}{\end{quote}}
\newcommand{\bquot}{\begin{quotation}}
\newcommand{\equot}{\end{quotation}}

\newcommand{\eqref}[1]{(\ref{#1})}
\newcommand{\beq}{\begin{equation}}
\newcommand{\eeq}{\end{equation}}
\newcommand{\ben}{\begin{eqnarray}}
\newcommand{\een}{\end{eqnarray}}
\newcommand{\be}{\begin{eqnarray*}}
\newcommand{\ee}{\end{eqnarray*}}
\newcommand{\bea}{\begin{eqalph}}
\newcommand{\eea}{\end{eqalph}}

\newcommand{\nw}[1]{[{#1}]}
\newcommand{\cB}{{\cal B}}
\newcommand{\gU}{{\got U}}
\newcommand{\gO}{{\got O}}
\newcommand{\cO}{{\cal O}}
\newcommand{\cA}{{\cal A}}

\newcommand{\gG}{{\got G}}
\newcommand{\cG}{{\got g}}
\newcommand{\gd}{{\got d}}

\newcommand{\gE}{{\got E}}

\newcommand{\gT}{{\got T}}

\newcommand{\gR}{{\got A}}
\newcommand{\gQ}{{\got Q}}

\newcommand{\cT}{{\cal T}}
\newcommand{\cP}{{\cal P}}
\newcommand{\cR}{{\cal R}}
\newcommand{\cL}{{\cal L}}
\newcommand{\cV}{{\cal V}}

\newcommand{\cQ}{{\cal Q}}
\newcommand{\cE}{{\cal E}}

\newcommand{\cF}{{\cal F}}

\newcommand{\cS}{{\cal S}}

\newcommand{\bL}{{\bf L}}
\newcommand{\bb}{{\bf 1}}

\newcommand{\bA}{{\bf A}}

\newcommand{\al}{\alpha}
\newcommand{\bt}{\beta}
\newcommand{\dl}{\delta}
\newcommand{\la}{\lambda}
\newcommand{\La}{\Lambda}
\newcommand{\f}{\phi}
\newcommand{\vf}{\varphi}

\newcommand{\p}{\pi}
\newcommand{\s}{\psi}
\newcommand{\x}{\xi}
\newcommand{\om}{\omega}

\newcommand{\m}{\mu}
\newcommand{\n}{\nu}
\newcommand{\g}{\gamma}
\newcommand{\G}{\Gamma}
\newcommand{\e}{\epsilon}
\newcommand{\ve}{\varepsilon}
\newcommand{\th}{\theta}

\newcommand{\r}{\rho}

\newcommand{\up}{\upsilon}

\newcommand{\si}{\sigma}
\newcommand{\Si}{\Sigma}

\newcommand{\Y}{Y\to X}
\newcommand{\w}{\wedge}

\newcommand{\wt}{\widetilde}
\newcommand{\wh}{\widehat}
\newcommand{\ol}{\overline}

\newcommand{\dr}{\partial}
\newcommand{\pdr}{\partial}

\newcommand{\ar}{\op\longrightarrow}

\newcommand{\mto}{\mapsto}

\newcommand{\xx}{\times}
\newcommand{\ox}{\otimes}

\newcommand{\ot}{\otimes}
\newcommand{\ap}{\approx}

\let\ssection=\section
\renewcommand{\section}{\setcounter{equation}{0}\ssection}

\newcounter{eqalph}[section]
\newcounter{equationa}[section]
\newcounter{example}[section]
\newcounter{remark}[section]
\newcounter{theorem}[section]
\newcounter{proposition}[section]
\newcounter{lemma}[section]
\newcounter{corollary}[section]
\newcounter{definition}[section]

\def\theremark{\arabic{section}.\arabic{remark}}

\def\thedefinition{\arabic{section}.\arabic{definition}}

\newenvironment{proof}{\noindent {\bf Proof.}\small}{\hfill{\footnotesize\bf
QED} \bigskip }
\newenvironment{ex}{\refstepcounter{remark} \medskip\noindent{\bf Example
\theremark.}\small}{{\Large $\bullet$} \bigskip }
\newenvironment{rem}{\refstepcounter{remark} \medskip\noindent{\bf Remark
\theremark.}\small}{{\Large $\bullet$} \bigskip }
\newenvironment{theo}{\refstepcounter{definition} \bigskip\noindent{\sc
Theorem \thedefinition}.}{$\Box$\bigskip }
\newenvironment{prop}{\refstepcounter{definition} \bigskip\noindent{\sc
Proposition \thedefinition}.}{$\Box$ \bigskip }
\newenvironment{lem}{\refstepcounter{definition} \bigskip\noindent{\sc Lemma
\thedefinition}.}{ $\Box$\bigskip }
\newenvironment{cor}{\refstepcounter{definition} \bigskip\noindent{\sc
Corollary \thedefinition}.}{ $\Box$\bigskip }
\newenvironment{defi}{\refstepcounter{definition} \bigskip\noindent{\sc
Definition \thedefinition}.}{$\Box$ \bigskip }

\newenvironment{eqalph}{\stepcounter{equation}
\setcounter{equationa}{\value{equation}}
\setcounter{equation}{0}

\begin{eqnarray}}{\end{eqnarray}
\setcounter{equation}{\value{equationa}}}

\newcommand{\mar}[1]{}

\newcommand{\bs}{{\bf s}}

\begin{document}
\hbox{}

{\parindent=0pt

{\large\bf Ten Lectures on Jet Manifolds in Classical and Quantum Field
Theory} 
\bigskip

{\sc  Gennadi
Sardanashvily}
\bigskip

\begin{small}
Department of Theoretical Physics, Physics Faculty, Moscow State
University, 117234 Moscow, Russia

E-mail: sard@grav.phys.msu.su

\bigskip

{\bf Abstract.}

These Lectures summarize the relevant material on existent applications of
jet manifold techniques to classical and quantum field theory. The
following topics are included: {\it 1. Fibre bundles, 
2. Jet manifolds, 3. Connections, 4.
Lagrangian field theory, 5. Gauge theory of principal connections, 6.
Higher order jets, 7. Infinite order jets, 8. The variational bicomplex,
9. Geometry of simple graded manifolds, 10. Jets of ghosts and antifields.}

\end{small}

\bigskip
\bigskip

{\large\bf Introduction}
}

\bigskip
\bigskip

Finite order jet manifolds \cite{kol,sau} provide the adequate mathematical
formulation of classical field theory
\cite{book,sard93,sard95}. Infinite order jets and jets of
odd variables find applications to
quantum field theory, namely, to the field-antifield BRST model
\cite{barn,barn00,brandt,brandt01}. Therefore, we aim to modify our survey
hep-th/9411089 and to complete it with the relevant facts on infinite
order jets, the variational bicomplex, and jets of graded manifolds
\cite{lmp,jmp01,book00,sard01}. 

All morphisms throughout are smooth
and manifolds are real, smooth, and
finite-dimensional. Smooth manifolds are customarily
assumed to be Hausdorff and second-countable
topological space (i.e., have a countable
base for topology). Consequently, they are paracompact, separable
(i.e.,
have a countable dense subset), and locally compact topological spaces, which
are countable at infinity.
Unless otherwise stated,
manifolds are assumed to be connected, i.e., are also arcwise 
connected.

\section{Fibre bundles}

A fibred manifold (or a fibration) 
over an $n$-dimensional base $X$ is defined as a manifold surjection
\mar{z10}\beq
\pi:Y\to X, \label{z10}
\eeq
where
$Y$ admits an atlas of fibred coordinates $(x^\la,y^i)$ such that
$(x^\la)$ are coordinates on
the base $X$, i.e.,
\be
\pi: Y\ni (x^\la,y^i)\mapsto (x^\la)\in X.
\ee
This condition is equivalent to $\pi$
being a submersion, i.e., the tangent map $T\pi:TY\to TX$ is a surjection.
It follows that $\pi$ is also an open map.

\glos{A. Smooth fibre bundles}

A fibred manifold
$Y\to X$ is said to be a (smooth) fibre bundle if
there exist a  manifold $V$, called a typical fibre,
and an open cover
$\gU=\{U_\xi\}$ of $X$  such that $Y$ is locally diffeomorphic
to the splittings
\mar{mos02}\beq
\psi_\xi:\pi^{-1}(U_\xi) \to U_\xi\times V, \label{mos02}
\eeq
glued together by means of transition functions
\mar{mos271}\beq
\rho_{\xi\zeta}=\psi_\xi\circ\psi_\zeta^{-1}:
U_\xi\cap U_\zeta\times V \to  U_\xi\cap U_\zeta\times V
\label{mos271}
\eeq
on overlaps $U_\xi\cap U_\zeta$. It follows that fibres
$Y_x=\pi^{-1}(x)$, $x\in X$, of a fibre bundle are
its closed imbedded submanifolds. Transition functions
$\rho_{\xi\zeta}$
fulfil the cocycle
condition
\mar{+9}\beq
\rho_{\xi\zeta}\circ\rho_{\zeta\iota}=\rho_{\xi\iota} \label{+9}
\eeq
on all overlaps $U_\xi\cap U_\zeta\cap U_\iota$.
We will also use the notation
\mar{sp21,2}\ben
&&\psi_\xi(x): Y_x\to V, \qquad x\in U_\xi,\label{sp21}\\
&& \rho_{\xi\zeta}(x):V\to V, \qquad x\in U_\xi\cap U_\zeta. \label{sp22}
\een
Trivialization charts
$(U_\xi, \psi_\xi)$ together with transition functions
$\rho_{\xi\zeta}$ (\ref{mos271})
constitute a bundle atlas
\mar{sp5}\beq
\Psi = \{(U_\xi, \psi_\xi), \rho_{\xi\zeta}\} \label{sp5}
\eeq
of a fibre bundle $Y\to X$.
Two bundle atlases are said to be equivalent
if their union is also a bundle
atlas, i.e., there exist transition functions between trivialization
charts of different atlases.
A fibre bundle $Y\to X$
is uniquely defined by a bundle atlas, and all its atlases are equivalent.

Throughout, only proper coverings of manifolds are considered, i.e.,
$U_\xi\neq U_\zeta$ if $\zeta\neq \xi$. A cover
$\gU'$ is said to be a refinement of a cover
$\gU$ if, for each $U'\in\gU'$, there exists $U\in\gU$ such that
$U'\subset U$. Of course, if a fibre bundle $Y\to X$ has a bundle
atlas over a cover $\gU$ of
$X$, it admits a bundle atlas over any refinement of $\gU$.
The following two theorems describe the particular covers which one can
choose for a bundle atlas.

\begin{theo} \label{sp1} \mar{sp1}
Every smooth fibre bundle $Y\to X$ admits a bundle atlas over a
countable cover
$\gU$ of $X$ where each member $U_\xi$ of $\gU$ is a domain
(i.e., a contractible open subset) whose
closure $\ol U_\xi$ is compact  \cite{greub}.
\end{theo}

\begin{proof}
The statement at once follows from the fact that, for any cover
$\gU$ of an
$n$-dimensional smooth manifold $X$, there exists a countable atlas
$\{(U'_i,\f_i)\}$ of $X$ such that: (i) the cover $\{U'_i\}$ refines $\gU$,
(ii) $\f_i(U'_i)=\Bbb R^n$, and (iii) $\ol U'_i$ is compact, $i\in\Bbb
N$.
\end{proof}

If the base $X$ is compact, there is a bundle atlas of $Y$ over a
finite cover of
$X$ which obeys the condition of Theorem \ref{sp1}. In general, every
smooth fibre bundle admits a bundle atlas over a finite cover of its
base $X$, but its members need not be
contractible and connected as follows.

\begin{theo} \label{sp2} \mar{sp2}
Every smooth fibre bundle $Y\to X$ admits a bundle atlas over a finite
cover $\gU$ of $X$.
\end{theo}

\begin{proof}
Let $\Psi$ (\ref{sp5}) be a bundle atlas of $Y\to X$ over a
cover $\gU$ of $X$.
For any cover $\gU$ of a manifold $X$, there exists its refinement
$\{U_{ij}\}$, where $j\in\Bbb N$ and $i$ runs through a finite set such that
$U_{ij}\cap U_{ik}=\emptyset$, $j\neq k$. Let
$\{(U_{ij}, \psi_{ij})\}$ be the corresponding bundle atlas of the
fibre bundle $Y\to X$. Then $Y$ has the finite bundle atlas
\be
U_i\op=^\df\op\cup_j U_{ij}, \qquad \psi_i(x)\op=^\df\psi_{ij}(x), \qquad x\in
U_{ij}\subset U_i.
\ee
\end{proof}

Without a loss of generality, we will further assume
that a cover $\gU$ for a bundle atlas of $Y\to X$ is also a
cover for a manifold atlas of the base $X$.
Then, given a bundle atlas $\Psi$ (\ref{sp5}), a fibre bundle $Y$ is
provided with the
associated  bundle coordinates
\be
x^\la(y)=(x^\la\circ \pi)(y), \qquad y^i(y)=(y^i\circ\psi_\xi)(y), \qquad
y\in \pi^{-1}(U_\xi),
\ee
where $x^\la$ are coordinates on $U_\xi\subset X$ and
$y^i$ are coordinates on the typical fibre $V$.

Morphisms of fibre bundles, by definition, preserve their fibrations,
i.e., send a fibre
to a fibre.
Namely, a  bundle morphism of a fibre bundle
$\pi:Y\to X$ to a fibre bundle $\pi': Y'\to X'$ is defined as a pair
$(\Phi,f)$ of
manifold morphisms which make up the commutative diagram
\be
\begin{array}{rcccl}
& Y &\ar^\Phi & Y'&\\
_\pi& \put(0,10){\vector(0,-1){20}} & & \put(0,10){\vector(0,-1){20}}&_{\pi'}\\
& X &\ar^f & X'&
\end{array}, \qquad \pi'\circ\Phi=f\circ\pi,
\ee
i.e., $\Phi$ is a
fibrewise morphism over $f$ which sends a fibre $Y_x$, $x\in X$, to a
fibre $Y'_{f(x)}$. A bundle diffeomorphism is called an isomorphism,
or an automorphism if it is an isomorphism to itself. In field
theory, any
automorphism of a fibre bundle is treated as a gauge
transformation. 
For the sake of brevity, a bundle morphism over $f=\id X$ is often said to be
a bundle morphism over $X$, and is denoted by $Y\ar_XY'$.
In particular, an
automorphism over $X$ is called a vertical automorphism
or a vertical gauge
transformation. Two different fibre bundles over the same
base $X$ are said to be
equivalent if there exists their
isomorphism over $X$.
A bundle monomorphism $\Phi:Y\to Y'$ over $X$ is
called a subbundle of the fibre bundle $Y'\to X$ if
$\Phi(Y)$ is a submanifold of $Y'$.

In particular,
a fibre bundle $Y\to X$ is said to be trivial if it is equivalent to
the Cartesian product of manifolds
\be
X\times V\ar^{\pr_1} X.
\ee
It should be emphasized that a trivial fibre bundle admits different
trivializations $Y\cong X\times V$
which differ from each other in surjections $Y\to V$.

\begin{theo} \label{spr280} \mar{spr280} A fibre
bundle over a contractible base is always trivial \cite{ste}.
\end{theo}

Classical fields are described by sections of
fibre bundles.
A section (or a global section)
of a fibre bundle $Y\to X$ is defined
as a manifold injection
$s:X\to Y$ such that $\pi\circ s=\id X$, i.e., a section sends any
point $x\in X$ into the fibre $Y_x$ over this point.
A section $s$ is
an imbedding, i.e., $s(X)\subset Y$ is both a
submanifold and a
topological subspace of $Y$. It is also a closed map, which
sends closed subsets of $X$ onto closed subsets of $Y$. In particular,
$\pi(X)$ is a closed submanifold of $Y$.
Similarly, a section of a fibre bundle $Y\to
X$ over a submanifold of $X$ is defined. Let us note that by a local
local section is customarily meant a section over an
open subset of $X$. A fibre bundle admits a local section
around each point of its base, but need not have a global section.

\begin{theo} \label{mos9} \mar{mos9}
A fibre bundle $Y\to X$ whose typical fibre is diffeomorphic to an
Euclidean space $\Bbb
R^m$ has a global section. More generally, its section over a closed
imbedded submanifold (e.g., a point) of $X$ is extended to a
global section \cite{ste}.
\end{theo}

Given a bundle atlas $\Psi$ and associated
bundle coordinates $(x^\la,y^i)$, a
section $s$ of a fibre bundle $Y\to X$ is represented by collections of
local functions $\{s^i=y^i\circ\psi_\xi \circ s\}$ on trivialization sets
$U_\xi$.

In conclusion, let us describe two standard constructions of new fibre
bundles from old ones.
\begin{itemize}
\item Given a fibre bundle $\pi:Y\to X$ and a manifold morphism $f:
X'\to X$, the pull-back of $Y$ by $f$ is defined as
a fibre bundle
\mar{mos106}\beq
f^*Y =\{(x',y)\in X'\times Y \,: \,\, \pi(y) =f(x')\} \label{mos106}
\eeq
over $X'$ provided with the natural surjection $(x',y)\mapsto x'$.
Roughly speaking, its fibre
over a point
$x'\in X'$ is that of
$Y$ over the point $f(x')\in X$.
\item Let $Y$ and $Y'$ be fibre bundles
over the same base $X$.
Their fibred product $Y\times Y'$
is a fibre bundle
over $X$ whose fibres are the Cartesian products $Y_x\times Y'_x$ of
those of fibre bundles $Y$ and $Y'$.
\end{itemize}

\glos{B. Vector and affine bundles}

Vector and affine bundles provide a standard framework in classical and
quantum field
theory. Matter fields are sections of vector bundles, while gauge
potentials are sections of an affine bundle.

A typical fibre and fibres of a smooth
vector bundle $\pi:Y\to X$  are
vector spaces of some finite dimension (called the fibre dimension
$\fdi Y$ of $Y$), and $Y$ admits a bundle atlas
$\Psi$ (\ref{sp5}) where trivialization morphisms
$\psi_\xi(x)$ (\ref{sp21}) and transition functions
$\rho_{\xi\zeta}(x)$ (\ref{sp22})
are linear isomorphisms of vector spaces.
The corresponding bundle coordinates
$(y^i)$ possess a linear coordinate transformation law
\be
y'^i=\rho^i_j(x)y^j.
\ee
We have the decomposition $y=y^ie_i(\pi(y))$, where
\be
\{e_i(x)\}=\psi_\xi^{-1}(x)\{v_i\}, \qquad x=\pi(y)\in U_\xi,
\ee
are fibre bases (or frames) for fibres $Y_x$ of
$Y$ and $\{v_i\}$ is a fixed basis for the
typical fibre $V$ of $Y$.

By virtue of Theorem \ref{mos9}, a vector bundle has a global
section, e.g., the canonical zero section $\wh 0(X)$ which sends every
point $x\in X$ to the origin 0 of the fibre $Y_x$ over $x$.

The following are the standard constructions of new vector bundles from
old ones.
\begin{itemize}
\item Given two vector bundles $Y$ and $Y'$ over the same base $X$,
their Whitney sum $Y\oplus Y'$ is a vector
bundle over $X$ whose fibres are the direct sums of
those of the vector bundles $Y$ and $Y'$.
\item Given two vector bundles $Y$ and $Y'$ over the same base $X$,
their tensor product $Y\ot Y'$ is a vector
bundle over $X$ whose fibres are the tensor products of
those of the vector bundles $Y$ and $Y'$.
Similarly, the exterior product
$Y\w Y$ of vector bundles
is defined. We call
\mar{spr880}\beq
\w Y=X\times \Bbb R\op\oplus_X Y\op\oplus_X \op\w^2 Y\op\oplus_X \cdots
\op\oplus_X \op\w^m Y, \qquad m=\fdi Y,\label{spr880}
\eeq
the exterior bundle of $Y$.
\item Let $Y\to X$ be a vector bundle. By
$Y^*\to X$ is denoted the dual vector bundle whose fibres are the duals
of those of $Y$. 
The interior product (or contraction) 
of $Y$ and $Y^*$ is defined as a bundle morphism
\be
\rfloor: Y\otimes Y^*\ar_X X\times \Bbb R.
\ee
\end{itemize}

Vector bundles are subject to linear bundle
morphisms, which are linear fibrewise maps.
They possess the following property.
Given vector bundles $Y'$ and $Y$ over the same base $X$, every linear
bundle morphism
\be
\Phi: Y'_x\ni \{e'_i(x)\}\mapsto \{\Phi^k_i(x)e_k(x)\}\in Y_x
\ee
over $X$ defines a global section
\be
\Phi: x\mapsto \Phi^k_i(x)e_k(x)\ot e'^i(x)
\ee
of the tensor product $Y\ot Y'^*$, and {\it vice versa}.

Given a linear bundle morphism $\Phi: Y'\to Y$ of vector bundles over
$X$, its kernel Ker$\,\Phi$ is
defined as the inverse image $\Phi^{-1}(\wh 0(X))$ of the canonical zero
section $\wh 0(X)$ of $Y$. If $\Phi$ is of constant rank, its
kernel Ker$\,\Phi$ and its image $\im\Phi$ are subbundles of the vector bundles
$Y'$ and $Y$, respectively. For instance, monomorphisms and
epimorphisms of vector bundles fulfil this condition.
If $Y'$ is a subbundle of the vector bundle
$Y\to X$, the factor bundle $Y/Y'$ over $X$ is
defined as a vector bundle whose fibres  are the
quotients $Y_x/Y'_x$, $x\in X$.

Let us consider a sequence
\be
Y'\ar^i Y\ar^j Y''
\ee
of vector bundles over $X$. It is called
exact at $Y$ if
Ker$\,j=\im i$.  Let
\mar{sp10}\beq
0\to Y'\ar^i Y\ar^j Y'' \to 0 \label{sp10}
\eeq
be a sequence of vector bundles over $X$, where $0$ denotes the
zero-dimensional vector bundle over $X$. This sequence is
called a short exact sequence
if it is exact at all terms $Y'$, $Y$, and $Y''$. This
means that $i$ is a bundle monomorphism, $j$ is a bundle
epimorphism, and Ker$\,j=\im i$. Then $Y''$ is the
factor bundle $Y/Y'$. One says that the short exact sequence
(\ref{sp10}) admits a 
splitting if
there exists a bundle monomorphism $s:Y''\to Y$ such that $j\circ s=\id Y''$
or, equivalently,
\be
Y=i(Y')\oplus s(Y'')\cong Y'\oplus Y''.
\ee

\begin{theo} \label{sp11} \mar{sp11}
Every exact sequence of vector bundles (\ref{sp10}) is split \cite{hir}.
\end{theo}

Given an exact sequence of vector bundles (\ref{sp10}), we have the 
(dual) exact sequence
of the dual bundles
\be
0\to Y''^*\ar^{j^*} Y^*\ar^{i^*} Y'^* \to 0.
\ee

Let us turn to affine bundles.
Given a vector bundle $\ol Y\to X$,
an affine bundle modelled over
$\ol Y\to X$ is a fibre bundle $Y\to X$ whose
fibres $Y_x$, $x\in X$, are affine spaces modelled
over the
corresponding fibres $\ol Y_x$ of the vector bundle $\ol Y$, and $Y$
admits a bundle atlas $\Psi$ (\ref{sp5}) whose trivialization morphisms
$\psi_\xi(x)$ and transition functions
functions $\rho_{\xi\zeta}(x)$ are affine maps.
The corresponding bundle coordinates $(y^i)$ possess an affine coordinate
transformation law
\be
y'^i=\rho^i_j(x)y^j+ \rho^i(x).
\ee
There are the bundle morphisms
\be
&&Y\op\times_X\ol Y\ar_X Y,\qquad (y^i, \ol y^i)\mapsto  y^i +\ol y^i,\\
&&Y\op\times_X Y\ar_X \ol Y,\qquad (y^i, y'^i)\mapsto  y^i - y'^i,
\ee
where $(\ol y^i)$ are linear bundle coordinates on the vector bundle
$\ol Y$.
For instance, every vector bundle has a natural structure of an affine
bundle.

By virtue of Theorem \ref{mos9}, every affine bundle has a global
section.

One can define a direct sum $Y\oplus \ol Y'$ of a vector bundle
$\ol Y'\to X$ and an affine bundle $Y\to X$ modelled over a vector
bundle $\ol Y\to X$. This is an affine bundle modelled over the Whitney
sum of vector bundles $\ol Y'\oplus \ol Y$.

Affine bundles are subject to affine bundle
morphisms which are affine fibrewise maps.
Any affine bundle morphism $\Phi:Y\to Y'$ from an affine bundle $Y$ modelled
over a vector bundle $\ol Y$ to an affine bundle
$Y'$ modelled over a vector bundle $\ol Y'$,
yields the linear bundle morphism of these vector bundles
\mar{1355'}\beq
\ol \Phi: \ol Y\to \ol Y', \qquad
\ol y'^i\circ \ol\Phi= \frac{\dr\Phi^i}{\dr y^j}\ol y^j. \label{1355'}
\eeq

The analogues of Theorems \ref{sp1}, \ref{sp2} on a particular cover
for atlases of vector and affine
bundles hold.

\glos{C. Tangent and cotangent bundles}

Tangent and cotangent bundles exemplify vector bundles.
The fibres of the tangent bundle
\be
\pi_Z:TZ\to Z
\ee
of a manifold $Z$ are tangent spaces to $Z$. The peculiarity of the
tangent bundle $TZ$ in comparison with other vector bundles over $Z$
lies in the fact that, given an atlas $\Psi_Z =\{(U_\xi,\phi_\xi)\}$
of a manifold $Z$, the tangent bundle of $Z$ is provided with the 
holonomic atlas
$\Psi =\{(U_\xi, \psi_\xi = T\phi_\xi)\}$,
where by $T\phi_\xi$ is meant the tangent map to $\f_\xi$.
Namely, given coordinates $(z^\la)$ on a manifold $Z$, the associated
bundle coordinates on $TZ$ are
holonomic coordinates $(\dot z^\la)$
with respect to the holonomic frames  $\{\dr_\la\}$ for
tangent spaces
$T_zZ$, $z\in Z$. Their transition functions read
\be
\dot z'^\la=\frac{\dr z'^\la}{\dr z^\m}\dot z^\m.
\ee
Every manifold morphism $f:Z\to Z'$ yields the linear bundle morphism
over $f$ of the tangent bundles
\mar{spr823}\beq
Tf: TZ\ar_f TZ',\qquad \dot z'^\la\circ Tf=\frac{\dr f^\la}{\dr z^\m}
\dot z^\m. \label{spr823}
\eeq
It is called the tangent map to $f$.

The cotangent bundle of a manifold $Z$
is the dual
\be
\pi_{*Z}:T^*Z\to Z
\ee
of the tangent bundle $TZ\to Z$. It is equipped with the holonomic
coordinates $(z^\la,\dot z_\la)$ with respect to the
coframes   $\{dz^\la\}$ for $T^*Z$ which are the duals of
$\{\dr_\la\}$. Their transition functions read
\be
\dot z'_\la=\frac{\dr z^\m}{\dr z'^\la}\dot z_\m.
\ee

A tensor product
\mar{sp20}\beq
T=(\op\ot^mTZ)\ot(\op\ot^kT^*Z), \qquad m,k\in \Bbb N, \label{sp20}
\eeq
over $Z$ of tangent and cotangent bundles is called a tensor bundle.

Tangent, cotangent and tensor bundles belong to the category of natural
fibre bundles which admit the canonical 
lift of any
diffeomorphism $f$ of a base to a bundle automorphism, called the natural
automorphism \cite{kol}.
For instance, the natural automorphism of the tangent bundle $TZ$
over a diffeomorphism $f$ of its base $Z$ is the tangent map $Tf$
(\ref{spr823}) to $f$. 
In view of the expression (\ref{spr823}), natural
automorphisms are also called  holonomic transformations
or general covariant transformations (in gravitation theory).

Let us turn now to peculiarities of tangent and cotangent bundles of
fibre bundles.

Let $\pi_Y:TY\to Y$ be the tangent bundle of a fibre bundle
$\pi: Y\to X$.
Given bundle coordinates $(x^\la,y^i)$ on $Y$, the tangent bundle $TY$
is equipped with the holonomic coordinates
$(x^\la,y^i,\dot x^\la, \dot y^i)$.
The tangent bundle $TY\to Y$ has the subbundle
$VY = \Ker T\pi$ which
consists of the vectors tangent to fibres of $Y$. It is called
the vertical tangent bundle of $Y$,
and is
provided with the holonomic coordinates $(x^\la,y^i,\dot y^i)$ with
respect to the vertical frames $\{\dr_i\}$.

Let $T\Phi$ be the tangent map to a bundle morphism $\Phi:Y\to Y'$.
Its restriction $V\Phi$ to $VY$ is a linear bundle morphism $VY\to VY'$
such that
\be
\dot y'^i\circ V\Phi =\dot y^j\dr_j\Phi^i.
\ee
It is called the vertical tangent map to $\Phi$.

In many important cases, the vertical tangent
bundle $VY\to Y$ of a fibre bundle $Y\to X$ is trivial, and is equivalent
to the fibred product
\mar{48'}\beq
VY\cong Y\op\times_X\ol Y \label{48'}
\eeq
of $Y$ and some vector bundle $\ol Y\to X$. This means that $VY$ can be
provided
with bundle coordinates $(x^\la,y^i,\ol y^i)$ such that a transformation
law of coordinates $\ol y^i$ is independent of coordinates
$y^i$. One calls (\ref{48'}) the vertical splitting. 

For instance, every affine bundle $Y\to X$ modelled over a
vector bundle $\ol Y\to X$
admits the canonical vertical splitting (\ref{48'}) with respect to the
holonomic coordinates $\dot y^i$
on $VY$, whose
transformation law coincides with that of the linear coordinates $\ol y^i$
on the vector
bundle $\ol Y$. If $Y$ is a vector bundle, the vertical splitting (\ref{48'})
reads
\mar{1.10}\beq
VY\cong Y\op\times_X Y. \label{1.10}
\eeq

The vertical cotangent bundle $V^*Y\to Y$ of a fibre bundle $Y\to X$
is defined as the dual
of the vertical tangent bundle $VY\to Y$. It is not a subbundle of the
cotangent bundle $T^*Y$, but there is
the canonical surjection
\mar{z11}\beq
\zeta: T^*Y\ni \dot x_\la dx^\la +\dot y_i dy^i \mapsto \dot y_i \ol dy^i\in
V^*Y, \label{z11}
\eeq
where $\{\ol dy^i\}$ are the  bases for the fibres of $V^*Y$ which
are duals of the holonomic frames $\{\dr_i\}$ for the vertical tangent
bundle $VY$. It should be emphasized that coframes $\{dy^i\}$ for 
$T^*Y$ and $\{\ol
dy^i\}$ for
$V^*Y$ are transformed in a different way.

With $VY$ and $V^*Y$, we have the following two exact sequences of
vector bundles over $Y$:
\mar{1.8ab}\bea
&& 0\to VY\hookrightarrow TY\op\to^{\pi_T} Y\op\times_X TX\to 0,
\label{1.8a} \\
&& 0\to Y\op\times_X T^*X\hookrightarrow T^*Y\op\to^\zeta V^*Y\to 0.
\label{1.8b}
\eea
In accordance with Theorem \ref{sp11}, they have a splitting which, by
definition, is a connection on a
fibre bundle $Y\to X$.

\glos{D. Composite fibre bundles}

Let us consider the composition
\mar{1.34}\beq
\pi: Y\to \Si\to X, \label{1.34}
\eeq
of fibre bundles
\mar{z275,6}\ben
&& \pi_{Y\Si}: Y\to\Si, \label{z275}\\
&& \pi_{\Si X}: \Si\to X. \label{z276}
\een
It is called the composite fibre
bundle. 
It is provided with bundle coordinates $(x^\la,\si^m,y^i)$, where
$(x^\la,\si^m)$ are bundle coordinates on the fibre bundle (\ref{z276}), i.e.,
transition functions of coordinates $\si^m$ are independent of
coordinates $y^i$.

The following two assertions make composite fibre bundles  useful for
numerous physical
applications \cite{book,book00}.

\begin{prop}\label{comp10} \mar{comp10}
Given a composite fibre bundle (\ref{1.34}), let $h$ be a global section
of the fibre bundle $\Si\to X$. Then the restriction
\mar{S10}\beq
Y_h=h^*Y \label{S10}
\eeq
of the fibre bundle $Y\to\Si$ to $h(X)\subset \Si$ is a subbundle
of the fibre bundle $Y\to X$.
\end{prop}

\begin{prop} \label{mos61} \mar{mos61}  (i) Given
a section $h$ of the fibre bundle
$\Si\to X$ and a section $s_\Si$ of the fibre bundle $Y\to\Si$, their
composition $s=s_\Si\circ h$
is a section of the composite fibre  bundle $Y\to X$ (\ref{1.34}).

(ii) Conversely, every section $s$ of the fibre bundle $Y\to X$ is a
composition of the section $h=\pi_{Y\Si}\circ s$ of the
fibre bundle $\Si\to X$ and some section $s_\Si$ of the fibre
bundle $Y\to \Si$ over the closed imbedded submanifold $h(X)\subset \Si$.
\end{prop}

In field theory, sections $h$ of the fibre bundle $\Si\to X$
play the role, e.g., of a Higgs field and a gravitational
field.

\glos{E. Vector fields}

A vector field on a manifold $Z$ is defined
as a global section of the tangent bundle $TZ\to Z$.
The set $\cT_1(Z)$ of vector fields on $Z$ is  a real Lie
algebra with respect to the  Lie bracket
\be
[v,u] = (v^\la\dr_\la u^\m - u^\la\dr_\la v^\m)\dr_\m,
\quad v=v^\la\dr_\la, \quad u=u^\la\dr_\la.
\ee

Every vector field on a manifold $Z$ can be seen as an
infinitesimal generator of a
local one-parameter group of
diffeomorphisms of $Z$ as follows \cite{kob}.
Given an open subset
$U\subset Z$ and an interval $(-\e,\e)$ of $\Bbb R$, by a
local one-parameter group of  diffeomorphisms of  $Z$
defined on
$(-\e,\e)\times U$ is meant a map
\be
G: (-\e,\e)\times U \ni (t,z)\mapsto G_t(z)\in Z
\ee
such that:
\begin{itemize}
\item for each $t\in (-\e,\e)$, the map $G_t$ is a
diffeomorphism of $U$
onto the open subset $G_t(U)\subset Z$;
\item $G_{t+t'}(z) = (G_t\circ G_{t'})(z)$ if $t,t',t+t'\in (-\e,\e)$
and $G_{t'}(z),z\in U$.
\end{itemize}
If $G$ is defined on $(-\e,\e)\times Z$, it can be
prolonged onto $\Bbb R\times Z$, and is
called a one-parameter group of diffeomorphisms of $Z$. 
Any local one-parameter group of  diffeomorphisms
$G$ on $U\subset Z$ defines a local vector field $u$ on $U$ by setting $u(z)$
to be the tangent vector to the curve
$z(t)=G_t(z)$ at $t=0$. Conversely, if $u$ is a vector field on a
manifold $Z$, there exists a unique local one-parameter
group $G_u$ of  diffeomorphisms on a neighbourhood of every point
$z\in Z$ which defines $u$. We will call $G_u$ a flow 
of the vector field $u$. A vector field $u$ on a
manifold $Z$ is called complete if its flow is a one-parameter group of
diffeomorphisms of $Z$.
For instance, every vector field on a compact manifold is complete \cite{kob}.

A vector field $u$
on a fibre bundle $Y\to X$ is an
infinitesimal generator of a local one-parameter
group $G_u$ of isomorphisms of $Y\to X$ if and only if it is a
projectable vector field on $Y$.
A vector field $u$
on a fibre bundle $Y\to X$ is called projectable
if it projects onto a vector field
on $X$, i.e., there exists a vector field $\tau$ on $X$ which makes up
the commutative diagram
\be
\begin{array}{rcccl}
& Y &\ar^u & TY&\\
_\pi& \put(0,10){\vector(0,-1){20}} & & \put(0,10){\vector(0,-1){20}}&_{T\pi}\\
& X &\ar^\tau & TX&
\end{array}, \qquad  \tau\circ\pi= T\pi\circ u.
\ee
A projectable vector field has the coordinate expression
\be
u=u^\la(x^\m) \dr_\la + u^i(x^\m,y^j) \dr_i, \qquad \tau=u^\la\dr_\la,
\ee
where $u^\la$ are local functions on $X$.
A projectable vector field
is said to be vertical if it
projects onto the zero vector field $\tau=0$ on $X$, i.e., $u=u^i\dr_i$
takes its values in the vertical tangent bundle $VY$.

In field theory, projectable vector fields on fibre bundles play
a role of infinitesimal generators of local
one-parameter groups of gauge transformations.

In general, a vector field $\tau=\tau^\la\dr_\la$ on a base $X$ of a
fibre bundle $Y\to X$ gives rise to a vector field on $Y$ by means
of a connection on this fibre bundle
(see the formula (\ref{b1.85}) below).
Nevertheless, every natural fibre bundle $Y\to X$
admits the canonical lift
$\wt \tau$ onto $Y$ of any vector field $\tau$ on $X$.
For instance, if $Y$ is the tensor bundle (\ref{sp20}), the above mentioned
canonical lift reads
\mar{l28}\beq
\wt\tau = \tau^\m\dr_\m + [\dr_\nu\tau^{\al_1}\dot
x^{\nu\al_2\cdots\al_m}_{\bt_1\cdots\bt_k} + \ldots
-\dr_{\bt_1}\tau^\nu \dot x^{\al_1\cdots\al_m}_{\nu\bt_2\cdots\bt_k}
-\ldots]\frac{\dr}{\dr \dot
x^{\al_1\cdots\al_m}_{\bt_1\cdots\bt_k}}. \label{l28}
\eeq
In particular, we have the
canonical lift
\mar{l27}\beq
\wt\tau = \tau^\m\dr_\m +\dr_\nu\tau^\al\dot x^\nu\frac{\dr}{\dr\dot x^\al}
\label{l27}
\eeq
onto the tangent bundle $TX$, and that
\mar{l27'}\beq
\wt\tau = \tau^\m\dr_\m -\dr_\bt\tau^\nu\dot x_\nu\frac{\dr}{\dr\dot x_\bt}
\label{l27'}
\eeq
onto the cotangent bundle $T^*X$. 

\glos{F. Exterior forms}

An exterior $r$-form on a manifold $Z$ is a
section
\be
\f =\frac{1}{r!}\f_{\la_1\dots\la_r}
dz^{\la_1}\w\cdots\w dz^{\la_r}
\ee
of the exterior product $\op\w^r T^*Z\to Z$. Let
$\cO^r(Z)$ denote the vector space of exterior $r$-forms on a manifold
$Z$. By definition, $\cO^0(Z)=C^\infty(Z)$ is the ring of smooth real
functions on $Z$.
All exterior forms on $Z$ constitute the ${\Bbb N}$-graded exterior
algebra $\cO^*(Z)$ of global sections of the exterior bundle $\w T^*Z$
(\ref{spr880}) with respect to the exterior product $\w$.
This algebra is provided with the
exterior differential
\be
&& d: \cO^r(Z) \to \cO^{r+1}(Z), \\
&& d\f= dz^\m\w \dr_\m\f=\frac{1}{r!}
\dr_\m\f_{\la_1\ldots\la_r} dz^\m\w dz^{\la_1}\w\cdots dz^{\la_r},
\ee
which is nilpotent, i.e., $d\circ
d=0$, and obeys the relation
\be
d(\f\w\si)= d(\f)\w \si +(-1)^{\nm\f}\f\w d(\si).
\ee
The symbol $|\f|$ stands for the form degree.

Given a manifold morphism $f:Z\to Z'$, any exterior $k$-form
$\f$ on $Z'$ yields the pull-back exterior form $f^*\f$ on $Z$ by the
condition
\be
f^*\f(v^1,\ldots,v^k)(z) = \f(Tf(v^1),\ldots,Tf(v^k))(f(z))
\ee
for an arbitrary collection of tangent vectors $v^1,\cdots, v^k\in T_zZ$.
The following relations hold:
\be
f^*(\f\w\si) =f^*\f\w f^*\si, \qquad  df^*\f =f^*(d\f).
\ee

In particular, given a fibre bundle
$\pi:Y\to X$, the pull-back onto $Y$ of exterior forms on $X$ by $\pi$
provides the
monomorphism of exterior algebras
\be
\pi^*:\cO^*(X)\to \cO^*(Y).
\ee
Elements of its image $\pi^*\cO^*(X)$ are called basic forms.
Exterior forms
on $Y$ such that $u\rfloor\f=0$ for an arbitrary vertical
vector field $u$ on $Y$ are said to be
horizontal forms. They are generated by
horizontal one-forms $\{dx^\la\}$. For instance, basic forms are
horizontal forms with coefficients in $C^\infty(X)\subset C^\infty(Y)$.
A horizontal form of degree $n=\dim X$ is called a density.
For instance, Lagrangians in field theory are densities.
We will use the notation
\mar{gm141}\beq
\om=dx^1\w\cdots\w dx^n, \qquad \om_\la=\dr_\la\rfloor\om,
\qquad \om_{\m\la}=\dr_\m\rfloor\dr_\la\rfloor\om. \label{gm141}
\eeq

The interior product (or contraction) of a vector field
$u = u^\m\dr_\m$ and an exterior $r$-form $\f$ on a manifold $Z$ is
given by the coordinate expression
\mar{d000}\ben
&&u\rfloor\f = \op\sum_{k=1}^r \frac{(-1)^{k-1}}{r!} u^{\la_k}
\f_{\la_1\ldots\la_k\ldots\la_r}
dz^{\la_1}\w\cdots\w\wh {dz}^{\la_k}\w\cdots \w dz^{\la_r} = \label{d000}\\
&& \qquad \frac{1}{(r-1)!}u^\m\f_{\m\al_2\ldots\al_r} dz^{\al_2}\w\cdots\w
dz^{\al_r}, \nonumber
\een
where the caret $\,\wh{}\,$ denotes omission.
The following relations hold:
\mar{gm11}\ben
&& \f(u_1,\ldots,u_r)=u_r\rfloor\cdots u_1\rfloor\f,\label{gm11}\\
&& u\rfloor(\f\w\si)= u\rfloor\f\w\si +(-1)^{\nm\f}\f\w u\rfloor\si,
\label{gm12}\\
&& [u,u']\rfloor\f=u\rfloor d(u'\rfloor\f)- u'\rfloor d(u\rfloor\f)
-u'\rfloor u\rfloor d\f, \qquad \f\in\cO^1(Z). \label{d1}
\een\mar{gm12,d1}

The Lie derivative of an exterior form
$\f$ along a vector field $u$ is defined as
\be
\bL_u\f = u\rfloor d\f +d(u\rfloor\f),
\ee
and fulfils the relation
\be
\bL_u(\f\w\si)= \bL_u\f\w\si +\f\w\bL_u\si.
\ee
In particular, if $f$ is a function, then
\be
\bL_u f =u(f)=u\rfloor d f.
\ee
It is important for physical applications that
an exterior form $\f$
is invariant under a local one-parameter group of  diffeomorphisms $G_t$
of $Z$ (i.e., $G_t^*\f=\f$) if and only if its
Lie derivative $\bL_u\f$ along the vector field $u$, generating 
$G_t$, vanishes.

\glos{G. Tangent-valued forms}

A tangent-valued $r$-form on a
manifold $Z$ is a section
\mar{spr611}\beq
\phi = \frac{1}{r!}\phi_{\la_1\ldots\la_r}^\m dz^{\la_1}\w\cdots\w
dz^{\la_r}\ot\dr_\m \label{spr611}
\eeq
of the tensor bundle $\op\w^r T^*Z\ot TZ\to Z$.
Tangent-valued forms
play a prominent role in jet formalism and theory of connections on
fibre bundles.

In particular,
there is one-to-one correspondence between the tangent-valued one-forms
$\f$ on a manifold $Z$ and the linear bundle endomorphisms
\mar{29b,b'}\ben
&& \wh\f:TZ\to TZ,\qquad
\wh\f: T_zZ\ni v\mapsto v\rfloor\f(z)\in T_zZ, \label{29b} \\
&&\wh\f^*:T^*Z\to T^*Z,\qquad
\wh\f^*: T_z^*Z\ni v^*\mapsto \f(z)\rfloor v^*\in T_z^*Z, \label{29b'}
\een
over $Z$.
For instance, the canonical tangent-valued one-form
\mar{b1.51}\beq
\th_Z= dz^\la\ot \dr_\la \label{b1.51}
\eeq
on $Z$ corresponds to the identity morphisms (\ref{29b}) and (\ref{29b'}).

The space $\cO^*(Z)\ox \cT_1(Z)$ of
tangent-valued forms is provided with the Fr\"olicher--Nijenhuis
bracket
\mar{1149}\ben
&& [\; ,\; ]_{\rm FN}:\cO^r(Z)\ot \cT_1(Z)\times \cO^s(Z)\ot \cT_1(Z)
\to\cO^{r+s}(Z)\ot \cT_1(Z), \nonumber \\
&& [\phi,\si]_{\rm FN} = \frac{1}{r!s!}(\phi_{\la_1
\dots\la_r}^\nu\dr_\n\si_{\la_{r+1}\dots\la_{r+s}}^\m - \si_{\la_{r+1}
\dots\la_{r+s}}^\nu\dr_\nu\phi_{\la_1\dots\la_r}^\m - \label{1149} \\
&&  r\phi_{\la_1\ldots\la_{r-1}\nu}^\m\dr_{\la_r}\si_{\la_{r+1}
\dots\la_{r+s}}^\nu + s \si_{\nu\la_{r+2}\ldots\la_{r+s}}^\m
\dr_{\la_{r+1}}\phi_{\la_1\ldots\la_r}^\nu)dz^{\la_1}\wedge\cdots
\wedge dz^{\la_{r+s}}\otimes\dr_\m.\nonumber
\een
The following relations hold:
\mar{1150,'}\ben
&& [\f,\s]_{\rm FN}=(-1)^{\nm\f\nm\s+1}[\s,\f]_{\rm FN}, \label{1150} \\
&& [\f, [\s, \th]_{\rm FN}]_{\rm FN} = [[\f, \s]_{\rm FN}, \th]_{\rm
FN} +(-1)^{\nm\f\nm\s}  [\s, [\f,\th]_{\rm FN}]_{\rm FN}. \label{1150'}
\een

Given a tangent-valued form  $\th$, the Nijenhuis
differential on $\cO^*(Z)\ot\cT_1(Z)$ 
along $\th$ is defined as
\mar{spr282}\beq
d_\th\si = [\th,\si]_{\rm FN}. \label{spr282}
\eeq
By virtue of the relation (\ref{1150'}), it has the property
\be
d_\f[\psi,\th]_{\rm FN} = [d_\phi\psi,\th]_{\rm FN}+
(-1)^{\mid\f\mid\mid\psi\mid} [\psi,d_\f\th]_{\rm FN}.
\ee
In particular, if $\th=u$ is a vector field, the Nijenhuis
differential is the  Lie derivative of tangent-valued
forms
\mar{1215}\ben
&& \bL_u\si= d_u\si=[u,\si]_{\rm FN} =(u^\n\dr_\n\si_{\la_1\ldots\la_s}^\m -
\si_{\la_1\dots\la_s}^\n\dr_\n u^\m + \label{1215}\\
&& \qquad s\si^\m_{\nu\la_2\ldots\la_s}\dr_{\la_1}u^\nu)dx^{\la_1}
\wedge\cdots\wedge dx^{\la_s}\otimes\dr_\m, \qquad \si\in\cO^s(M)\ot\cT(M).
\nonumber
\een

Let $Y\to X$ be a fibre bundle. In the sequel, we will deal with
the following classes of tangent-valued forms on $Y$:
\begin{itemize}
\item  tangent-valued horizontal forms
\be
&& \phi : Y\to\op\w^r T^*X\op\otimes_Y TY,\\
&& \phi =\frac{1}{r!}dx^{\la_1}\wedge\dots\wedge dx^{\la_r}\otimes
[\phi_{\la_1\dots\la_r}^\m(y) \dr_\m +\phi_{\la_1\dots\la_r}^i(y) \dr_i];
\ee
\item vertical-valued horizontal forms
\be
&&\phi : Y\to\op\w^r T^*X\op\otimes_Y VY,\\
&&\phi =\frac{1}{r!}\phi_{\la_1\dots\la_r}^i(y)dx^{\la_1}\wedge\dots
\wedge dx^{\la_r}\otimes\dr_i;
\ee
\item vertical-valued horizontal one-forms,
called soldering forms, 
\mar{1290}\beq
\si = \si_\la^i(y) dx^\la\otimes\dr_i; \label{1290}
\eeq
\item  basic vertical-valued horizontal forms 
\be
\phi =\frac{1}{r!}\phi_{\la_1\dots\la_r}^i(x)dx^{\la_1}\wedge\dots
\wedge dx^{\la_r}\otimes\dr_i
\ee
on an affine bundle which are constant along
its fibres.
\end{itemize}

Any tangent valued form $\f$ (\ref{spr611}) on a manifold $Z$ defines
the vertical-valued form
\be
\phi = \frac{1}{r!}\phi_{\la_1\ldots\la_r}^\m dz^{\la_1}\w\cdots\w
dz^{\la_r}\ot\dot\dr_\m, \qquad \dot\dr_\m=\frac{\dr}{\dr\dot z^\m},
\ee
on the tangent bundle $TZ$. For instance, the canonical tangent-valued
form $\th_Z$ (\ref{b1.51}) on a manifold $Z$ yields the canonical
vertical-valued form
\mar{z117'}\beq
\dot\th_Z= dz^\la\ot \dot\dr_\la \label{z117'}
\eeq
on the tangent bundle $TZ$. By this reason, tangent-valued one-forms on a
manifold $Z$ are also called soldering forms.

\section{Jet manifolds}

Jet manifolds provide the standard language for theory of
(non-linear) differential operators, the calculus of variations, Lagrangian
and Hamiltonian formalisms \cite{gol,book,vinb,palais}. 
Here, we restrict our consideration to the notion of
jets of sections of fibre bundles.

\glos{A. First order jet manifolds}

Given a fibre bundle $Y\to X$ with bundle coordinates $(x^\la,y^i)$,
let us consider the equivalence classes
$j^1_xs$ of its sections $s$, which are identified by their values
$s^i(x)$ and the values of their first order derivatives
$\dr_\mu s^i(x)$ at a point $x\in X$. They are
called the first order jets of sections
at $x$. One can justify that the definition of jets is
coordinate-independent.  The key point is that the set  $J^1Y$ of 
first order jets
$j^1_xs$, $x\in X$, is a
smooth manifold with respect to the adapted coordinates
$(x^\la,y^i,y_\la^i)$ such that
\mar{50}\ben
&& y_\la^i(j^1_xs)=\dr_\la s^i(x),\nonumber\\
&&{y'}^i_\la = \frac{\dr x^\m}{\dr{x'}^\la}(\dr_\m
+y^j_\m\dr_j)y'^i.\label{50}
\een
It is called the first order jet manifold of
the fibre bundle $Y\to X$.

The jet manifold $J^1Y$ admits the natural fibrations
\mar{1.14,5}\ben
&&\pi^1:J^1Y\ni j^1_xs\mapsto x\in X, \label{1.14}\\
&&\pi^1_0:J^1Y\ni j^1_xs\mapsto s(x)\in Y. \label{1.15}
\een
A glance at the transformation law (\ref{50}) shows that $\pi^1_0$
is an affine bundle modelled over the vector
bundle
\mar{cc9}\beq
T^*X \op\otimes_Y VY\to Y.\label{cc9}
\eeq
It is convenient to call $\pi^1$ (\ref{1.14})
the jet bundle, while
$\pi^1_0$ (\ref{1.15}) is said to be 
the affine jet bundle. 

Let us note that, if $Y\to X$ is a vector or an affine bundle, the
jet bundle $\pi_1$ (\ref{1.14}) is so.

Jets can be expressed in terms of familiar tangent-valued forms as follows.
There are the canonical imbeddings
\mar{18,24}\ben
&&\la_1:J^1Y\op\hookrightarrow_Y
T^*X \op\otimes_Y TY,\qquad \la_1=dx^\la
\otimes (\dr_\la + y^i_\la \dr_i)=dx^\la\otimes d_\la, \label{18}\\
&&\th_1:J^1Y \hookrightarrow T^*Y\op\otimes_Y VY,\qquad
\th_1=(dy^i- y^i_\la dx^\la)\otimes
\dr_i=\th^i \otimes \dr_i,\label{24}
\een
where $d_\la$ are said to be total derivatives, and
$\th^i$ are called  contact forms. 
Identifying the jet manifold
$J^1Y$ to its images
under the canonical morphisms (\ref{18}) and (\ref{24}),
one can represent jets $j^1_xs=(x^\la,y^i,y^i_\m)$ by tangent-valued forms
\mar{cc4}\beq
dx^\la \otimes (\dr_\la + y^i_\la \dr_i) \quad {\rm and} \quad
(dy^i- y^i_\la dx^\la)\otimes\dr_i. \label{cc4}
\eeq

Sections and morphisms of fibre bundles admit prolongations to jet
manifolds as follows.

Any section $s$ of a fibre bundle $Y\to X$ has the jet prolongation
to the section
\be
(J^1s)(x)\op =^\df j_x^1s, \qquad
y_\la^i\circ J^1s= \dr_\la s^i(x),
\ee
of the jet bundle $J^1Y\to X$. A section
$\ol s$ of the jet bundle $J^1Y\to X$ is called  holonomic
or integrable
if it is the jet prolongation of some section
of the fibre bundle $Y\to X$.

Any bundle morphism $\Phi:Y\to Y'$ over a diffeomorphism $f$
admits a  jet prolongation
to a bundle morphism over $\Phi$ of affine jet bundles
\be
J^1\Phi : J^1Y \ar_\Phi J^1Y',\qquad {y'}^i_\la\circ
J^1\Phi=\frac{\dr(f^{-1})^\m}{\dr x'^\la}d_\m\Phi^i.
\ee

Any projectable vector field $u = u^\la\dr_\la + u^i\dr_i$
on a fibre bundle $Y\to X$ has a jet prolongation
to the projectable vector field
\mar{1.21}\ben
&&J^1u =r_1\circ J^1u: J^1Y\to J^1TY\to TJ^1Y,\nonumber \\
&& J^1u =u^\la\dr_\la + u^i\dr_i + (d_\la u^i
- y_\m^i\dr_\la u^\m)\dr_i^\la, \label{1.21}
\een
on the jet manifold $J^1Y$.
In order to obtain (\ref{1.21}), the canonical
bundle morphism
\be
  r_1: J^1TY\to TJ^1Y,\qquad
\dot y^i_\la\circ r_1 = (\dot y^i)_\la-y^i_\m\dot x^\m_\la
\ee
is used.
In particular, there is the canonical isomorphism
\mar{d020}\beq
VJ^1Y=J^1VY, \qquad \dot y^i_\la=(\dot y^i)_\la.\label{d020}
\eeq

\glos{B. Second order jet manifolds}

Taking the first order jet manifold  of the jet bundle
$J^1Y\to X$, we obtain the repeated jet manifold $J^1J^1Y$
provided with the adapted coordinates
$(x^\la ,y^i,y^i_\la ,\wh y_\m^i,y^i_{\m\la})$, with transition functions
\be
\wh y'^i_\la = \frac{\dr x^\al}{\dr{x'}^\la} d_\al y'^i, \qquad
{y'}_{\m\la}^i= \frac{\dr x^\al}{\dr{x'}^\m}d_\al {y'}^i_\la, \qquad
d_\al = \dr_\al +\wh y^j_\al\dr_j
+y^j_{\nu\al}\dr^\nu_j.
\ee

There exist two different affine fibrations of $J^1J^1Y$ over $J^1Y$:
\begin{itemize}
\item the familiar affine jet bundle (\ref{1.15}):
\mar{gm213}\beq
\pi_{11}:J^1J^1Y\to J^1Y, \qquad y_\la^i\circ\p_{11} = y_\la^i, \label{gm213}
\eeq
\item and the affine bundle
\mar{gm214}\beq
J^1\pi^1_0:J^1J^1Y\to J^1Y,\qquad y_\la^i\circ J^1\pi_0^1 = \wh y_\la^i.
\label{gm214}
\eeq
\end{itemize}
In general, there is no canonical identification of these fibrations.
The points $q\in J^1J^1Y$, where $\pi_{11}(q)=J^1\pi^1_0(q)$,
form the affine subbundle $\wh J^2Y\to J^1Y$ of $J^1J^1Y$ called the
sesquiholonomic jet manifold. It
is given by the coordinate conditions $\wh y^i_\la= y^i_\la$,
and is coordinated  by  $(x^\la ,y^i, y^i_\la,y^i_{\m\la})$.

The second order jet manifold
$J^2Y$ of a fibre bundle
$Y\to X$ can be defined as the affine subbundle
of the fibre bundle
$\wh J^2Y\to J^1Y$ given by the
coordinate conditions
$y^i_{\la\m}=y^i_{\m\la}$. It is coordinated by
$(x^\la ,y^i, y^i_\la,y^i_{\la\m}=y^i_{\m\la})$.
The second order jet manifold $J^2Y$  can also be introduced
as the set of the
equivalence classes $j_x^2s$ of sections $s$ of the fibre bundle $Y\to X$,
which are identified by their values and the values of their first and
second order partial derivatives at points $x\in X$, i.e.,
\be
y^i_\la (j_x^2s)=\dr_\la s^i(x),\qquad
y^i_{\la\m}(j_x^2s)=\dr_\la\dr_\m s^i(x).
\ee

Let $s$ be a section of a fibre bundle $Y\to X$, and let $J^1s$ be its jet
prolongation to a section of the jet bundle $J^1Y\to X$. The latter
gives rise to the section $J^1J^1s$ of the repeated jet bundle
$J^1J^1Y\to X$. This section takes its values into the second order jet
manifold
$J^2Y$. It is called the second order jet prolongation of the section
$s$, and is denoted by
$J^2s$.

\begin{prop}\label{1.5.3} \mar{1.5.3} Let $\ol s$ be a section of the
jet bundle
$J^1Y\to X$, and let $J^1\ol s$ be its jet prolongation to
the section of the repeated
jet bundle $J^1J^1Y\to X$. The following three facts are equivalent:
(i) $\ol s=J^1s$ where $s$ is a section of the fibre bundle $Y\to X$,
(ii) $J^1\ol s$ takes its values into $\wh J^2Y$,
(iii) $J^1\ol s$ takes its values into $J^2Y$.
\end{prop}

\glos{C. Higher order jet manifolds}

The notion of first and second order jet manifolds is naturally
extended to higher order jets (see Lecture 6 for a detailed
exposition).
The $r$-order jet manifold  $J^rY$  of
a fibre bundle $Y\to X$ 
is defined as the disjoint  union
of the equivalence classes $j^r_xs$ of sections $s$ of $Y\to X$
identified by the
$r+1$ terms of their Taylor series at points of $X$.
It is a smooth manifold endowed with the adapted coordinates
$(x^\la, y^i_\La)$, $0\leq\nm\La \leq r$, where $\La=(\la_k\ldots\la_1)$
denotes a multi-index modulo permutations and
\be
  y^i_{\la_k\cdots\la_1}(j^r_xs)= \dr_{\la_k}\cdots \dr_{\la_1}s^i(x),
\qquad 0\leq k\leq r.
\ee
The transformation law of these coordinates  reads
\mar{55.21}\beq
{y'}^i_{\la+\La}=\frac{\dr x^\m}{\dr'x^\la}d_\m y'^i_\La, \label{55.21}
\eeq
where $\la+\La=(\la\la_k\ldots\la_1)$ and
\be
d_\la = \dr_\la + \op\sum_{|\La|< r} y^i_{\la+\La}\dr_i^\La
=\dr_\la  +y^i_\la\dr_i +y^i_{\la\m}\dr_i^\m  +\cdots
\ee
are higher order total derivatives. 
These derivatives act on exterior forms on $J^rY$ and
obey the relations
\be
d_\la(\f\w\si)=d_\la(\f)\w\si +\f\w d_\la(\si),\qquad
d_\la(d\f)=d(d_\la(\f)).
\ee
For instance,
\be
d_\la(dx^\m)=0, \qquad d_\la(dy^i_\La)=
dy^i_{\la+\La}.
\ee

Let us also mention the following two operations:
  the horizontal projection $h_0$ given
by the relations
\mar{mos40}\beq
h_0(dx^\la)=dx^\la, \qquad h_0(dy^i_{\la_k\cdots\la_1})=
y^i_{\m\la_k\ldots\la_1}dx^\m,
\label{mos40}
\eeq
and the horizontal differential
\mar{mos41}\ben
&& d_H(\f)=dx^\la\w d_\la(\f), \label{mos41} \\
&& d_H\circ d_H=0,\qquad h_0\circ d=d_H\circ h_0. \nonumber
\een

\glos{D. Differential equations and differential operators}

Let us now formulate the notions of a (non-linear)
differential equation and a differential
operator in terms of jets.

\begin{defi} \label{sp30} \mar{sp30}
A $k$-order differential equation
on a fibre bundle $Y\to X$ is defined as a closed subbundle $\gE$ of the jet
bundle $J^kY\to X$.
Its classical solution is a (local) section
$s$ of $Y\to X$ whose
$k$-order jet prolongation $J^ks$ lives in $\gE$.
\end{defi}

One usually considers differential equations associated to differential
operators.

\begin{defi} \label{sp31} \mar{sp31}
Let $E\to X$ be a
vector bundle coordinated by $(x^\la,v^A)$, $A=1,\ldots,m$.
A bundle morphism
\mar{gm2}\beq
\cE: J^kY\op\to_X E,  \qquad
v^A\circ \cE= \cE^A(x^\la, y^i, y^i_\la,\ldots, y^i_{\la_k\cdots \la_1}),
\label{gm2}
\eeq
is called a $k$-order differential operator
on a fibre bundle $Y\to X$.
It sends each section
$s$ of $Y\to X$ onto the section
\be
(\cE\circ J^ks)^A(x)= \cE^A(x^\la, s^i(x), \dr_\la s^i(x),\ldots,
\dr_{\la_k}\cdots \dr_{\la_1}s^i(x))
\ee
  of the vector
bundle $E\to X$.
\end{defi}

Let us suppose that the canonical zero section $\wh 0(X)$ of the vector bundle
$E\to X$ belongs to the image $\cE(J^kY)$.
Then the kernel
operator of a differential operator $\cE$ is defined as
\mar{z60}\beq
\Ker\cE =\cE^{-1}(\wh 0(X))\subset J^kY. \label{z60}
\eeq
If $\Ker\cE$ (\ref{z60})
is a closed subbundle of the jet bundle $J^kY\to X$, it is a
$k$-order differential equation, associated to the differential
operator $\cE$. It is
written in the coordinate form
\be
\cE^A(x^\la, y^i, y^i_\la, \ldots,y^i_{\la_k\cdots \la_1}) =0, \qquad
A=1,\ldots,m.
\ee

\section{Connections on fibre bundles}

Connections play a prominent role in classical field theory
because they enable one to deal with invariantly defined
objects. Partial derivatives of sections of fibre bundles
(i.e., of classical fields) are
ill defined. One need connections in order to replace them with
covariant derivatives.
Gauge theory shows clearly that this is a basic physical principle.

We start from the traditional geometric notion of a connection
as a horizontal lift, but then
follow its equivalent definition as a jet field
\cite{book,kol,book00,sau}. It enables us to include
connections in an natural way in field
dynamics.

\glos{A. Connections as tangent-valued forms}

A connection on a fibre
bundle
$Y\to X$ is customarily defined as a linear bundle monomorphism
\mar{150}\beq
\G: Y\op\times_X TX\op\to_Y TY, \qquad
  \G: \dot x^\la\dr_\la \mapsto \dot x^\la(\dr_\la+\G^i_\la(y)\dr_i),
\label{150}
\eeq
which splits the
exact sequence (\ref{1.8a}), i.e.,
\be
\pi_T\circ \G=\id (Y\op\times_X TX).
\ee
The image $HY$ of $Y\op\times_X TX$ by a connection $\G$
is called the horizontal distribution. It splits the tangent bundle $TY$ as
\mar{152}\ben
&& TY=HY\op\oplus_Y VY, \label{152} \\
&& \dot x^\la \dr_\la + \dot y^i \dr_i = \dot
x^\la(\dr_\la + \G_\la^i \dr_i) + (\dot y^i - \dot
x^\la\G_\la^i)\dr_i.  \nonumber
\een
By virtue of Theorem \ref{sp11}, a connection on a fibre bundle always exists.

A connection $\G$ (\ref{150}) defines  the
horizontal tangent-valued one-form
\mar{154}\beq
\G = dx^\la\otimes (\dr_\la +
\G_\la^i\dr_i) \label{154}
\eeq
on $Y$ such that $\G(\dr_\la)= \dr_\la\rfloor\G$.
Conversely, every horizontal tangent-valued one-form on a fibre bundle
$Y\to X$
which projects onto the canonical tangent-valued form $\th_X$
(\ref{b1.51}) on $X$ defines a connection on $Y\to X$.

In an equivalent way, the horizontal splitting (\ref{152}) is given
by the vertical-valued form
\mar{b1.223}\beq
\G= (dy^i -\G^i_\la dx^\la)\ot\dr_i, \label{b1.223}
\eeq
which determines the epimorphism
\be
\G: TY\ni \dot x^\la \dr_\la + \dot y^i \dr_i\to (\dot x^\la \dr_\la +
\dot y^i \dr_i)\rfloor \G=(\dot y^i - \dot
x^\la\G_\la^i)\dr_i\in VY.
\ee

Given a connection $\G$,
a vector field $u$ on a fibre bundle $Y\to X$ is called horizontal
if it lives in the horizontal
distribution $HY$, i.e.,
takes the form
\mar{mos167}\beq
u=u^\la(y)(\dr_\la +\G^i_\la(y)\dr_i). \label{mos167}
\eeq
Any
vector field $\tau$ on the base $X$ of a fibre bundle $Y\to X$ admits
the horizontal lift
\mar{b1.85}\beq
\G \tau=\tau\rfloor\G=\tau^\la(\dr_\la +\G^i_\la\dr_i) \label{b1.85}
\eeq
onto $Y$ by means of a connection $\G$ (\ref{154}) on $Y\to X$.

Given the splitting (\ref{150}), the dual splitting of the exact
sequence (\ref{1.8b}) is
\mar{cc3}\beq
\G: V^*Y\ni \ol dy^i\mapsto \G\rfloor \ol dy^i=dy^i-\G^i_\la dx^\la\in
T^*Y, \label{cc3}
\eeq
where $\G$ is the vertical-valued form (\ref{b1.223}).

\glos{B. Connections as jet fields}

There is one-to-one correspondence between
the connections on a fibre bundle
$Y\to X$ and the jet fields,
i.e.,  global sections
of the affine jet bundle $J^1Y\to Y$  \cite{book,sau}.
Indeed, given a global section $\G$ of $J^1Y\to Y$, the
tangent-valued form
\be
\la_1\circ \G= dx^\la\ot(\dr_\la +\G^i_\la \dr_i)
\ee
provides the
horizontal splitting (\ref{152}) of $TY$.
Accordingly, the vertical-valued form
\be
\th_1\circ \G=(dy^i-\G^i_\la dx^\la)\ot \dr_i
\ee
leads to the dual
splitting (\ref{cc3}).

It follows immediately from this definition that
connections on a fibre bundle $Y\to X$ constitute an affine space
modelled over
the vector space of soldering forms $\si$ (\ref{1290}).
One also
deduces from (\ref{50}) the coordinate transformation law
of connections
\be
\G'^i_\la = \frac{\dr x^\m}{\dr{x'}^\la}(\dr_\m
+\G^j_\m\dr_j)y'^i.
\ee

The following are two standard constructions of new connections from
old ones.
\begin{itemize}
\item
Let $Y$ and $Y'$ be fibre bundles over the same base $X$.
Given a connection $\G$ on $\Y$ and a connection $\G'$ on $Y'\to
X$, the fibred product $Y\op\times_X Y'$  is provided with the 
product connection
\mar{b1.96}\beq
\G\times\G' = dx^\la\ot(\dr_\la +\G^i_\la\frac{\dr}{\dr y^i} +
\G'^j_\la\frac{\dr}{\dr y'^j}). \label{b1.96}
\eeq
\item Given a fibre bundle $Y\to X$, let $f:X'\to X$ be a manifold
morphism and $f^*Y$
the pull-back of $Y$ over $X'$. Any connection $\G$ (\ref{b1.223})
on $Y\to X$ yields the
pull-back connection
\mar{mos82}\beq
f^*\G=(dy^i-\G^i_\la(f^\m(x'^\nu),y^j)\frac{\dr f^\la}{\dr
x'^\m}dx'^\m)\ot\dr_i
\label{mos82}
\eeq
on the pull-back fibre bundle $f^*Y\to X'$.
\end{itemize}

The key point for physical applications lies in the fact that every connection
$\G$ on a fibre bundle $Y\to X$ yields
the first order
differential operator
\mar{2116}\ben
&& D^\G:J^1Y\op\to_Y T^*X\op\otimes_Y VY, \label{2116}\\
&& D^\G=\la_1- \G\circ \pi^1_0 =(y^i_\la -\G^i_\la)dx^\la\otimes\dr_i,
\nonumber
\een
called the covariant differential relative to the connection $\G$. 
If $s:X\to Y$ is a section, one defines its
covariant differential
\mar{+190}\beq
\nabla^\G s \op=^\df D_\G\circ J^1s = (\dr_\la s^i - \G_\la^i\circ s)
dx^\la\ox \dr_i
\label{+190}
\eeq
and its covariant derivative
\mar{sp32}\beq
\nabla_\tau^\G s
=\tau\rfloor\nabla^\G s \label{sp32}
\eeq
  along a vector field $\tau$ on $X$.
  A (local) section $s$ of $Y\to X$ is said to be an 
integral section of a connection $\G$ 
(or parallel with respect to $\G$) if
$s$ obeys the  equivalent conditions
\mar{b1.86}\beq
\nabla^\G s=0 \quad {\rm or} \quad J^1s=\G\circ s. \label{b1.86}
\eeq
Furthermore, if $s:X\to Y$ is a global section, there
exists a connection
$\G$ such that $s$ is an integral section of $\G$. This connection is defined
as an extension of the local section $s(x)\mapsto J^1s(x)$ of the
affine jet bundle $J^1Y\to Y$ over the closed imbedded submanifold $s(X)\subset
Y$ in accordance with Theorem \ref{mos9}.

\glos{C. Curvature and torsion}

Let $\G$ be a connection on a fibre bundle $Y\to X$.
Given vector fields $\tau$, $\tau'$ on $X$ and their horizontal lifts
$\G\tau$ and $\G\tau'$ (\ref{b1.85}) on $Y$, let us compute
the vertical vector field
\mar{160a,160}\ben
&& R(\tau,\tau')=\G [\tau,\tau'] - [\G \tau, \G \tau']= \tau^\la
\tau'^\m R_{\la\m}^i\dr_i, \label{160a} \\
&& R_{\la\m}^i = \dr_\la\G_\m^i - \dr_\m\G_\la^i +
\G_\la^j\dr_j \G_\m^i - \G_\m^j\dr_j \G_\la^i. \label{160}
\een
It can be seen as the contraction of vector fields $\tau$ and $\tau'$
with the vertical-valued
horizontal two-form
\mar{161'}\beq
R = \frac12 R_{\la\m}^i dx^\la\wedge dx^\m\otimes\dr_i \label{161'}
\eeq
on $Y$, called the curvature of the connection $\G$. 
In an equivalent way, the curvature (\ref{161'}) is defined as the
Nijenhuis differential
\mar{1178a}\beq
R=\frac{1}{2} d_\G\G=\frac{1}{2} [\G,\G]_{\rm FN}:Y\to \op\w^2T^*X\ox VY.
\label{1178a}
\eeq
Then we at once obtain from (\ref{1150}) -- (\ref{1150'}) the identities
\mar{1179,80}\ben
&& [R,R]_{\rm FN}=0, \label{1179}\\
&& d_\G R= [\G, R]_{\rm FN}=0. \label{1180}
\een

Given a soldering form $\si$ (\ref{1290}) on $Y\to X$,
one defines the soldered curvature
\mar{1186}\ben
&& \r =\frac{1}{2} d_\si \si = \frac{1}{2} [\si,\si]_{\rm FN}:Y\to
\op\w^2T^*X\ox VY, \label{1186} \\
&& \r =\frac{1}{2} \r_{\la\m}^i dx^\la\w dx^\m\ox \dr_i, \nonumber\\
&&
\r_{\la\m}^i =\si_\la^j\dr_j\si_\m^i - \si_\m^j\dr_j\si_\la^i,
\nonumber
\een
which fulfils the identities
\be
[\r,\r]_{\rm FN}=0,  \qquad d_\si \r =[\si,\r]_{\rm FN}=0.
\ee

Given a connection $\G$ and a soldering form $\si$, the 
torsion of $\G$ with respect to $\si$ is defined as
\mar{1190}\ben
&& T = d_\G \si = d_\si \G :Y\to \op\w^2 T^*X\ox VY, \nonumber\\
&& T = (\dr_\la\si_\m^i + \G_\la^j\dr_j\si_\m^i -
\dr_j\G_\la^i\si_\m^j) dx^\la\w dx^\m\ox \dr_i. \label{1190}
\een
In particular, if $\G' =\G + \si$, we have the important relations
\mar{1193a,}\ben
&& T' =T + 2\r, \label{1193a} \\
&& R' = R + \r +T. \label{1193}
\een

\glos{D. Linear connections}

Any vector bundle $Y\to X$
admits a linear connection.  
This is defined as a section of the affine jet bundle $J^Y\to Y$
which is a linear morphism of vector
bundles over $X$. A linear connection is given by the tangent-valued form
\mar{167}\beq
\G =dx^\la\ot(\dr_\la + \G_\la{}^i{}_j(x) y^j\dr_i). \label{167}
\eeq

There are the following standard constructions of new linear
connections from old ones.
\begin{itemize}
\item  Let $Y\to X$ be a vector bundle, coordinated by $(x^\la,y^i)$, and
$Y^*\to X$ its dual, coordinated by $(x^\la,y_i)$. Any
linear connection $\G$ (\ref{167}) on the vector bundle
$Y\to X$ defines the dual linear connection
\mar{spr300}\beq
\G^*=dx^\la\ot(\dr_\la - \G_\la{}^j{}_i(x) y_j\dr^i) \label{spr300}
\eeq
on $Y^*\to X$.
\item Let $\G$ and $\G'$ be, respectively, linear connections on vector
bundles
$Y\to X$ and $Y'\to X$ over the same base $X$. The
direct sum connection
$\G\oplus\G'$ on the Whitney sum $Y\oplus
Y'$ of these vector bundles is defined as the product
connection (\ref{b1.96}).
\item Let $Y$ coordinated by $(x^\la,y^i)$ and $Y'$ coordinated by
$(x^\la,y^a)$ be vector bundles
over the same base $X$. Their tensor product $Y\otimes Y'$ is endowed with
the bundle coordinates $(x^\la,y^{ia})$. Any linear connections
$\G$ and $\G'$ on $Y\to X$ and $Y'\to X$ define the linear
tensor product
connection
\mar{b1.92}\beq
\G\otimes\G'=dx^\la\ot[\dr_\la +(\G_\la{}^i{}_j y^{ja}+\G'_\la{}^a{}_b
y^{ib}) \frac{\dr}{\dr y^{ia}}] \label{b1.92}
\eeq
on $Y\ot Y'\to X$.
\end{itemize}

The curvature of a linear connection $\G$ (\ref{167}) on a vector
bundle $Y\to X$ is usually written as a $Y$-valued two-form
\mar{mos4}\ben
&&R=\frac12 R_{\la\m}{}^i{}_j(x)y^j dx^\la\w dx^\m\ot e_i,\nonumber\\
&&R_{\la\m}{}^i{}_j = \dr_\la \G_\m{}^i{}_j - \dr_\m
\G_\la{}^i{}_j + \G_\la{}^h{}_j \G_\m{}^i{}_h - \G_\m{}^h{}_j
\G_\la{}^i{}_h, \label{mos4}
\een
due to the canonical vertical splitting (\ref{1.10}), where
$\{\dr_i\}=\{e_i\}$.
For any two vector fields
$\tau$ and $\tau'$ on $X$, this curvature yields the 0-order
differential operator
\mar{+98}\beq
R(\tau,\tau')\circ
s=(\nabla_{[\tau,\tau']}^\G -[\nabla_\tau^\G,\nabla_{\tau'}^\G])s \label{+98}
\eeq
on section $s$ of the vector bundle $Y\to X$.

\glos{E. World connections}

An important example of linear connections is a  connection
\mar{B}\beq
K= dx^\la\otimes (\dr_\la +K_\la{}^\m{}_\n \dot x^\n
\dot\dr_\m) \label{B}
\eeq
on
the tangent bundle $TX$ of a manifold $X$. It is called a world
connection or, simply, a connection
on a manifold  $X$. 
The dual connection (\ref{spr300}) on the cotangent bundle
$T^*X$ is
\mar{C}\beq
K^*= dx^\la\otimes (\dr_\la -K_\la{}^\m{}_\n\dot x_\m
\dot\dr^\n). \label{C}
\eeq
Then, using the tensor product connection
(\ref{b1.92}), one can introduce the corresponding linear connection on
an arbitrary tensor bundle (\ref{sp20}).

A world
connection (\ref{B}) is called symmetric if
$K_\m{}^\n{}_\la = K_\la{}^\n{}_\m$. Of course, this property
is coordinate-independent. Let us note that, given a world
connection $K$ (\ref{B}), the tangent-valued form
\mar{spr833}\beq
K_r= dx^\la\otimes (\dr_\la +(r K_\la{}^\m{}_\n +
(1-r) K_\n{}^\m{}_\la)\dot x^\n
\dot\dr_\m), \qquad 0\leq r\leq 1, \label{spr833}
\eeq
is also a world connection. For instance, $K_{1/2}$ is a symmetric
connection, called the symmetric part of the connection $K$.

\begin{rem} \label{mos7} \mar{mos7}
  It should be emphasized that the expressions (\ref{B}) --
(\ref{C}) for a world connection
  differ in a minus sign
from those usually used in the physics literature.
\end{rem}

Due to the canonical vertical splitting
\mar{spr610}\beq
VTX\cong TX\op\times TX, \label{spr610}
\eeq
the curvature of a world connection $K$
(\ref{B}) on the
tangent bundle $TX$ can be written as the $TX$-valued two-form (\ref{mos4})
on $X$:
\mar{1203}\ben
&& R=\frac12R_{\la\m}{}^\al{}_\bt\dot x^\bt dx^\la\w dx^\m\ot\dr_\al,
\nonumber\\
&& R_{\la\m}{}^\al{}_\bt = \dr_\la K_\m{}^\al{}_\bt - \dr_\m
K_\la{}^\al{}_\bt + K_\la{}^\g{}_\bt K_\m{}^\al{}_\g -
K_\m{}^\g{}_\bt K_\la{}^\al{}_\g.  \label{1203}
\een
Its Ricci tensor
$R_{\la\bt}=R_{\la\m}{}^\m{}_\bt$ is introduced.

A torsion of a world connection is defined as the torsion
(\ref{1190}) of the connection $\G$ (\ref{B}) on the tangent bundle
$TX$ with respect to the canonical
vertical-valued form $\dot\th_X$ (\ref{z117'}). Due to the vertical
splitting (\ref{spr610}), it is also written as a tangent-valued
two-form
\mar{191}\ben
&& T =\frac12
T_\m{}^\n{}_\la  dx^\la\w dx^\m\ot\dr_\n, \label{191} \\
&& T_\m{}^\n{}_\la  = K_\m{}^\n{}_\la - K_\la{}^\n{}_\m,
\nonumber
\een
on $X$. A world connection is symmetric if and only if its torsion (\ref{191})
vanishes.

For instance, every manifold $X$ can be provided with a non-degenerate
fibre metric
\be
g\in\op\vee^2\cO^1(X), \qquad g=g_{\la\m}dx^\la\ot dx^\m,
\ee
in the tangent bundle $TX$, and with the dual metric
\be
g\in\op\vee^2\cT^1(X), \qquad g=g^{\la\m}\dr_\la\ot \dr_\m
\ee
in the cotangent bundle $T^*X$. It is called a world metric on $X$.
For any world metric $g$,
there exists a unique symmetric world connection
\mar{b1.400}\beq
K_\la{}^\n{}_\m = \{_\la{}^\n{}_\m\}=-\frac{1}{2}g^{\n\r}(\dr_\la g_{\r\m} +
\dr_\m g_{\r\la}-\dr_\r g_{\la\m}) \label{b1.400}
\eeq
such that $g$ is an integral section of $K$, i.e.
\be
\nabla_\la g^{\al\bt}=\dr_\la\, g^{\al \bt} -
g^{\al\g}\{_\la{}^\bt{}_\g\} - g^{\bt\g}\{_\la{}^\al{}_\g\}=0.
\ee
This is the Levi--Civita connection,
and its components (\ref{b1.400}) are called
Christoffel symbols. 

\glos{F. Affine connections}

Any affine bundle $Y\to X$ modelled over a vector bundle $\ol Y\to X$
admits an affine connection. 
This is defined as a section of the affine jet bundle $J^1Y\to Y$
which is an affine morphism of affine
bundles over $X$. An affine connection is given by the tangent-valued form
\mar{184}\beq
\G_\la^i=\G_\la{}^i{}_j(x) y^j + \si_\la ^i(x). \label{184}
\eeq
For any affine connection $\G:Y\to J^1Y$ (\ref{184}), the corresponding
linear derivative $\ol \G:\ol Y\to J^1\ol Y$ (\ref{1355'}) defines a unique
linear connection
\mar{mos032}\beq
\ol\G_\la^i=\G_\la{}^i{}_j(x) \ol y^j, \label{mos032}
\eeq
on the vector bundle $\ol Y\to X$,
where $(x^\la,\ol y^i)$ are the associated linear bundle coordinates on
$\ol Y$.

Of course, since every vector bundle has a natural structure of an
affine bundle,
any linear connection on a vector bundle is also an affine connection.

Affine connections on an affine bundle $Y\to X$ constitute an affine space
modelled over basic soldering forms on $Y\to X$. In view of the vertical
splitting (\ref{48'}), these soldering forms can be seen as global sections of
the vector bundle $T^*X\ot\ol Y\to X$. If $Y\to X$ is a vector bundle,
both the affine connection $\G$ (\ref{184}) and the associated linear
connection $\ol\G$ are connections on the same vector bundle $Y\to X$, and
their difference is also a  basic soldering form
on $Y$. Thus, every affine
connection on a vector bundle $Y\to X$ is the sum $\G=\ol\G+\si$ of a 
linear connection
$\ol G$ and a basic soldering form $\si$ on $Y\to X$.
Furthermore, let $R$ and $\ol R$ be the curvatures of
an affine connection $\G$ and the associated linear connection $\ol \G$,
respectively.  It is readily observed that $R = \ol R + T$,
where the $VY$-valued two-form
\mar{mos036}\ben
&& T=d_\G\si=d_\si\G :X\to \op\wedge^2 T^*X\op\otimes_X VY, \nonumber \\
&& T =\frac12 T_{\la
\m}^i dx^\la\wedge dx^\m\otimes \dr_i, \label{mos036} \\
&& T_{\la \m}^i = \dr_\la\si_\m^i - \dr_\m\si_\la^i +
\si_\la^h \G_\m{}^i{}_h - \si_\m^h \G_\la{}^i{}_h, \nonumber
\een
is the torsion (\ref{1190}) of the connection $\G$
with respect to the basic soldering form $\si$.

In particular, let us consider the tangent bundle $TX$ of a manifold
$X$ and the canonical soldering form $\si=\dot\th_X$ (\ref{z117'}) on
$TX$. Given an arbitrary
world connection $\G$ (\ref{B}) on $TX$, the corresponding
affine connection
\mar{b1.97}\beq
  A=\G +\th_X, \qquad
A_\la^\m=\G_\la{}^\m{}_\n \dot x^\n +\dl^\m_\la, \label{b1.97}
\eeq
on $TX$ is called the Cartan connection. 
Since the soldered curvature $\rho$ (\ref{1186}) of $\dot\th_X$ equals to zero,
the torsion (\ref{1193a}) of the Cartan connection coincides with the torsion
$T$ (\ref{191}) of the  world connection $\G$, while its curvature
(\ref{1193}) is the sum $R+T$ of the curvature and the torsion of $\G$.

\glos{G. Composite connections}

Let us consider a composite fibre bundle $Y\to\Si\to X$ (\ref{1.34}),
coordinated by $(x^\la, \si^m, y^i)$.
We aim studying the relations between connections
on fibre bundles $Y\to X$,
$Y\to\Si$ and $\Si\to X$.
These connections are given
respectively by the tangent-valued forms
\mar{spr290-2}\ben
&& \g=dx^\la\ot (\dr_\la +\g_\la^m\dr_m + \g_\la^i\dr_i), \label{spr290}\\
&&  A_\Si=dx^\la\ot (\dr_\la + A_\la^i\dr_i) +d\si^m\ot (\dr_m + A_m^i\dr_i),
\label{spr291}\\
&& \G=dx^\la\ot (\dr_\la + \G_\la^m\dr_m). \label{spr292}
\een

A connection $\g$ (\ref{spr290}) on the fibre bundle $Y\to X$ is said
to be projectable over a
connection $\G$ (\ref{spr292}) on the fibre bundle $\Si\to X$ if, for
any vector field $\tau$ on $X$, its horizontal lift $\g\tau$ on $Y$ by
means of the connection $\g$
is a projectable vector field over the horizontal lift $\G\tau$ of
$\tau$ on $\Si$ by means of the connection $\G$. This property takes place if
and only if $\g_\la^m=\G_\la^m$, i.e., components $\g_\la^m$ of the
connection $\g$ (\ref{spr290}) must be independent of the fibre
coordinates $y^i$.

A connection $A_\Si$ (\ref{spr291}) on the fibre bundle $Y\to\Si$ and
a connection $\G$ (\ref{spr292}) on the fibre bundle $\Si\to X$ define
a connection  on the composite fibre bundle $Y\to X$ as the composition
of bundle morphisms
\be
\g:Y\op\xx_XTX\ar^{(\id,\G)} Y\op\xx_\Si T\Si\ar^{A_\Si} TY.
\ee
This composite connection reads
\mar{b1.114}\beq
\g=dx^\la\ot (\dr_\la +\G_\la^m\dr_m + (A_\la^i + A_m^i\G_\la^m)\dr_i).
\label{b1.114}
\eeq
It is projectable over $\G$.  Moreover, this is a unique connection
such that the
horizontal lift $\g\tau$ on $Y$ of a
vector field $\tau$ on $X$
by means of the composite connection
$\g$ (\ref{b1.114}) coincides with the composition $A_\Si(\G\tau)$ of
horizontal lifts of $\tau$ on $\Si$ by means of the connection $\G$ and
then on $Y$ by means of the connection $A_\Si$.
For the sake of brevity, let us write
$\g=A_\Si\circ\G$.

Given a composite fibre bundle $Y$ (\ref{1.34}), there are the exact
sequences
\mar{63}\ben
&& 0\to V_\Si Y\hookrightarrow VY\to Y\op\times_\Si V\Si\to 0, \label{63a}\\
&& 0\to Y\op\times_\Si V^*\Si \hookrightarrow V^*Y\to V^*_\Si Y\to 0
\label{63b}
\een
of vector bundles over $Y$, where $V_\Si Y$ and $V_\Si^*Y$ are
the vertical tangent  and the vertical cotangent
bundles of the fibre bundle $Y\to\Si$ which are coordinated by
$(x^\la,\si^m,y^i, \dot y^i)$ and $(x^\la,\si^m,y^i, \dot y_i)$,
respectively. Let us consider a splitting of these exact sequences
\mar{63c,d}\ben
&& B: VY\ni \dot y^i\dr_i +\dot \si^m\dr_m \to (\dot y^i\dr_i +\dot
\si^m\dr_m)\rfloor B= \label{63c}\\
&&\qquad (\dot y^i -\dot \si^m B^i_m)\dr_i\in V_\Si Y,
\nonumber\\
&& B: V_\Si^*Y\ni \ol dy^i\to B\rfloor \ol dy^i= \ol dy^i- B^i_m\ol
d\si^m\in V^*Y, \label{63d}
\een
given by the form
\mar{mmm}\beq
B=(\ol dy^i-B^i_m\ol d\si^m)\ot\dr_i. \label{mmm}
\eeq
Then a connection $\g$ (\ref{spr290}) on $Y\to X$ and a splitting $B$
(\ref{63c}) define the connection
\mar{nnn}\ben
&& A_\Si=B\circ \g: TY\to VY \to V_\Si Y, \nonumber\\
&& A_\Si=dx^\la\ot(\dr_\la +(\g^i_\la - B^i_m\g^m_\la)\dr_i) +
d\si^m\ot (\dr_m +B^i_m\dr_i), \label{nnn}
\een
on the fibre bundle $Y\to\Si$.

Conversely, every connection
$A_\Si$ (\ref{spr291}) on the fibre bundle $Y\to\Si$
provides the splitting
\mar{46a}\beq
A_\Si: TY\supset VY \ni \dot y^i\dr_i + \dot\si^m\dr_m \to
(\dot y^i -A^i_m\dot\si^m)\dr_i \label{46a}
\eeq
of the exact sequence (\ref{63a}).
Using this splitting, one can construct the first order
differential operator
\mar{7.10}\beq
\wt D: J^1Y\to T^*X\op\otimes_Y V_\Si Y,
\qquad \wt D= dx^\la\otimes(y^i_\la-
A^i_\la -A^i_m\si^m_\la)\dr_i, \label{7.10}
\eeq
called the vertical covariant differential, 
on the composite fibre bundle $Y\to X$.
It can also be defined as the composition
\be
\wt D=\pr_1\circ D^\g: J^1Y\to T^*X\op\ot_YVY\to T^*X\op\otimes_Y VY_\Si,
\ee
where $D^\g$ is the covariant differential (\ref{2116}) relative to some
composite connection $A_\Si\circ\G$ (\ref{b1.114}), but $\wt D$ does
not depend on the choice of the connection $\G$ on the fibre bundle
$\Si\to X$.

The vertical covariant differential (\ref{7.10}) possesses the following
important property. Let $h$ be a section of the fibre bundle $\Si\to X$
and $Y_h\to X$ the restriction (\ref{S10}) of the fibre bundle
$Y\to\Si$ to $h(X)\subset \Si$.  This is a
subbundle $i_h:Y_h\hookrightarrow Y$ of the fibre bundle $Y\to X$.
Every connection $A_\Si$ (\ref{spr291}) induces the pull-back
connection
\mar{mos83}\beq
A_h=i_h^*A_\Si=dx^\la\ot[\dr_\la+((A^i_m\circ h)\dr_\la h^m
+(A\circ h)^i_\la)\dr_i] \label{mos83}
\eeq
on $Y_h\to X$.
Then the
restriction of the vertical covariant differential $\wt D$
(\ref{7.10}) to $J^1i_h(J^1Y_h)\subset J^1Y$ coincides with the
familiar covariant differential $D^{A_h}$ (\ref{2116}) on $Y_h$ relative
to the pull-back connection $A_h$ (\ref{mos83}).

\section{Lagrangian field theory}

Let us apply the above mathematical formalism to
formulation of Lagrangian field theory on fibre bundles.
Here, we restrict our consideration to first order Lagrangian
formalism since the most
contemporary field models are of this type.

The configuration
space
of first order Lagrangian field theory on a fibre bundle $Y\to X$,
coordinated by $(x^\la,y^i,y^i_\la)$,
is the
first order jet manifold $J^1Y$ of $Y\to X$, coordinated by
$(x^\la,y^i,y^i_\la)$. Accordingly, a first order Lagrangian
is defined as
a density
\mar{cc201}\beq
L=\cL(x^\la,y^i,y^i_\la)\om: J^1Y\to\op\w^nT^*X,
\qquad n=\dim X, \label{cc201}
\eeq
on $J^1Y$ (see the notation (\ref{gm141})).
Let us follow the
standard formulation of the variational problem on fibre bundles where
deformations of sections of a fibre bundle $Y\to X$ are induced by
local one-parameter groups of automorphisms of $Y\to X$ over $X$ (i.e.,
vertical gauge transformations).
Here, we will not study the calculus of variations in depth,
but apply in a straightforward manner the first variational formula.

Since a projectable vector field $u$ on a
fibre bundle $Y\to X$ is an infinitesimal
generator of a local one-parameter group of gauge transformations
of $Y\to X$, one can think of its jet prolongation $J^1u$ (\ref{1.21})
as being the infinitesimal generator of gauge transformations of the
configuration space $J^1Y$. Let
\mar{04}\beq
\bL_{J^1u}L=[\dr_\la u^\la\cL +(u^\la\dr_\la+
u^i\dr_i +(d_\la u^i -y^i_\m\dr_\la u^\m)\dr^\la_i)\cL]\om \label{04}
\eeq
be the Lie derivative of a Lagrangian $L$
along $J^1u$.
The first variational formula provides
its canonical decomposition in accordance
with the variational problem.
This decomposition reads
\mar{bC30'}\ben
&& \bL_{J^1u}L=
  u_V\rfloor \cE_L + d_Hh_0(u\rfloor H_L) \label{bC30'} \\
&& \qquad   =(u^i-y^i_\m u^\m )(\dr_i-d_\la \dr^\la_i)\cL\om -
d_\la[\pi^\la_i(u^\m y^i_\m -u^i) -u^\la\cL]\om, \nonumber
\een
where $u_V=(u\rfloor\th^i)\dr_i$,
\mar{305}\ben
&& \cE_L: J^2Y\to T^*Y\w(\op\w^nT^*X), \nonumber \\
&& \cE_L= (\dr_i\cL- d_\la\pi^\la_i) \th^i\w\om,
\qquad \pi^\la_i=\dr^\la_i\cL, \label{305}
\een
is the Euler--Lagrange operator
associated to the Lagrangian
$L$, and
\mar{N41,303}\ben
&& H_L: J^1Y\to Z_Y=T^*Y\w(\op\w^{n-1}T^*X), \label{N41}\\
&& H_L=L +\pi^\la_i\th^i\w\om_\la=\pi^\la_i dy^i\w\om_\la +(\cL- \pi^\la_i
y^i_\la)\om,
\label{303}
\een
is the Poincar\'e--Cartan form
(see the notation (\ref{gm141}), (\ref{mos40}) and (\ref{mos41})).

The kernel of the Euler-Lagrange operator $\cE_L$ (\ref{305})
defines the system of second order
Euler--Lagrange equations, 
given by the coordinate equalities
\mar{b327}\beq
(\dr_i- d_\la\dr^\la_i)\cL=0, \label{b327}
\eeq
A solution  of these equations is a section
$s$ of the fibre bundle $X\to Y$, whose second order jet prolongation $J^2s$
lives in (\ref{b327}), i.e.,
\mar{2.29}\beq
\dr_i\cL\circ s-(\dr_\la+\dr_\la s^j\dr_j
+\dr_\la\dr_\m s^j \dr^\m_j)\dr^\la_i\cL\circ s=0.\label{2.29}
\eeq

\begin{rem}\label{vareq} \mar{vareq} The kernel (\ref{b327})
of the Euler--Lagrange
operator $\cE_L$ need not be a closed subbundle of the second order jet bundle
$J^2Y\to X$. Therefore, it may happen that
the Euler--Lagrange
equations are not differential equations in a strict sense.
\end{rem}

\begin{rem}\label{mos10} \mar{mos10}
Different Lagrangians $L$ and $L'$ can lead to the same
Euler--Lagrange operator if their difference $L_0=L-L'$ is a 
variationally trivial Lagrangian, 
whose Euler--Lagrange operator vanishes identically.
A Lagrangian $L_0$ is variationally trivial if and only if
\mar{mos11}\beq
L_0=h_0(\vf) \label{mos11}
\eeq
where
$\vf$ is a closed $n$-form on $Y$ (see Lecture 8). We have at least locally
$\vf=d\xi$, and then
\be
L_0=h_0(d\xi)=d_H(h_0(\xi))=d_\la h_0(\xi)^\la\om, \qquad
h_0(\xi)=h_0(\xi)^\la\om_\la.
\ee
\end{rem}

The Poincar\'e--Cartan form
$H_L$ (\ref{303}) is a particular Lepagean equivalent
of a Lagrangian $L$
(i.e.,
$h_0(H_L)=L$). In contrast with other Lepagean forms,
it is a horizontal form on the affine jet bundle $J^1Y\to
Y$. The fibre bundle $Z_Y$ (\ref{N41}), called
the homogeneous Legendre bundle, 
is endowed with holonomic
coordinates $(x^\la,y^i,p^\la_i,p)$, possessing the transition functions
\mar{2.3}\beq
{p'}^\la_i = \det (\frac{\dr x^\ve}{\dr {x'}^\nu}) \frac{\dr
y^j}{\dr{y'}^i}
\frac{\dr {x'}^\la}{\dr x^\m}p^\m_j, \qquad
p'=\det (\frac{\dr x^\ve}{\dr {x'}^\nu})
(p-\frac{\dr y^j}{\dr {y'}^i}\frac{\dr {y'}^i}{\dr x^\mu}p^\mu_j).
\label{2.3}
\eeq
Relative to these coordinates, the morphism (\ref{N41}) reads
\be
(p^\m_i, p)\circ H_L =(\pi^\m_i, \cL-\pi^\m_i y^i_\m ).
\ee
A glance at the transition functions (\ref{2.3}) shows that $Z_Y$ is a
one-dimensional affine bundle
\mar{b418'}\beq
\pi_{Z\Pi}:Z_Y\to \Pi\label{b418'}
\eeq
over the Legendre bundle
\mar{00}\beq
\Pi=\op\w^nT^*X\op\ot_YV^*Y\op\ot_YTX= V^*Y\op\w_Y(\op\w^{n-1}_YT^*X),
\label{00}
\eeq
endowed with
holonomic coordinates $(x^\la,y^i,p^\la_i)$. Then the composition
\mar{m11}\beq
\wh L=\pi_{Z\Pi}\circ H_L:J^1Y \op\to_Y \Pi,
\qquad (x^\la, y^i,p^\la_i)\circ\wh L =(x^\la, y^i,\pi^\la_i), \label{m11}
\eeq
is the well-known Legendre map. 
One can think of $p^\la_i$ as being the covariant momenta
of field functions, and
the Legendre bundle $\Pi$ (\ref{00}) plays the role of a
finite-dimensional momentum phase space of fields in the covariant Hamiltonian
field theory \cite{book,jpa99,sard95}.

The first variational formula (\ref{bC30'}) provides the standard
procedure for the study of differential conservation laws in Lagrangian
field theory as follows.

Let $u$ be a projectable vector field on a fibre bundle $Y\to X$
treated as the infinitesimal  generator of a local one-parameter group $G_u$
of gauge
transformations.
On-shell, i.e., on
the kernel (\ref{b327}) of the Euler--Lagrange operator $\cE_L$,
the first variational formula (\ref{bC30'}) leads to the weak
identity
\mar{J4}\beq
\bL_{J^1u}L \ap -d_\la \gT^\la\om, \label{J4}
\eeq
where
\mar{Q30}\beq
\gT =\gT^\la\om_\la, \qquad
\gT^\la =\pi^\la_i(u^\m y^i_\m-u^i )-u^\la\cL, \label{Q30}
\eeq
is the symmetry current
along the vector field $u$.
Let a Lagrangian $L$ be invariant under the gauge group $G_u$. This 
implies that
the  Lie derivative $\bL_{J^1u}L$ (\ref{04})
vanishes. Then we obtain the
weak conservation law 
\mar{K4}\beq
d_\la\gT^\la\ap 0 \label{K4}
\eeq
of the symmetry current $\gT$ (\ref{Q30}).

\begin{rem}
It should be emphasized that,  the first variational formula defines the
symmetry current (\ref{Q30})
modulo the terms $d_\m(c^{\m\la}_i(y^i_\nu u^\nu - u^i))$,
where $c^{\m\la}_i$ are arbitrary skew-symmetric functions on $Y$
\cite{book}. Here, we set aside these boundary terms which are
independent of a Lagrangian.
\end{rem}

The weak conservation law (\ref{K4}) leads to the differential
conservation law
\mar{b3118'}\beq
\dr_\la(\gT^\la\circ s)=0 \label{b3118'}
\eeq
on solutions of the Euler--Lagrange equations (\ref{2.29}).
It implies the integral conservation
law
\mar{b3118}\beq
\op\int_{\dr N} s^*\gT =0, \label{b3118}
\eeq
where $N$ is a compact $n$-dimensional submanifold of $X$ with the boundary
$\dr N$.

\begin{rem}\label{superpt} \mar{superpt}
In gauge theory, a symmetry current $\gT$ (\ref{Q30}) takes the
form
\mar{b381}\beq
\gT= W+ d_HU=(W^\la +d_\m U^{\m\la})\om_\la, \label{b381}
\eeq
where the term $W$ depends only on the
variational derivatives
\mar{1222}\beq
\dl_i\cL=(\dr_i- d_\la\dr^\la_i)\cL, \label{1222}
\eeq
i.e., $W\ap 0$ and
\be
U=U^{\m\la}\om_{\m\la}:J^1Y\to \op\w^{n-2}T^*X
\ee
is a horizontal $(n-2)$-form on $J^1Y\to X$. Then one says that
$\gT$ reduces to the superpotential $U$ 
\cite{fat,book,sard97}.
On-shell, such a
symmetry current reduces to a $d_H$-exact form (\ref{b381}). Then the
differential conservation law (\ref{b3118'}) and the integral
conservation law (\ref{b3118}) become tautological. At the same time, the
superpotential form (\ref{b381}) of $\gT$ implies the following integral
relation
\mar{b3119}\beq
\op\int_{N^{n-1}} s^*\gT = \op\int_{\dr N^{n-1}}
s^*U, \label{b3119}
\eeq
where $N^{n-1}$ is a compact oriented $(n-1)$-dimensional submanifold of $X$
with the boundary
$\dr N^{n-1}$. One can think of this relation as being a part of the
Euler--Lagrange equations written in an integral form.
\end{rem}

\begin{rem}\label{gensym}\mar{gensym}
Let us consider conservation laws in the case of gauge transformations
which  preserve the Euler--Lagrange operator $\cE_L$, but not necessarily a
Lagrangian $L$. 
Let $u$ be a projectable vector field on $Y\to X$,
which is the infinitesimal generator of a local one-parameter group of
such transformations, i.e.,
\be
\bL_{J^2u}\cE_L=0,
\ee
where $J^2u$ is the second order jet prolongation of the vector field $u$.
There is  the useful relation
\mar{mos12}\beq
\bL_{J^2u}\cE_L=\cE_{\bL_{J^1u}L} \label{mos12}
\eeq
\cite{book}.
Then, in accordance with (\ref{mos11}), we have locally
\mar{mos12'}\beq
\bL_{J^1u}L=d_\la h_0(\xi)^\la\om. \label{mos12'}
\eeq
In this case, the weak identity (\ref{J4}) reads
\mar{b3141}\beq
0\ap d_\la(h_0(\xi)^\la -\gT^\la), \label{b3141}
\eeq
where $\gT$ is the symmetry current (\ref{Q30}) along the vector field $u$.
\end{rem}

\begin{rem}\label{bakgr} \mar{bakgr}
Background fields, which do not live in the dynamic shell (\ref{b327}),
violate
conservation laws as follows. Let us consider the product
\mar{C41}\beq
Y_{\rm tot}=Y\op\times_X Y'\label{C41}
\eeq
of a fibre bundle $Y$, coordinated by $(x^\la, y^i)$, whose sections are
dynamic fields and of a fibre bundle
$Y'$, coordinated by $(x^\la, y^A)$, whose sections are background
fields which take the background values
\be
y^B=\f^B(x), \qquad y^B_\la= \dr_\la\f^B(x).
\ee
A Lagrangian $L$ of dynamic and background fields is defined
on the total configuration space $J^1Y_{\rm tot}$.
Let $u$ be a projectable vector field on $Y_{\rm tot}$ which also
projects onto $Y'$ because gauge
transformations of background fields do not depend on dynamic fields. This
vector field takes the coordinate form
\mar{l68}\beq
u=u^\la(x^\m)\dr_\la + u^A(x^\mu,y^B)\dr_A + u^i(x^\mu,y^B, y^j)\dr_i.
\label{l68}
\eeq
Substitution of $u$ (\ref{l68}) in the formula (\ref{bC30'}) leads to
the first variational formula in the presence of background
fields: 
\mar{gm555}\ben
&&\dr_\la u^\la\cL +[u^\la\dr_\la+  u^A\dr_A +
u^i\dr_i +(d_\la u^A -y^A_\m\dr_\la
u^\m)\dr^\la_A + \label{gm555}\\
&&\qquad  (d_\la u^i -y^i_\m\dr_\la
u^\m)\dr^\la_i]\cL =
(u^A-y^A_\la u^\la)\dr_A\cL + \pi^\la_Ad_\la (u^A-y^A_\mu u^\mu)  +\nonumber\\
&&\qquad (u^i-y^i_\la u^\la)\dl_i\cL
-d_\la[\pi^\la_i(u^\m y^i_\m -u^i) -u^\la\cL]. \nonumber
\een
Then the weak identity
\be
&&\dr_\la u^\la\cL +[u^\la\dr_\la+  u^A\dr_A +
u^i\dr_i +(d_\la u^A -y^A_\m\dr_\la
u^\m)\dr^\la_A + \\
&& \qquad  (d_\la u^i -y^i_\m\dr_\la u^\m)\dr^\la_i]\cL\ap
(u^A-y^A_\la u^\la)\dr_A\cL + \pi^\la_Ad_\la (u^A-y^A_\mu u^\mu) - \\
&&\qquad d_\la[\pi^\la_i(u^\m y^i_\m -u^i)-u^\la\cL]
\ee
holds on the shell (\ref{b327}).
If a total Lagrangian $L$ is invariant under gauge
transformations of $Y_{\rm tot}$, we obtain  the
weak identity
\mar{l70}\beq
(u^A-y^A_\m u^\m)\dr_A\cL + \pi^\la_A d_\la (u^A-y^A_\mu
u^\mu) \ap d_\la\gT^\la, \label{l70}
\eeq
which is the transformation law of the symmetry current $\gT$ in the 
presence of
background fields.
\end{rem}

\section{Gauge theory of principal connections}

The reader is referred, e.g., to \cite{kob} for the standard 
exposition of geometry of
principal bundles and to \cite{mara} for the customary geometric
formulation of gauge theory.  Here, gauge theory
of principal connections is phrased in terms of jet manifolds on the 
same footing
as other Lagrangian field theories on fibre bundles \cite{book,book00}.

\glos{A. Principal bundles}

Let $\pi_P :P\to X$ be
a principal bundle  with a real structure
Lie group $G$ of finite non-zero dimension. For the sake of brevity, 
we call $P$
  a principal $G$-bundle.
By definition, a principal bundle $P\to X$ is provided with
the free transitive right action
\mar{1}\ben
&&R_G:P\op\times_X G \to P, \label{1}\\
&& R_g : p\mto pg, \quad p\in P,\quad g\in G, \nonumber
\een
of its structure group $G$ on $P$. It follows that the typical fibre 
of a principal $G$-bundle
is isomorphic to the group space of $G$, and that $P/G=X$.
The structure group $G$
acts on the typical fibre by left multiplications which do not preserve the
group structure of $G$. Therefore,
the typical fibre of a principal bundle is only a group space, but not a group
(cf. the group bundle $P^G$ (\ref{b3130}) below). Since the left 
action of transition functions
on the typical fibre $G$ commutes with its right multiplications,
a principal bundle admits the global right action (\ref{1}) of the 
structure group.

A principal $G$-bundle $P$ is equipped with a bundle atlas
\mar{1120}\beq
\Psi_P=\{(U_\al,\psi^P_\al,\rho_{\al\bt})\}, \label{1120}
\eeq
  whose trivialization morphisms
\be
\s_\al^P :\p_P^{-1}(U_\al)\to
U_\al\times G
\ee
obey the equivariance condition
\mar{1119}\beq
(\pr_2\circ \s_\al^P)(pg)=(\pr_2\circ \s_\al^P)(p)g, \qquad \forall g\in G,
\qquad \forall p\in\pi^{-1}_P(U_\al). \label{1119}
\eeq
Due to this property, every trivialization morphism $\psi^P_\al$
determines a unique local section $z_\al$ of $P$ over $U_\al$ such that
\be
\pr_2\circ \psi^P_\al\circ z_\al=\bb,
\ee
where $\bb$ is the unit element of $G$. The
transformation rules for $z_\al$ read
\mar{b1.202}\beq
z_\bt(x)=z_\al(x)\rho_{\al\bt}(x),\qquad x\in U_\al\cap U_\bt,\label{b1.202}
\eeq
where $\rho_{\al\bt}(x)$
are $G$-valued transition functions (\ref{sp22}) of the atlas
$\Psi_P$. Conversely, the family $\{(U_\al,z_\al)\}$ of local sections
of
$P$ with the transition functions (\ref{b1.202}) determines a unique 
bundle atlas
of $P$.

In particular, it follows that only trivial principal bundles have 
global sections.

Let us note that the pull-back of a principal bundle is also a principal
bundle with the same
structure group.

The quotient of the tangent bundle $TP\to P$ and that of the vertical tangent
bundle $VP$ of $P$ by the tangent
prolongation
$TR_G$ of the canonical action $R_G$ (\ref{1}) are
vector bundles
\mar{b1.205}\beq
  T_GP=TP/G, \qquad   V_GP=VP/G \label{b1.205}
\eeq
over $X$. Sections of $T_GP\to X$ are naturally identified with
$G$-invariant vector fields on $P$, while those of
$V_GP\to X$ are
$G$-invariant vertical vector fields on $P$.
Accordingly, the Lie bracket of $G$-invariant vector fields on $P$ goes to the
quotients (\ref{b1.205}), and induces
the Lie brackets of their sections. Let us write these brackets in an 
explicit form.

Owing to the equivariance condition (\ref{1119}), any bundle atlas (\ref{1120})
of $P$ yields the
associated bundle atlases $\{U_\al,T\psi^P_\al/G)\}$ of $T_GP$ and
$\{U_\al,V\psi^P_\al/G)\}$ of $V_GP$. Given a basis $\{\ve_p\}$ for the
right Lie algebra $\cG_r$, let $\{\pdr_\la, e_p\}$ and $\{e_p\}$,
where $e_p=(\psi^P_\al/G)^{-1}(\ve_p)$, be the corresponding
local fibre bases for the vector bundles $T_GP$ and $V_GP$, respectively.
Relative to these bases, the Lie
bracket of sections
\be
\x =\x^\la\pdr_\la + \x^p e_p,\qquad \eta = \eta^\m \pdr_\m +
\eta^q e_q
\ee
of the vector bundle $T_GP\to X$ reads
\mar{1129}\beq
  [\x,\eta ]=(\x^\m\pdr_\m\eta^\la - \eta^\m\pdr_\m\x^\la)\pdr_\la
  +(\x^\la\pdr_\la\eta^r - \eta^\la\pdr_\la\x^r +
c_{pq}^r\x^p\eta^q) e_r. \label{1129}
\eeq
Putting $\xi^\la=0$ and $\eta^\m=0$, we obtain the Lie bracket
\mar{1129'}\beq
  [\x,\eta]= c_{pq}^r\x^p\eta^q e_r \label{1129'}
\eeq
of sections of the vector bundle $V_G\to P$.

A glance at the expression (\ref{1129'}) shows
that $V_GP\to X$ is a finite-dimensional Lie algebra
bundle,  whose typical fibre is the right
Lie algebra ${\got g}_r$ of the group
$G$. The structure group $G$ acts on this typical fibre by the 
adjoint representation. In the physical literature, $V_GP$ is often called
the gauge algebra bundle
because,  if the base $X$ is compact, a suitable Sobolev completion
of the space of sections of $V_GP\to X$ is the Lie algebra
of the gauge Lie group.

Let $J^1P$ be the first order jet manifold of a principal $G$-bundle $P\to
X$. Its quotient
\mar{B1}\beq
C=J^1P/G\label{B1}
\eeq
by the jet prolongation of the canonical action $R_G$ (\ref{1}) is a 
fibre bundle
over $X$.
Bearing in mind the canonical imbedding
\be
\la_1: J^1P\to T^*X\ot TP
\ee
(\ref{18}) and passing to the quotient by $G$, we obtain the
corresponding canonical imbedding
\mar{1216}\beq
\la_C: C\to T^*X\ot T_GP \label{1216}
\eeq
of the fibre bundle $C$ (\ref{B1}).
It follows that $C$ is an affine bundle
modelled over the vector bundle
\mar{B1'}\beq
\ol C=T^*X\op\ot_X V_GP\to X. \label{B1'}
\eeq
Given a bundle atlas of $P$ and the associated bundle atlas of $V_GP$,
the affine bundle $C$ is provided with affine bundle coordinates
$(x^\la,a^q_\la)$, and its elements are
represented by $T_GP$-valued one-forms
\mar{1217}\beq
a=dx^\la\ot(\dr_\la + a^q_\la e_q) \label{1217}
\eeq
on $X$. One calls $C$ (\ref{B1}) the connection bundle
because its sections are naturally identified with
principal connections
on the principal bundle $P\to X$ as follows.

\glos{B. Principal connections}

In the case of a principal bundle $P\to X$,
the exact sequence (\ref{1.8a}) reduces to the exact sequence
\mar{1.33}\beq
0\to V_GP\op\hookrightarrow_X T_GP\to TX\to 0 \label{1.33}
\eeq
over $X$
by taking the quotient with respect to the right action of the group $G$.
The exact sequence of vector bundles (\ref{1.33}) yields the
exact sequence of the structure modules of their sections
\mar{1.33'}\beq
0\to V_GP(X)\ar T_GP(X)\ar \cT_1(X)\to 0. \label{1.33'}
\eeq
Principal connections split these exact sequences as follows.

A principal connection $A$ on a principal
bundle $P\to X$ is defined as a global section $A$ of the affine
jet bundle $J^1P\to P$ which is
equivariant under the right action (\ref{1}) of the group $G$ on $P$, i.e.,
\mar{b1.210}\beq
J^1R_g\circ A= A\circ R_g, \qquad \forall g\in G. \label{b1.210}
\eeq
Due to this equivariance condition, there is
one-to-one correspondence between the principal connections
on a principal bundle $P\to X$ and the global sections of the
affine bundle $C\to X$ (\ref{B1}). In  view of the imbedding (\ref{1216}),
a principal connection
splits the exact sequence (\ref{1.33}),
and is represented by a $T_GP$-valued form
\mar{1131}\beq
A=dx^\la\ot (\pdr_\la + A_\la^q(x) e_q)  \label{1131}
\eeq
on $X$. Since the connection bundle $C\to X$ is affine,
principal connections on a principal
bundle always exist.

Hereafter, we agree to identify gauge potentials
in gauge theory on a principal $G$-bundle $P$ to
global sections of the connection bundle $C\to X$ (\ref{B1}).

\begin{rem} Let us relate the $T_GP$-valued connection form 
(\ref{1131}) with the familiar
connection form on $P$ and the local connection form on $X$, associated to a
principal connection in \cite{kob}. Let us first recall that,
since the tangent bundle of a Lie group admits the canonical trivialization
along left-invariant vector fields, the vertical tangent bundle
$VP\to P$ of a principal bundle $P\to X$ also possesses the canonical 
trivialization
\be
\al:VP\cong P\times{\got g}_l
\ee
such that
$\al^{-1}(\e_m)$
are the familiar fundamental
vector fields
on $P$ corresponding to the basis elements
$\e_m$ of the left Lie algebra ${\got g}_l$ of the Lie group $G$.
Let a principal connection on a principal bundle $P\to X$ be
represented by the vertical-valued form $A$ (\ref{b1.223}). Then
\mar{+231}\beq
\wt A: P\ar^A T^*P\op\ot_P VP\ar^{\id\ot\al} T^*P\ot {\got g}_l \label{+231}
\eeq
is the above mentioned ${\got g}_l$-valued connection form
on the principal bundle
$P$. Given a trivialization chart $(U_\zeta,\psi^P_\zeta,z_\zeta)$ of $P$,
this form reads
\mar{mos166}\beq
\wt A=\psi_\zeta^{P*}(\th_l - \wt A^q_\la dx^\la\ot\e_q), \label{mos166}
\eeq
where $\th_l$ is the canonical ${\got g}_l$-valued one-form 
on $G$
and $\wt A^p_\la$ are equivariant functions on $P$ such that
\be
\wt A^q_\la(pg)\e_q =\wt A^q_\la(p){\rm Ad}\, g^{-1}(\e_q).
\ee
The pull-back $A_\zeta=z^*_\zeta\wt A$ of the connection form
$\wt A$ onto $U_\zeta$ is the above-mentioned ${\got g}_l$-valued
local connection one-form
\mar{b1.225}\beq
A_\zeta=- A^q_\la dx^\la\ot\e_q=A^q_\la dx^\la\ot \ve_q \label{b1.225}
\eeq
on $X$, where $A^q_\la=\wt A^q_\la\circ z_\zeta$ are
coefficients of the form (\ref{1131}).
We have $A_\zeta=\psi_\zeta^P \bA$,
where
\mar{b1.225'}\beq
\bA=A-\th_X=A^q_\la dx^\la\ot e_q \label{b1.225'}
\eeq
is the local $V_GP$-valued part of the form $A$ (\ref{1131}).
We will refer to $\bA$ (\ref{b1.225'})
as a local connection form. 
\end{rem}

The following theorems \cite{kob} state the pull-back and push-forward
operations of principal connections.

\begin{theo} \label{mos252} \mar{mos252}
Let $P\to X$ be a principal fibre bundle and $f^*P\to X'$ (\ref{mos106})
the pull-back principal bundle with the same structure group.
If
$A$ is a principal connection on $P\to X$, then the pull-back connection $f^*A$
(\ref{mos82}) on $f^*P\to X'$ is a principal connection.
\end{theo}

\begin{theo} \label{mos253} \mar{mos253}
Let $P'\to X$ and $P\to X$ be principle bundles with structure groups
$G'$ and $G$, respectively. Let $\Phi: P'\to P$ be a principal bundle morphism
over
$X$ with the corresponding homomorphism $G'\to G$. For every principal
connection
$A'$ on $P'$, there exists a unique principal connection
$A$ on $P$  such that the tangent map $T\Phi$ to $\Phi$ sends the
horizontal subspaces relative to $A'$ into those relative to $A$.
\end{theo}

\glos{C. The strength of a principal connection}

Given a principal $G$-bundle $P\to X$, the
Fr\"olicher--Nijenhuis
bracket (\ref{1149}) on the space $\cO^*(P)\ot\cT_1(P)$ of tangent-valued
forms on $P$ is compatible with
the canonical action $R_G$ (\ref{1}) of $G$ on $P$, and yields the
induced Fr\"olicher--Nijenhuis bracket on the space
$\cO^*(X)\ot T_GP(X)$ of $T_GP$-valued forms on $X$.
Its coordinate form issues from the Lie bracket (\ref{1129}).

Then any principal connection
$A\in \cO^1(X)\ot T_GP(X)$ (\ref{1131}) sets the Nijenhuis differential
\mar{1159b}\ben
&& d_A : \cO^r(X)\ot T_GP(X)\to \cO^{r+1}(X)\ot
V_GP(X),\nonumber \\
&& d_A\f = [A,\f]_{\rm FN}, \quad \f\in \cO^r(X)\ot T_GP(X),
\label{1159b}
\een
on the space $\cO^*(X)\ot T_GP(X)$. Similarly to
the curvature $R$ (\ref{1178a}), let us put
\mar{mos36,'}\ben
&& F_A = \frac{1}{2} d_AA = \frac{1}{2} [A, A]_{\rm FN}
\in \cO^2(X)\ot V_GP(X), \label{mos36}\\
&& F_A
=\frac12 F^r_{\la\m} dx^\la\w dx^\m\ot e_r, \nonumber \\
&& F_{\la\m}^r = \dr_\la A_\m^r -
\dr_\m A_\la^r + c_{pq}^rA_\la^p A_\m^q, \label{mos36'}
\een
It is called the strength of
a principal connection $A$, and is given locally by
the expression
\mar{+619}\beq
F_A=d\bA +\frac12 [\bA,\bA]=d\bA + \bA\w \bA, \label{+619}
\eeq
where $\bA$ is the local connection form (\ref{b1.225'}). By 
definition, the strength
(\ref{mos36}) of a principal connection obeys the second Bianchi identity
\mar{1255}\beq
d_AF_A=[A,F_A]_{\rm FN}=0. \label{1255}
\eeq

It should
be emphasized
that the strength $F_A$ (\ref{mos36}) is not the standard curvature 
(\ref{161'})
of a connection on $P$, but there are the local relations
$\psi^P_\zeta F_A=z^*_\zeta\Theta$, where
\mar{1200}\beq
\Theta=d\wt A +\frac12[\wt A,\wt A] \label{1200}
\eeq
is the
${\got g}_l$-valued  curvature form
on $P$ (see the expression (\ref{mos118}) below). In particular, a principal
connection is flat
if and only if its strength vanishes.

\glos{D. Associated bundles}

Given a principal $G$-bundle $\pi_P:P\to X$,
let $V$ be a manifold provided with an effective left action
\be
G\times V\ni (g,v)\mapsto gv\in V
\ee
of the Lie group $G$.
Let us consider the quotient
\mar{b1.230}\beq
Y=(P\times V)/G \label{b1.230}
\eeq
of the product $P\times V$ by identification of
elements
$(p,v)$ and $(pg,g^{-1}v)$ for all $g\in G$.
We will use the notation $(pG,G^{-1}v)$ for its points.
  Let $[p]$ denote
the restriction of the canonical surjection
\mar{mos75}\beq
P\times V\to (P\times V)/G \label{mos75}
\eeq
to the subset $\{p\}\times V$ so that
\be
[p](v)=[pg](g^{-1}v).
\ee
Then the map
\be
Y\ni[p](V)\mapsto \pi_P(p)\in X,
\ee
makes the quotient $Y$ (\ref{b1.230}) to a fibre bundle
over $X$. 

Let us note that, for any $G$-bundle, there exists an associated
principal $G$-bundle \cite{ste}. The peculiarity of the $G$-bundle
$Y$ (\ref{b1.230}) is that it appears canonically associated
  to a principal bundle $P$. Indeed,
  every bundle atlas $\Psi_P=\{(U_\al,z_\al)\}$ of
$P$ determines a unique
associated bundle atlas
\be
\Psi=\{(U_\al,\psi_\al(x)=[z_\al(x)]^{-1})\}
\ee
of the quotient $Y$ (\ref{b1.230}), and each automorphism of $P$
also yields
the corresponding
automorphism (\ref{024}) of $Y$.

Therefore, unless otherwise stated,
by a fibre bundle associated to a principal bundle
$P\to X$ (or, simply,
a $P$-associated fibre bundle) we will mean
the quotient (\ref{b1.230}).

Every principal connection $A$ on a principal bundle
$P\to X$ induces a unique
connection on the associated fibre bundle $Y$ (\ref{b1.230}).
  Given the horizontal splitting of the
tangent bundle $TP$ relative to $A$,
the tangent map to the canonical morphism (\ref{mos75}) defines the
horizontal splitting of the tangent bundle $TY$ of $Y$ and, consequently,
a connection on $Y\to X$
\cite{kob}. This is called the associated principal
connection or, simply, a principal
connection on a $P$-associated bundle $Y\to X$.
If $Y$ is a vector bundle, this connection takes the form
\mar{A}\beq
A=dx^\la\ot(\dr_\la -A^p_\la I_p{}^i_jy^j\dr_i), \label{A}
\eeq
where $I_p$ are generators of the linear representation of the Lie
algebra ${\got g}_r$ in the vector space $V$. The curvature (\ref{161'})
of this connection
reads
\mar{mos118}\beq
R=-\frac12 F_{\la\m}^pI_p{}^i_jy^j dx^\la\w dx^\m\ot\dr_i, \label{mos118}
\eeq
where $F_{\la\m}^p$ are coefficients (\ref{mos36'}) of the strength of
a principal connection $A$.

In particular, any principal connection $A$ yields the associated
linear connection on the gauge algebra bundle $V_GP\to X$.
The corresponding covariant differential $\nabla^A\x$ (\ref{+190}) of 
its sections
$\x=\x^pe_p$ reads
\mar{+430}\ben
&&\nabla^A\x : X\to T^*X\ot V_GP, \nonumber \\
&& \nabla^A\x = (\dr_\la\x^r + c_{pq}^r A_\la
^p\x^q)dx^\la\ot e_r.  \label{+430}
\een
It coincides with the Nijenhuis differential
\mar{1218}\beq
d_A\xi=[A,\xi]_{\rm FN}= \nabla^A\xi \label{1218}
\eeq
(\ref{1159b}) of $\xi$ seen as a $V_GP$-valued 0-form,
and is given by the local expression
\mar{+775}\beq
\nabla^A\xi= d\xi +[\bA,\xi], \label{+775}
\eeq
where $\bA$ is the local connection form (\ref{b1.225'}).

\glos{E. The configuration space of classical gauge theory}

Since gauge potentials
are represented by global sections of the connection bundle $C\to X$
(\ref{B1}), its first order jet manifold $J^1C$ plays the role of
a configuration space of classical gauge theory. The key point is that
the jet manifold $J^1C$ admits the canonical splitting over $C$
which leads to a unique canonical Yang--Mills Lagrangian of gauge theory
on $J^1C$.

Let us describe this splitting. One can show that the principal $G$-bundle
\mar{1285}\beq
J^1P\to J^1P/G=C \label{1285}
\eeq
is canonically isomorphic to the trivial pull-back bundle
\mar{b1.251}\beq
P_C=C\op\times_X P\to C, \label{b1.251}
\eeq
and that the latter admits the canonical
principal connection
\mar{266}\beq
\cA =dx^\la\ot(\dr_\la +a_\la^p e_p) + da^r_\la\ot\dr^\la_r
\in \cO^1(C)\ot T_G(P_C)(C)
\label{266}
\eeq
\cite{gar77,book,book00}.
Since $C$ (\ref{B1}) is an affine bundle
modelled over the vector bundle $\ol C$ (\ref{B1'}), the vertical 
tangent bundle
  of $C$ possesses the canonical trivialization
\mar{1219}\beq
VC =C\op\times_X T^*X\ot V_GP, \label{1219}
\eeq
while
\be
V_GP_C=V_G(C\op\times_XP)=C\op\times_X V_GP.
\ee
Then the strength $F_\cA$ of the connection (\ref{266}) is the $V_GP$-valued
horizontal two-form
\mar{267}\ben
&& F_\cA =\frac{1}{2} d_\cA \cA = \frac{1}{2} [\cA,\cA]_{\rm FN}
\in \cO^2(C)\ot V_GP(X), \nonumber \\
&& F_\cA =(da_\m^r\w dx^\m + \frac{1}{2} c_{pq}^r a_\la^p
a_\m^q dx^\la\w dx^\m)\ot e_r, \label{267}
\een
on $C$. It is readily observed that, given a global section  connection
$A$ of the connection bundle $C\to X$, the pull-back $A^*F_\cA = F_A$
is the strength (\ref{mos36}) of the principal connection $A$.

Let us take the pull-back of the form $F_\cA$ onto $J^1C$ with respect to the
fibration (\ref{1285}), and consider the
$V_GP$-valued horizontal two-form
\mar{295}\ben
&& \cF=h_0(F_\cA)=\frac{1}{2} \cF_{\la\m}^r dx^\la\w dx^\m\ot
e_r, \nonumber \\
&& \cF_{\la\m}^r = a_{\la\m}^r -
a_{\m\la}^r +c_{pq}^r a_\la^p a_\m^q, \label{295}
\een
where $h_0$ is the horizontal projection (\ref{mos40}).
It is readily observed that
\mar{294}\beq
\cF/2:J^1C\to C\op\times_X\op\w^2T^*X\ot V_GP \label{294}
\eeq
is an affine morphism over $C$ of constant rank. Hence, its kernel
$C_+=\Ker\, \cF$
is the affine subbundle of $J^1C\to C$, and we have a desired
canonical
splitting
\mar{296,'}\ben
&& J^1C =C_+\op\oplus_C C_-=C_+\op\oplus (C\op\times_X\op\w^2T^*X\ot V_GP),
\label{296}\\
&&a_{\la\m}^r =
\frac{1}{2}(a_{\la\m}^r + a_{\m\la}^r
  - c_{pq}^r a_\la^p a_\m^q) + \frac{1}{2}
(a_{\la\m}^r - a_{\m\la}^r +
c_{pq}^r a_\la^p a_\m^q), \label{296'}
\een
over $C$ of the jet manifold $J^1C$.
The corresponding canonical projections are
\mar{299}\beq
\pr_1=\cS:J^1C\to C_+, \qquad
  \cS^r_{\la\m}=\frac12(a_{\la\m}^r + a_{\m\la}^r
  - c_{pq}^r a_\la^p a_\m^q), \label{299}
\eeq
and $\pr_2=\cF/2$ (\ref{294}).

\glos{F. Gauge transformations}

In classical gauge theory, several classes of gauge transformations
are examined \cite{book,mara,soc}. A most general  gauge transformation is
defined as an automorphism $\Phi_P$
of a principal $G$-bundle $P$
which is equivariant under the canonical
action (\ref{1}) of the group $G$ on $G$, i.e.,
\be
R_g\circ\Phi_P=\Phi_P\circ R_g, \qquad \forall g\in G.
\ee
Such an automorphism of $P$ yields the corresponding
automorphism
\mar{024}\beq
\Phi_Y: (pG,G^{-1}v) \to (\Phi_P(p)G,G^{-1}v) \label{024}
\eeq
of the $P$-associated bundle $Y$ (\ref{b1.230}) and
the corresponding automorphism
\mar{b3105}\beq
\Phi_C:J^1P/G\to J^1\Phi_P(J^1P)/G \label{b3105}
\eeq
of the connection bundle $C$.

Hereafter, we deal with only
vertical automorphisms of the principal bundle $P$, and agree to call them 
gauge transformations
in gauge theory.  Accordingly,
the group $\gG(P)$ of vertical
automorphisms of a principal $G$-bundle $P$ is called the
gauge group. 

Every vertical automorphism of a principal bundle $P$ is represented as
\mar{b3111}\beq
\Phi_P(p)=pf(p), \qquad p\in P, \label{b3111}
\eeq
where $f$ is a $G$-valued equivariant function on $P$, i.e.,
\mar{b3115}\beq
f(pg)=g^{-1}f(p)g, \qquad \forall g\in G. \label{b3115}
\eeq
There is one-to-one correspondence
between these
functions and the global sections
$s$ of the group bundle
\mar{b3130}\beq
P^G =(P\times G)/G, \label{b3130}
\eeq
whose typical fibre is the group $G$, subject to
the adjoint representation of the structure group $G$.
Therefore, $P^G$ (\ref{b3130}) is also called the
adjoint bundle. 
There is the canonical
fibre-to-fibre action of the group bundle $P^G$ on any
$P$-associated bundle $Y$ by the law
\be
&& P^G\op\times_X Y\to Y, \\
&& ((pG, G^{-1}gG), (pG, G^{-1}v)) \mapsto (pG, G^{-1}gv).
\ee
Then the above-mentioned correspondence is given by the relation
\be
P^G\op\times_X P\ni (s(\pi_P(p)), p)\mapsto pf(p)\in P,
\ee
where $P$ is considered as a $G$-bundle associated to itself.
Hence, the gauge group $\gG(P)$ of
vertical automorphisms of a principal $G$-bundle $P\to X$
is isomorphic to the group of global sections of the $P$-associated 
group bundle
(\ref{b3130}).

In order to study the gauge invariance of one or another object
in gauge theory , it suffices to examine its invariance under
an arbitrary one-parameter subgroup
$[\Phi_P]$ of the gauge group. Its infinitesimal generator is
a $G$-invariant vertical vector field $\xi$ on a principal bundle $P$ or,
equivalently, a section
\mar{b3106}\beq
\x=\x^p(x) e_p \label{b3106}
\eeq
of the gauge algebra bundle $V_GP\to X$ (\ref{b1.205}).
We will call
it a gauge vector field. 
One can think of its components $\x^p(x)$ as being gauge parameters.
Gauge vector fields (\ref{b3106}) are
transformed under the infinitesimal
generators of gauge transformations (i.e., other gauge vector fields)
$\xi'$
by the adjoint
representation
\be
\bL_{\xi'}\xi=[\x',\x] = c^p_{rq}{\x'}^r\x^q e_p,
\qquad \x,\x'\in V_GP(X).
\ee
Accordingly, gauge parameters are subject to the
coadjoint representation 
\mar{b3116}\beq
\x': \x^p \mto -c^p_{rq}{\x'}^r\x^q.\label{b3116}
\eeq

Given a gauge vector field $\x$ (\ref{b3106}) seen as the 
infinitesimal generator
of a one-parameter gauge group $[\Phi_P]$, let us obtain
the gauge vector fields on a $P$-associated bundle $Y$ and the 
connection bundle
$C$.

The corresponding gauge vector field on the
$P$-associated vector bundle $Y\to X$ issues from the relation (\ref{024}),
and reads
\be
\x_Y =\x^p I_p^i \dr_i,
\ee
where $I_p$ are generators of the group $G$, acting on the typical fibre
$V$ of $Y$.

The gauge vector field $\xi$ (\ref{b3106}) acts
on elements $a$ (\ref{1217}) of the connection bundle by the law
\be
\bL_\xi a=[\xi,a]_{\rm FN}=(-\dr_\la \xi^r+c^r_{pq}\xi^pa^q_\la)dx^\la\ot e_r.
\ee
In view of the vertical splitting (\ref{1219}), this quantity can be regarded
as the vertical vector field
\mar{C85}\beq
\x_C=(-\dr_\la \xi^r+c^r_{pq}\xi^pa^q_\la)\dr^\la_r \label{C85}
\eeq
on the connection bundle $C$, and is the infinitesimal generator of the
one-parameter group $[\Phi_C]$ of vertical automorphisms (\ref{b3105}) of
$C$, i.e., a desired gauge vector field on $C$.

\glos{G. Lagrangian gauge theory}

Classical gauge theory of unbroken symmetries on a principal 
$G$-bundle $P\to X$
deals with two types of fields. These are
gauge potentials identified to global sections of the connection bundle
$C\to X$ (\ref{B1}) and
matter fields represented by global sections of
a $P$-associated vector bundle $Y$ (\ref{b1.230}), called a  
matter bundle. Therefore, the total configuration space of 
classical gauge theory
is the product of jet bundles
\mar{C75}\beq
J^1Y_{\rm tot}=J^1Y\op\times_X J^1C. \label{C75}
\eeq
Let us study a gauge invariant Lagrangian on this configuration
space.

A total gauge vector field on the product
$C\op\times_XY$ reads
\mar{C76}\beq
\x_{YC}=
(-\dr_\la \xi^r+c^r_{pq}\xi^pa^q_\la)\dr^\la_r +\x^pI_p^i\dr_i
   = (u^{A\la}_p\dr_\la\x^p+ u^A_p\x^p)\dr_A, \label{C76}
\eeq
where we utilize the collective index $A$, and put the notation
\be
u_p^{A\la}\dr_A= -\dl_p^r\dr^\la_r, \qquad u^A_p\dr_A=
c^r_{qp}a^q_\la\dr^\la_r + I_p^i\dr_i.
\ee

A Lagrangian  $L$  on the configuration space  (\ref{C75}) is said
to be gauge-invariant if
its Lie derivative ${\bf L}_{J^1\x_{YC}} L$ along any
gauge vector field $\x$ (\ref{b3106})
vanishes. Then the first variational formula
(\ref{bC30'}) leads to the strong equality
\mar{b3109}\beq
0=(u^A_p\x^p + u^{A\m}_p\dr_\m\x^p)\dl_A\cL +
d_\la[(u^A_p\x^p + u^{A\m}_p\dr_\m\x^p)\pi^\la_A], \label{b3109}
\eeq
where $\dl_A\cL$ are the variational derivatives (\ref{1222}) of $L$ and
the total derivative reads
\be
d_\la =\dr_\la +a^p_{\la\m}\dr_p^\m +y_\la^i\dr_i.
\ee
Due to the arbitrariness of gauge parameters $\x^p$, this equality falls
into the system of strong equalities
\mar{D1}\bea
&& u^A_p\dl_A\cL + d_\m(u^A_p\pi^\m_A)=0, \label{D1a}\\
&& u^{A\m}_p\dl_A\cL
+ d_\la(u^{A\m}_p\pi^\la_A) + u^A_p\pi^\m_A =0,\label{D1b}\\
&& u^{A\la}_p\pi^\m_A+ u^{A\m}_p\pi^\la_A=0. \label{D1c}
\eea
Substituting (\ref{D1b}) and (\ref{D1c}) in
(\ref{D1a}), we obtain the well-known constraints
\be
u^A_p\dl_A\cL -d_\m(u^{A\m}_p\dl_A\cL)=0
\ee
for the variational
derivatives of a gauge-invariant Lagrangian $L$.

Treating the equalities (\ref{D1a}) -- (\ref{D1c}) as the equations for
a gauge-invariant
Lagrangian, let us solve these equations for a Lagrangian
\mar{b3127}\beq
L=\cL(x^\la,a^r_\m, a^r_{\la\m})\om: J^1C\to \op\w^nT^*X \label{b3127}
\eeq
without matter fields. In this case, the equations (\ref{D1a}) -- (\ref{D1c})
read
\mar{2110}\bea
&& c_{pq}^r(a_\m^p\pdr_r^\m\cL + a^p_{\la\m} \pdr_r^{\la\m}\cL)  =  0,
\label{2110a} \\
&& \pdr_q^\m \cL + c_{pq}^r a_\la^p \pdr_r^{\m\la}\cL  = 0, \label{2110b} \\
&& \pdr_p^{\m\la}\cL + \pdr_p^{\la\m}\cL = 0. \label{2110c}
\eea
Let rewrite them relative to the coordinates
$(a^q_\m, \cS^r_{\m\la}, \cF^r_{\m\la})$
(\ref{295}) and (\ref{299}),
associated to
the canonical splitting (\ref{296}) of the jet manifold
$J^1C$.  The equation (\ref{2110c}) reads
\mar{b3128}\beq
\frac{\dr\cL}{\dr \cS^r_{\m\la}}=0. \label{b3128}
\eeq
Then a simple computation brings the equation (\ref{2110b}) into the form
\mar{b3129}\beq
\dr^\m_q\cL =0. \label{b3129}
\eeq
A glance at the equations (\ref{b3128}) and (\ref{b3129}) shows that the
gauge-invariant Lagrangian  (\ref{b3127}) factorizes through the
strength
$\cF$ (\ref{295}) of gauge potentials.
As a consequence, the equation (\ref{2110a}) takes the form
\be
c^r_{pq}\cF^p_{\la\m}\frac{\dr\cL}{\dr \cF^r_{\la\m}}=0.
\ee
It admits a unique solution in the class of quadratic Lagrangians which is
the conventional Yang-Mills Lagrangian $L_{\rm YM}$ of gauge potentials on
the configuration space $J^1C$.  In the presence of a background world
metric
$g$ on the base
$X$, it  reads
\mar{5.1}\beq
L_{\rm YM}=\frac{1}{4\ve^2}a^G_{pq}g^{\la\m}g^{\bt\n}\cF^p_{\la
\bt}\cF^q_{\m\n}\sqrt{\nm g}\om, \qquad  g=\det(g_{\m\n}), \label{5.1}
\eeq
where  $a^G$ is a $G$-invariant bilinear form
on the Lie algebra of ${\got g}_r$ and
$\ve$ is a coupling constant.

\glos{H. Gauge conservation laws}

On-shell, the strong equality (\ref{b3109}) becomes the
weak Noether conservation law
\mar{C300}\beq
0\ap d_\la[(u^A_p\x^p + u^{A\m}_p\dr_\m\x^p)\pi^\la_A] \label{C300}
\eeq
of the Noether current
\mar{b3110}\beq
\gT^\la=-(u^A_p\x^p + u^{A\m}_p\dr_\m\x^p)\pi^\la_A. \label{b3110}
\eeq
Accordingly, the equalities (\ref{D1a}) -- (\ref{D1c}) on-shell lead to the
familiar Noether identities
\mar{D2}\bea
&&  d_\m(u^A_p\pi^\m_A)\ap 0, \label{D2a}\\
&& d_\la(u^{A\m}_p\pi^\la_A)
+ u^A_p\pi^\m_A \ap 0,\label{D2b}\\
&& u^{A\la}_p\pi^\m_A+
u^{A\m}_p\pi^\la_A=0 \label{D2c}
\eea
for a gauge-invariant Lagrangian  $L$.
They are equivalent to the weak equality (\ref{C300}) due to the
arbitrariness of the gauge parameters $\x^p(x)$.

A glance at the expressions (\ref{C300}) and (\ref{b3110}) shows that both the
Noether conservation law and the Noether current depend on gauge
parameters. The weak identities (\ref{D2a}) -- (\ref{D2c}) play the role of the
necessary and sufficient conditions in order that the Noether conservation law
(\ref{C300}) is maintained under changes of gauge parameters. This means
that, if the equality (\ref{C300}) holds for  gauge parameters
$\x$, it does so for arbitrary deviations $\x + \dl\x$ of $\x$.
In particular, the Noether conservation law (\ref{C300}) is maintained under
gauge transformations, when gauge parameters are transformed by the
coadjoint representation (\ref{b3116}).

It is easily seen that the equalities (\ref{D2a}) -- (\ref{D2c})
are not mutually independent, but (\ref{D2a}) is a corollary of
(\ref{D2b}) and (\ref{D2c}).
This property reflects the fact that, in accordance with the
strong equalities (\ref{D1b}) and
(\ref{D1c}), the Noether current (\ref{b3110}) is brought
into the superpotential
form
\be
\gT^\la =\x^p
u^{A\la}_p\dl_A\cL - d_\m(\x^p
u^{A\m}_p\pi^\la_A), \qquad U^{\m\la}= -\x^p
u^{A\m}_p\pi^\la_A,
\ee
(\ref{b381}).
Since a matter field Lagrangian is independent of the
jet coordinates $a^p_{\la\m}$, the
Noether superpotential
\be
U^{\m\la}=\x^p \pi^{\m\la}_p
\ee
depends only on gauge potentials. The corresponding integral
relation (\ref{b3119}) reads
\mar{b3223}\beq
\op\int_{N^{n-1}} s^*\gT^\la\om_\la = \op\int_{\dr N^{n-1}}
s^*( \x^p \pi^{\m\la}_p)\om_{\m\la}, \label{b3223}
\eeq
where $N^{n-1}$ is a compact oriented $(n-1)$-dimensional submanifold of $X$
with the boundary
$\dr N^{n-1}$. One can think of (\ref{b3223}) as being the integral relation
between the Noether current (\ref{b3110}) and the gauge field, generated by
this current. In electromagnetic theory seen as a $U(1)$ gauge theory,
the similar relation between an electric current and the electromagnetic
field generated by this current is well known. However, it is free from gauge
parameters due to the peculiarity of Abelian gauge models.

It should be emphasized that the Noether current
(\ref{b3110}) in gauge theory takes the superpotential form (\ref{b381}) o
because the infinitesimal generators of gauge transformations (\ref{C76})
depend on derivatives of gauge parameters.

\section{Higher order jets}

As was mentioned in Lecture 2, there is a natural higher order
generalization of the first and second order jet manifolds
\cite{book,kol,man,sau}. 
Recall the notation. Given bundle
coordinates
$(x^\la,y^i)$ of a fibre bundle $Y\to X$, by  $\La$,
$\nm\La=r$, is meant a collection of numbers $(\la_r...\la_1)$ modulo
permutations. By $\La+\Si$ we denote the collection
\be
\La+\Si=(\la_r\cdots\la_1\si_k\cdots\si_1) 
\ee
modulo permutations, while $\La\Si$ is the union of collections 
\be
\La\Si=(\la_r\cdots\la_1\si_k\cdots\si_1), 
\ee
where the indices $\la_i$ and $\si_j$ are not permuted.
Recall the symbol of the total derivative 
\mar{5.32}\beq
d_\la = \dr_\la + \op\sum_{|\La|=0}^k y^i_{\La+\la}\dr_i^\La.
\label{5.32}
\eeq
We will use the notation 
\be
\dr_\La=\dr_{\la_r}\circ\cdots\circ\dr_{\la_1}, \qquad
 d_\La=d_{\la_r}\circ\cdots\circ d_{\la_1}, \qquad
\La=(\la_r...\la_1).
\ee

The $r$-order jet manifold  $J^rY$  of
a fibre bundle $Y\to X$ is defined as the disjoint  union  
\mar{+400}\beq
J^rY=\op\bigcup_{x\in X}j^r_xs \label{+400}
\eeq
of the equivalence classes $j^r_xs$ of sections $s$ of $Y$ so that
sections
$s$ and $s'$ belong to the same equivalence class $j^r_xs$ if and only if
\be
s^i(x)={s'}^i(x), \qquad\dr_\La s^i(x)=\dr_\La {s'}^i(x),
\qquad 0< \mid\La\mid \leq r. 
\ee
In brief, one can say that sections of $Y\to X$ are identified by the
$r+1$ terms of their Taylor series at points of $X$. 
The particular choice of a coordinate atlas does not matter
for this definition.
Given an atlas of bundle coordinates $(x^\la,y^i)$ of a fibre bundle
 $Y\to X$, the set (\ref{+400}) is endowed with an atlas
of the adapted coordinates
\mar{55.1}\ben
&&(x^\la, y^i_\La),\qquad   0\leq\nm\La \leq r, \label{55.1}\\
&&(x^\la, y^i_\La)\circ s= (x^\la, \dr_\La s^i(x)), \nonumber
\een
together with transition functions (\ref{55.21}).
The coordinates (\ref{55.1}) bring the set $J^rY$  into a
smooth manifold of finite dimension 
\be
\di J^rY= n +m\op\sum_{i=0}^r \frac{(n+i-1)!}{i!(n-1)!}.
\ee

The coordinates (\ref{55.1}) are compatible with the natural surjections
\be
\pi_l^r: J^rY\to J^lY, \quad r>l,
\ee
which form the composite bundle
\be
\pi^r: J^rY\op\ar^{\pi^r_{r-1}} J^{r-1}Y\op\ar^{\pi^{r-1}_{r-2}} \cdots
\op\ar^{\pi^1_0} Y\op\ar^\pi X
\ee
with the properties
\be
\pi^k_h\circ\pi^r_k=\pi^r_h, \qquad \pi^h\circ\pi^r_h=\pi^r.
\ee
A glance at the transition functions (\ref{55.21}) when $\nm\La=r$ shows that
the fibration
\be
\pi_{r-1}^r: J^rY\to J^{r-1}Y 
\ee
is an affine bundle modelled over the vector bundle
\mar{5.117}\beq
\op\vee^rT^*X\op\ot_{J^{r-1}Y}VY \to J^{r-1}Y.
\label{5.117}
\eeq 

\begin{rem}
To introduce higher order jet manifolds, one can use
the construction of the repeated jet manifolds.
Let us consider the $r$-order jet manifold $J^rJ^kY$ of the jet
bundle $J^kY\to X$. It is coordinated by 
\be
(x^\m, y^i_{\Si\La}), \qquad \mid\La\mid \leq k, \qquad \mid\Si\mid \leq r.
\ee
There is the canonical monomorphism 
\be
\si_{rk}: J^{r+k}Y\hookrightarrow J^rJ^kY 
\ee
given by the coordinate relations
\be
y^i_{\Si\La}\circ \si_{rk}= y^i_{\Si+\La}.
\ee
\end{rem}

In the calculus in $r$-order jets, we have the $r$-order jet
prolongation functor 
such that,  given fibre bundles $Y$ and $Y'$ over $X$,
every bundle morphism $\Phi:Y\to Y'$ over a diffeomorphism $f$ of $X$ admits
the $r$-order jet prolongation to the morphism 
\mar{5.152}\beq
J^r\Phi: J^rY\ni j^r_xs\mapsto j^r_{f(x)}(\Phi\circ s\circ f^{-1})
\in  J^rY' \label{5.152}
\eeq
 of the $r$-order jet manifolds. 
The jet
prolongation functor is exact. If $\Phi$
is an injection [surjection], so is  $J^r\Phi$. 
It also preserves an
algebraic structure. In particular, if $Y\to X$ is a vector 
bundle, so is $J^rY\to X$. If $Y\to X$ is an affine bundle modelled over the
vector bundle $\ol Y\to X$, then $J^rY\to X$ is an affine bundle modelled over
the vector bundle $J^r\ol Y\to X$. 

Every section $s$ of a fibre bundle $Y\to X$
admits the $r$-order jet prolongation  to the 
section 
\be
(J^rs)(x)= j^r_xs
\ee
 of the jet bundle $J^rY\to X$. Such a section of
$J^rY\to X$ is called holonomic. 

Every exterior form $\f$ on the jet manifold $J^kY$ gives rise to the
pull-back form $\pi^{k+i}_k{}^*\f$ on the jet manifold $J^{k+i}Y$. Let
$\cO_k^*=\cO^*(J^kY)$ be the algebra of exterior forms on the jet
manifold $J^kY$. We  have the direct system of $\Bbb R$-algebras
\mar{5.7}\beq
\cO^*(X)\op\longrightarrow^{\pi^*} \cO^*(Y) 
\op\longrightarrow^{\pi^1_0{}^*} \cO_1^*
\op\longrightarrow^{\pi^2_1{}^*} \cdots \op\longrightarrow^{\pi^r_{r-1}{}^*}
 \cO_r^* \longrightarrow\cdots. \label{5.7}
\eeq
The subsystem of (\ref{5.7}) is the direct system  
\mar{+408}\beq
C^\infty(X)\op\longrightarrow^{\pi^*} C^\infty(Y) 
\op\longrightarrow^{\pi^1_0{}^*} \cO_1^0
\op\longrightarrow^{\pi^2_1{}^*} \cdots 
\op\longrightarrow^{\pi^r_{r-1}{}^*} \cO_r^0\longrightarrow\cdots
\label{+408}
\eeq
of the $\Bbb R$-rings of real smooth functions
$\cO^0_k=C^\infty(J^kY)$ on the jet manifolds $J^kY$.
Therefore, one can think of (\ref{5.7}) and (\ref{+408}) as being the direct
systems of $C^\infty(X)$-modules.

Given the $k$-order jet manifold $J^kY$ of $Y\to X$, there exists the
canonical bundle morphism 
\be
r_{(k)}: J^kTY\to TJ^kY
\ee
over $J^kY\op\times_XJ^kTX\to J^kY\op\times_XTX$
whose coordinate expression is
\be
(x^\la,y^i_\La,\dot x^\la,\dot y^i_\La)\circ r_{(k)}=(x^\la,y^i_\La,\dot
x^\la, (\dot y^i)_\La -\sum (\dot y^i)_{\m+\Si}(\dot x^\m)_\Xi), \qquad
0\leq\nm\La\leq k,
\ee
where the sum is taken over all partitions $\Si +\Xi=\La$ and $0<\nm\Xi$. In
particular,  we have the canonical isomorphism over $J^kY$
\mar{5.30}\beq
r_{(k)}:J^kVY\to VJ^kY, \qquad 
(\dot y^i)_\La = \dot y^i_\La\circ r_{(k)}. \label{5.30}
\eeq
As a consequence, every projectable vector field $u=u^\m\dr_\m +u^i\dr_i$
on a fibre bundle $Y\to X$ has the following $k$-order jet prolongation
to the vector field on $J^kY$:
\mar{55.5}\ben
&& J^k u =r_{(k)}\circ J^ku: J^kY\to TJ^kY, \label{55.5}\\
&&  J^k u =
u^\la\dr_\la + u^i\dr_i + u_\La^i\dr_i^\La, \qquad 0< \mid\La\mid \leq k,
\nonumber\\
&& u_{\la+\La}^i = d_\la u^i_\La - y_{\m+\La}^i\dr_\la u^\m, \qquad 
0<\nm\La < k,
\nonumber
\een
(cf. (\ref{1.21}) for $k=1$).
In particular, the $k$-order jet lift (\ref{55.5}) of a vertical vector field
on $Y\to X$ is a vertical vector field on $J^kY\to X$ due to the
isomorphism (\ref{5.30}).

A vector field $u_r$
on an $r$-order jet manifold  $J^rY$ is called projectable
if, for any $k< r$, there exists a
projectable vector field 
$u_k$ on $J^kY$ such that  
\be
u_k\circ \pi^r_k=T\pi^r_k\circ u_r.
\ee
A projectable vector field $\up$ on $J^rY$ has the coordinate expression
\be
\up= u_\la \dr_\la + u^i_\La \dr^\La_i, \qquad 0\leq |\La|\leq r,
\ee
such that $u_\la$ depends only on the coordinates $x^\m$ and every
component
$u^i_\La$ is independent of the coordinates $y^i_\Xi$, $|\Xi|>|\La|$.

  Let us
denote by
$\cP^k$   the vector space of projectable vector
fields on the jet manifold $J^kY$. It is easily seen that $\cP^r$
is a Lie algebra over $\Bbb R$ and that the morphisms $T\pi^r_k$, $k<r$, 
constitute the inverse system 
\mar{55.6}\beq
\cP^0 \op\longleftarrow^{T\pi^1_0}
\cP^1\op\longleftarrow^{T\pi^2_1}\cdots
\op\longleftarrow^{T\pi^{r-1}_{r-2}}\cP^{r-1} \op\longleftarrow^{T\pi^r_{r-1}}
\cP^r \longleftarrow\cdots
\label{55.6} 
\eeq
of these Lie algebras.

\begin{prop}\label{ch532} \mar{ch532} \cite{bau,tak1}.
The  $k$-order jet lift (\ref{55.5}) is the Lie algebra monomorphism of the Lie
algebra $\cP^0$ of projectable vector fields on $Y\to X$ to the Lie
algebra $\cP^k$ of projectable vector fields
on $J^kY$ such that  
\mar{5.33}\beq
T\pi^r_k(J^r u)=J^k u\circ \pi^r_k. \label{5.33}
\eeq
\end{prop}

The jet lift $J^k u$ (\ref{55.5}) is said to be an integrable vector
field
on $J^kY$. Every projectable vector field
$u_k$ on $J^kY$ is decomposed into the sum
\mar{+404}\beq
u_k= J^k(T\pi^k_0(u_k)) +v_k \label{+404}
\eeq
of the integrable vector field $J^k(T\pi^k_0(u_k))$ and the projectable
vector field
$v_r$ which is vertical  with respect to some fibration $J^kY\to Y$.

Similarly to the exact sequences (\ref{1.8a}) -- (\ref{1.8b}) over $J^0Y=Y$, we
have the exact sequences 
\mar{55.17,18}\ben
&& 0\to VJ^kY\hookrightarrow TJ^kY\to TX\op\times_XJ^kY\to 0, \label{55.17}\\
&& 0\to J^kY\op\times_X T^*X \hookrightarrow TJ^kY\to V^*J^kY\to 0,
\label{55.18}
\een
of vector bundles over $J^kY$. They do not admit a canonical
splitting. Nevertheless, their pull-backs onto $J^{k+1}Y$ are split canonically
due to 
 the following canonical bundle monomorphisms over $J^kY$:
\mar{5.16}\ben
&&\la_{(k)}: J^{k+1}Y\op\hookrightarrow T^*X\op\otimes_{J^kY}
TJ^kY,\nonumber\\
 &&\la_{(k)} =dx^\la\otimes d_\la^{(k)}, \label{5.16}
\een
\mar{5.15'}\ben
&&\th_{(k)}: J^{k+1}Y\hookrightarrow T^*J^kY\op\otimes_{J^kY}
VJ^kY,\nonumber\\ 
&&\th_{(k)} =\sum(dy^i_\La- y^i_{\la+\La}dx^\la)\otimes\dr_i^\La,
\label{5.15'}
\een
where the sum
is over all multi-indices $\La$, $\mid\La\mid\leq k$. 
The forms
\mar{55.15}\beq
\th^i_\La=dy^i_\La- y^i_{\La+\la}dx^\la \label{55.15}
\eeq
are also called 
the contact forms.
The monomorphisms (\ref{5.16}) and (\ref{5.15'}) yield the bundle
monomorphisms
over $J^{k+1}Y$
\mar{5.70,1}\ben
&& \wh\la_{(k)}: TX\op\times_XJ^{k+1}Y\op\hookrightarrow
TJ^kY\op\times_{J^kY}J^{k+1}Y, \label{5.70} \\
&&\wh\th_{(k)}: V^*J^kY\op\times_{J^kY}
\op\hookrightarrow T^*J^kY\op\times_{J^kY}J^{k+1}Y. \label{5.71} 
\een
These monomorphisms split 
the exact sequences  (\ref{55.17})  and (\ref{55.18}) over
$J^{k+1}Y$ and define the canonical horizontal splittings
of the pull-backs 
\mar{5.72}\ben
&&\pi^{k+1*}_kTJ^kY =
\wh\la_{(k)}(TX\op\times_XJ^{k+1}Y) \op\oplus_{J^{k+1}Y}VJ^kY,
\label{5.72}\\ 
&& \dot x^\la\dr_\la +\sum \dot y^i_\La\dr^\La_i =\dot
x^\la d_\la^{(k)} +\sum (\dot y^i_\La -\dot x^\la
y^i_{\la+\La}) \dr_i^\La,\nonumber
\een
\mar{55.19}\ben
&& \pi^{k+1*}_kT^*J^kY =T^*X
\op\oplus_{J^{k+1}Y} \wh\th_{(k)}(V^*J^kY\op\times_{J^kY}J^{k+1}Y),
\label{55.19} \\
&& \dot x_\la dx^\la +\sum \dot y_i^\La dy_\La^i =
(\dot x_\la +\sum \dot y_i^\La y_{\la+\La}^i)dx^\la +\sum \dot y_i^\La
\th_\La^i,\nonumber
\een
where summation are over all multi-indices $\nm\La\leq k$.

In accordance with the canonical horizontal splitting (\ref{5.72}), the
pull-back
\be
\ol u_k: J^{k+1}Y\ar^{\pi^{k+1}_k\times\Id} J^kY\times
J^{k+1}\ar^{u_k\times\Id} TJ^kY\op\times_{J^kY} J^{k+1}
\ee
onto $J^{k+1}Y$ of any vector field $u_k$ on $J^kY$ admits the canonical
horizontal splitting
\mar{+402}\beq
\ol u = u_H +u_V=
(u^\la d_\la^{(k)} + \sum y^i_{\la+\La}\dr_i^\La) +\sum (u^i_\La - u^\la
y^i_{\la+\La}) \dr_i^\La,  \label{+402}
\eeq
where the sums are over all multi-indices $|\La|\leq k$.
By virtue of the canonical horizontal splitting (\ref{55.19}), every
exterior 1-form $\f$ on $J^kY$ admits the canonical splitting of its pull-back
\mar{+403}\beq
\pi^{k+1*}_k\f =h_0\f +(\f- h_0(\f)), \label{+403}
\eeq
where $h_0$ is the horizontal projection (\ref{mos40}).

\section{Infinite order jets} 

The direct
system (\ref{5.7}) of  $\Bbb R$-algebras of exterior forms
and  the inverse system
(\ref{55.6}) of the real Lie algebras of projectable vector fields on jet
manifolds  admit
the limits for
$r\to\infty$ in the category of modules and that of Lie algebras, 
respectively.
Intuitively, one can think 
of elements of these limits as being the objects 
defined on the  projective limit of the inverse system
\mar{5.10}\beq
X\op\longleftarrow^\pi Y\op\longleftarrow^{\pi^1_0}\cdots \longleftarrow
J^{r-1}Y \op\longleftarrow^{\pi^r_{r-1}} J^rY\longleftarrow\cdots
\label{5.10}
\eeq
of finite order jet manifolds $J^rY$. 

\begin{rem}
Recall that, by
 a projective limit of the inverse system
(\ref{5.10}) is meant  a set
$J^\infty Y$ such that, for any $k$, there
exist surjections
\mar{5.74}\beq
\pi^\infty: J^\infty Y\to X, \quad \pi^\infty_0: J^\infty Y\to Y, \quad 
\quad 
\pi^\infty_k: J^\infty Y\to J^kY, \label{5.74}
\eeq
which make up the commutative diagrams
\be
\begin{array}{rcl}
& J^\infty Y & \\
_{\pi^\infty_k} & \swarrow \searrow &_{\pi^\infty_r}\\
J^kY & \ar_{\pi^k_r} & J^rY
\end{array}
\ee
for any admissible $k$ and $r<k$ \cite{massey}.
\end{rem}

The projective limit of the inverse system (\ref{5.10}) exists.  It is
called  the infinite order jet space. 
This space consists of those elements
\be
(\ldots,q_i,\ldots,q_j, \ldots), \qquad q_i\in J^iY, \qquad q_j\in J^jY,
\ee
of the Cartesian product $\op\prod_k J^kY$
which satisfy the relations $q_i=\pi^j_i(q_j)$ for all $j>i$. 
Thus, elements of the infinite order jet space $J^\infty Y$
really represent $\infty$-jets $j^\infty_xs$ 
of local sections of $Y\to X$. These sections  belong to the same jet 
$j^\infty_xs$ if and only if their Taylor series at a point $x\in X$ coincide
with each other.

\begin{rem}\label{c510} \mar{c510}
The space $J^\infty
Y$ is also the projective limit of the inverse subsystem of (\ref{5.10})  which
starts from any finite order $J^rY$. 
\end{rem}

The infinite order jet space $J^\infty Y$
is provided with the weakest
topology such that the surjections (\ref{5.74})
are continuous. The base of open sets of this topology in $J^\infty Y$
consists of the inverse images of open subsets of $J^kY$, $k=0,\ldots$, under
the mappings (\ref{5.74}). This topology makes $J^\infty Y$ a
paracompact Fr\'echet manifold \cite{tak2}.
A bundle coordinate atlas
$\{U,(x^\la,y^i)\}$ of $Y\to X$ yields the manifold
coordinate atlas
\mar{jet1}\beq
\{(\pi^\infty_0)^{-1}(U_Y), (x^\la, y^i_\La)\}, \qquad 0\leq|\La|,
\label{jet1}
\eeq
 of $J^\infty
Y$, together with the transition functions  
\mar{55.21'}\beq
{y'}^i_{\la+\La}=\frac{\dr x^\m}{\dr x'^\la}d_\m y'^i_\La, \label{55.21'}
\eeq
where $\La=(\la_k\ldots\la_1)$, $\la+\La=(\la\la_k\ldots\la_1)$ are
multi-indices and
$d_\la$ denotes the total derivative 
\mar{5.177}\beq
d_\la = \dr_\la + \op\sum_{|\La|\geq 0} y^i_{\la+\La}\dr_i^\La.\label{5.177}
\eeq

Though $J^\infty Y$ fails to be a smooth manifold, one can introduce
the differential calculus on $J^\infty Y$ as follows.

Let us consider the direct system (\ref{5.7}) of $\Bbb R$-modules
$\cO^*_k$ of exterior forms on finite order jet manifolds $J^kY$.
The limit 
$\cO^*_\infty$  of this direct system, by definition, obeys the following
conditions \cite{massey}:
\begin{itemize}
\item for any $r$, there
exists an injection $\cO_r^*\to \cO_\infty^*$;
\item the diagrams
\be
\begin{array}{rcl}
\cO_k^* & \ar^{\pi^{r*}_k} & \cO_r^*\\
_{\pi^{\infty *}_k} & \searrow \swarrow &_{\pi^{\infty *}_r}\\
& \cO_\infty^* & \\
\end{array}
\ee
 are commutative for any $r$ and $k<r$.
\end{itemize}
Such a direct limit exists. 
This is the quotient of the
direct sum $\op\bigoplus_k\cO_k^*$
with respect to identification of the pull-back forms
\be
 \pi^{\infty*}_r\f=\pi^{\infty*}_k\si, \qquad \f\in \cO_r^*, \qquad \si \in
\cO_k^*,
\ee
if $\f=\pi^{r*}_k\si$. In other words, $\cO_\infty^*$ consists of all the
exterior forms on finite order jet manifolds module pull-back identification.
Therefore, we will denote the image of $\cO_r^*$ in $\cO_\infty^*$
 by $\cO_r^*$ and the elements
$\pi^{\infty^*}_r\f$ of $\cO_\infty^*$ simply by $\f$.
 
\begin{rem}\label{c511} \mar{c511}
Obviously, $\cO_\infty^*$ is the direct limit  of the direct subsystem of
(\ref{5.7}) which starts from any finite order $r$. 
\end{rem}

 The $\Bbb R$-module $\cO_\infty^*$ possesses the structure of the
filtered module as follows \cite{vinb}. Let us consider the direct system
(\ref{+408}) of the commutative 
$\Bbb R$-rings
of  smooth functions on the jet manifolds $J^rY$.
Its direct limit $\cO^0_\infty$ consists of functions on finite order jet
manifolds modulo pull-back identification. This is the
$\Bbb R$-ring filtered 
by the
$\Bbb R$-rings $\cO^0_k\subset \cO^0_{k+i}$.
Then $\cO_\infty^*$ has the filtered $\cO^0$-module structure
given by the $\cO^0_k$-submodules $\cO_k^*$ of
$\cO^*_\infty$.

An endomorphism $\Delta$ of $\cO_\infty^*$ is called a filtered morphism
if there exists $i\in \Bbb N$ such that the
restriction of 
$\Delta$ to $\cO_k^*$ is the homomorphism of $\cO_k^*$ into  $\cO_{k+i}^*$
over the injection $\cO^0_k\hookrightarrow\cO^0_{k+i}$ for all $k$. 

\begin{theo} \label{+468} \mar{+468} \cite{massey}. 
Every direct system of endomorphisms
$\{\g_k\}$ of $\cO_k$
such that 
\be
\pi^j_i{}^*\circ\g_i = \g_j\circ\pi^j_i{}^* 
\ee
for all $j>i$ has the direct limit $\g_\infty$ 
in filtered endomorphisms of $\cO_\infty^*$. 
If all $\g_k$ are
monomorphisms (resp. epimorphisms), then $\g_\infty$ is also  a monomorphism
(resp. epimorphism). This result also remains true for the general case of 
 morphism between two different direct systems.  
\end{theo}

\begin{cor}\label{ch511} \mar{ch511} \cite{massey}. The operation of taking
homology groups of chain and cochain complexes commutes with the passage to
the direct limit.
\end{cor}

The operation of multiplication 
\be
\f\to f\f, \qquad  f\in
C^\infty(X), \qquad \f\in \cO^*_r
\ee
 has the direct limit, and $\cO^*_\infty$ possesses the structure of
$C^\infty(X)$-algebra. The operations of the exterior product
$\w$ and the exterior differential $d$ also have the direct limits on
$\cO_\infty^*$. We will denote them by the same symbols
$\w$ and $d$, respectively. They provide $\cO_\infty^*$ with the structure of a
$\Bbb Z$-graded exterior algebra:
\be
\cO_\infty^*=\op\bigoplus_{m=0}^\infty \cO^m_\infty,
\ee
where $\cO^m_\infty$  are the direct limits of the
direct systems 
\be
\cO^m(X)\op\longrightarrow^{\pi^*} \cO^m_0 \op\longrightarrow^{\pi^1_0{}^*}
\cO_1^m \longrightarrow
\cdots
 \cO_r^m \op\longrightarrow^{\pi^{r+1}_r{}^*} \cO_{r+1}^m \longrightarrow
\cdots
\ee
of $\Bbb R$-modules $\cO_r^m$ of exterior $m$-forms on
$r$-order jet manifolds $J^rY$.
We agree to call lements of $\cO^m_\infty$ the exterior $m$-forms on the
infinite order jet space. The familiar relations of an exterior algebra
take place:
\be
&&\cO^i_\infty\w\cO^j_\infty\subset \cO^{i+j}_\infty,\\
&& d: \cO^i_\infty\to \cO^{i+1}_\infty,  \\
&& d\circ d= 0.
\ee

As a consequence, we have the following De Rham complex
of exterior forms on the infinite
order  jet space
\mar{5.13} \beq
0\longrightarrow \Bbb
R\longrightarrow
\cO^0_\infty\op\longrightarrow^d\cO^1_\infty\op\longrightarrow^d
\cdots\,.
\label{5.13}
 \eeq
Let us consider the cohomology group $H^m(\cO_\infty^*)$
 of this complex.
By virtue of Corollary \ref{ch511}, this is isomorphic to the direct limit
of the direct system of homomorphisms
\be
H^m(\cO_r^*)\longrightarrow H^m(\cO_{r+1}^*)\longrightarrow \cdots
\ee
 of the cohomology groups  $H^m(\cO_r^*)$ of the cochain
complexes
 \be
0\longrightarrow\Bbb R
\longrightarrow \cO^0_r\op\longrightarrow^d\cO^1_r\op\longrightarrow^d
\cdots
\op\longrightarrow^d \cO^l_r\to 0, \qquad l=\di J^rY,
 \ee
i.e., of the De Rham cohomology groups $H^m(\cO_r^*)=H^m(J^rY)$ of jet
manifolds $J^rY$.
 The following assertion completes our consideration of cohomology of
the complex (\ref{5.13}).

\begin{prop}\label{ch512} \mar{ch512} The De Rham cohomology $H^*(J^rY)$ of jet
manifolds
$J^rY$ coincide with the De Rham cohomology $H^*(Y)$ of the fibre bundle
$Y\to X$ \cite{ander,bau}. 
\end{prop}

The proof is based on the fact that the fibre bundle $J^rY\to
J^{r-1}Y$ is affine, and it has the same De Rham cohomology than its
base. 
It follows that the cohomology groups $H^m(\cO_\infty^*)$, $m>0$, of the
cochain complex (\ref{5.13}) coincide with the De Rham cohomology
groups $H^m(Y)$ of $Y\to X$. 

Given a manifold coordinate atlas (\ref{jet1}) of $J^\infty Y$, 
the elements of the direct limit $\cO^*_\infty$ can be
written in the coordinate form as exterior forms on finite order jets.
In particular, the basic 1-forms $dx^\la$ and the contact 1-forms $\th^i_\La$ 
(\ref{55.15}) constitute the set of local generating elements of the 
$\cO^0_\infty$-module $\cO_\infty^1$ of 1-forms on $J^\infty Y$. Moreover, the
basic 1-forms $dx^\la$ and the contact 1-forms $\th^i_\La$ have independent
coordinate transformation laws. It follows that there is the
canonical splitting 
\mar{+416}\beq
\cO_\infty^1= \cO_\infty^{0,1} \oplus \cO_\infty^{1,0} \label{+416}
\eeq
of the module $\cO_\infty^1$ in the 
$\cO^0_\infty$-submodules $\cO_\infty^{0,1}$ and $\cO_\infty^{1,0}$
generated separately by basic and contact forms.
The splitting (\ref{+416}) provides the canonical
decomposition 
\be
\cO^*_\infty =\op\oplus_{k,s}\cO^{k,s}_\infty, \qquad 0\leq k, \qquad
0\leq s\leq n,
\ee
of $\cO^*_\infty$ into $\cO^0_\infty$-modules $\cO^{k,s}_\infty$
of $k$-contact and $s$-horizontal forms, together with the corresponding
projections
\be
h_k:\cO^*_\infty\to \cO^{k,*}_\infty, \quad 0\leq k, \qquad
h^s:\cO^*_\infty\to \cO^{*,s}_\infty, \quad 0\leq s
\leq n.
\ee 
Accordingly, the
exterior differential on $\cO_\infty^*$ is split
into the sum $d=d_H+d_V$ of horizontal and vertical
differentials such that
\be
&& d_H\circ h_k=h_k\circ d\circ h_k, \qquad d_H(\f)=
dx^\la\w d_\la(\f), \\ 
&& d_V \circ h^s=h^s\circ d\circ h^s, \qquad
d_V(\f)=\th^i_\La \w \dr^\La_i\f, \qquad \f\in\cO^*_\infty.
\ee

The operators $d_H$ and $d_V$ obey the familiar relations 
\be
&& d_H(\f\w\si) = d_H(\f)\w\si +(-1)^{\mid\f\mid}\f\w d_H(\si),\qquad
\f,\si\in \cO_\infty^*, \\
&& d_V(\f\w\si) = d_V(\f)\w\si +(-1)^{\mid\f\mid}\f\w d_V(\si),
\ee
and possess the nilpotency property
\mar{5.175}\beq
d_H\circ d_H=0, \qquad d_V\circ d_V=0, \qquad d_V\circ d_H +d_H\circ d_V=0.
\label{5.175}
\eeq
Recall also the relation
\be
h_0\circ d= d_H\circ h_0.
\ee
The horizontal differential can be written in the form
\mar{+417}\beq
 d_H\f= dx^\la\w d_\la(\f), \label{+417}
\eeq
where $d_\la$ (\ref{5.177})
are the total derivatives in infinite order
jets. 
It should be emphasized that, though the sum in the expression
(\ref{5.177}) is taken with respect to an infinite number of collections $\La$,
the operator (\ref{5.177}) is well defined since, given
any form $\f\in \cO^*_\infty$, the expression $d_\la(\f)$ involves only a
finite number of the terms $\dr_i^\La$.
The total derivatives 
satisfy the relations 
\be
&& d_\la(\f\w\si)=d_\la(\f)\w\si +\f\w d_\la(\si),\\
&& d_\la(d\f)=d(d_\la(\f)), \\
&& [d_\la, d_\al]=0,
\ee
and, in contrast with partial derivatives $\dr_\la$, they have the coordinate
transformation law
\be
d'_\la =\frac{\dr x^\m}{\dr {x'}^\la}d_\m.
\ee
The operators
$d_H$ and $d_V$ take the following coordinate form:
\be
\begin{array}{lcll}
d_H f= d_\la f dx^\la, & \qquad\qquad &  d_Vf= \dr_i^\La f\th^i_\La, \qquad &
f\in \cO^0_\infty,\\ 
d_H(dx^\m) =0, & &  d_V(dx^i)=0, &\\
d_H(\th^i_\La)= dx^\la\w \th^i_{\la+\La},
& & d_V(\th^i_\La)= 0 & 0\leq\nm\La.
\end{array}
\ee

One can think of the
splitting (\ref{+416})
as being the canonical horizontal splitting. It is similar both to
the horizontal
splitting (\ref{cc3}) of the cotangent bundle of a fibre bundle
by means of a connection and the
canonical horizontal splittings (\ref{+403}) of 1-forms on finite order jet
manifolds. Therefore, one can say that the splitting (\ref{+416})
defines the canonical connection
 on the infinite order jet space $J^\infty Y$.

Given a vector field $\tau$ on $X$, let us consider
the map
\mar{+419}\beq
\nabla_\tau:\cO_\infty^0\ni f\to \tau\rfloor (d_H f)\in \cO_\infty^0.
\label{+419}
\eeq 
This is a derivation of the ring $\cO_\infty^0$. Moreover, if
$\cO_\infty^0$ is regarded as a $C^\infty(X)$-ring, the map
(\ref{+419}) satisfies the Leibniz rule.
Hence, the assignment
\mar{+418}\beq
\nabla: \tau\mapsto \tau\rfloor d_H= \tau^\la d_\la \label{+418}
\eeq
is the canonical connection on the
$C^\infty(X)$-ring
$\cO_\infty^0$ \cite{book00}.  One can think
of $\nabla_\tau$ (\ref{+418}) as being the horizontal lift
$\tau^\la \dr_\la \mapsto \tau^\la d_\la$
onto $J^\infty Y$ of a vector field $\tau$ on $X$ by means of a canonical
connection on the (topological) fibre bundle $J^\infty Y\to X$. 

However, $\nabla_\tau$ (\ref{+418}) on $J^\infty Y$ is not a 
projectable vector
field on $J^\infty Y$, though
they projected over vector fields on $X$. Projectable vector fields on
$J^\infty Y$ (their definition is a repetition of that for finite order jet
manifolds) are elements of the projective limit
$\cP^\infty$ of the inverse system (\ref{55.6}). This projective limit exists.
Its definition is a repetition of that of $J^\infty Y$. This is a Lie algebra
such that the surjections
\be
T\pi^\infty_k: \cP^\infty \to \cP^k
\ee
are Lie algebra morphisms which constitute the commutative diagrams
\be
\begin{array}{rcl}
& \cP^\infty  & \\
_{T\pi^\infty_k} & \swarrow \searrow &_{T\pi^\infty_r}\\
\cP^k & \ar_{T\pi^k_r} & \cP^r
\end{array}
\ee
for any $k$ and $r<k$. In brief, we
will say that elements of $\cP^\infty$ are vector fields
on the
infinite order jet space $J^\infty Y$.

In particular, let $u$ be a projectable
vector field on $Y$. There exists an element
$J^\infty u\in \cP^\infty$ such that 
\be
T\pi^\infty_k(J^\infty u)=J^ku, \qquad \forall k\geq 0.
\ee 
One can think of $J^\infty u$ as being the $\infty$-order jet prolongation of
the vector field $u$ on $Y$. It is given by the recurrence formula
(\ref{55.5}) where $0\leq |\La|$. Then any element of $\cP^\infty$ is
decomposed into the sum similar to (\ref{+404}) where $k=\infty$. Of course,
it is not the horizontal decomposition. Given a vector field $\up$ on
$J^\infty Y$, projected onto a vector field $\tau$ on $X$, we have its
horizontal splitting 
\be
\up=\up_H+\up_V= \tau^\la d_\la + (\up -\tau^\la d_\la)
\ee
by means of the canonical connection $\nabla$ (\ref{+418}) (cf. (\ref{+402})).
Note that the component $\up_V$ of this splitting is not a projectable
vector field on $J^\infty Y$, but is a vertical vector field with respect to
the fibration
$J^\infty Y\to X$.

Though $\nabla_\tau$ (\ref{+418}) on $J^\infty Y$ is not an element of
the projective limit $\cP^\infty$, it is also a vector
field on $J^\infty Y$ as follows.

A real function $f: J^\infty Y\to \Bbb R$
is said to be smooth if, for every $q\in
J^\infty Y$, 
there exists a neighbourhood $U$ of $q$ and a smooth function 
$f^{(k)}$ on  $J^kY$
for some $k$ such that
\be
f|_U= f^{(k)}\circ \pi^\infty_k\mid_U.
\ee
Then the same equality takes place for any $r>k$. 
Smooth functions on
$J^\infty Y$ constitute an $\Bbb R$-ring $C^\infty(J^\infty Y)$. 
In particular, the pull-back $\pi^{\infty*}_rf$ of any smooth function on
$J^rY$ is a smooth function on $J^\infty Y$, and there is a monomorphism
$\cO^0_\infty\to C^\infty(J^\infty Y)$. The key point is that the
paracompact space $J^\infty Y$ admits partition of unity by elements of
the ring $C^\infty(J^\infty Y)$ \cite{tak2}.

Vector fields on  $J^\infty Y$ can be defined as derivations of the ring
$C^\infty(J^\infty Y$.
Since a derivation of $\cO^0_\infty$
is a local operation and
$J^\infty Y$ admits a smooth partition of unity, the derivations (\ref{+419})
can be extended to the ring
$C^\infty(J^\infty Y)$ of smooth functions on the infinite order jet space
$J^\infty Y$. Accordingly, the connection $\nabla$ (\ref{+418}) is extended
to the canonical connection  on the $C^\infty(X)$-ring $C^\infty(J^\infty Y)$.
Extended to $C^\infty(J^\infty Y)$, the derivations (\ref{+419}), by
definition, are vector fields 
on the infinite order jet space $J^\infty Y$.

\section{The variational bicomplex}

The algebra $\cO^*_\infty$ together with the horizontal differential
$d_H$ and the variational operator $\dl$ constitute the variational
bicomplex of exterior forms on $J^\infty Y$. Cohomology of this
bicomplex provide solution of the global inverse problem of the
calculus of variations in field theory. Moreover, extended to the jet
space of ghosts and antifields, the algebra $\cO^*_\infty$ is the main
ingredient in the field-antfield BRST theory for studying BRST
cohomology modulo $d_H$. Passing to the direct limit of the de Rham
complexes of exterior forms on finite order jet manifolds, the de Rham
cohomology has been found in Proposition \ref{ch512}. However, this is
not a way of studying other cohomology of the algebra $\cO^*_\infty$.

To solve this problem, we enlarge $\cO^*_\infty$ to the graded
differential algebra $\cQ^*_\infty$ of exterior forms
which are 
locally the pull-back of exterior forms on finite order jet manifolds. 
The de Rham cohomology, $d_H$- and $\dl$-cohomology of
$\cQ^*_\infty$ have been investigated in \cite{ander80,tak2}. 
Then one can show that the graded differential algebra 
$\cO^*_\infty$ has the same $d_H$- and $\dl$-cohomology as $\cQ^*_\infty$
\cite{jpa99,lmp}.

We follow the terminology of
\cite{bred,hir}, where a sheaf $S$ is a particular topological bundle,
$\ol S$ denotes the canonical presheaf of
sections of the sheaf $S$, and 
$\G(S)$ is the group of global sections of $S$. 

\glos{A. The variational bicomplex}

Let $\gO^*_r$ be a sheaf
of germs of exterior forms on the $r$-order jet manifold $J^rY$ and 
$\ol\gO^*_r$ its canonical presheaf.  There is the direct system of canonical
presheaves
\be
\ol\gO^*_X\op\longrightarrow^{\pi^*} \ol\gO^*_0 
\op\longrightarrow^{\pi^1_0{}^*} \ol\gO_1^*
\op\longrightarrow^{\pi^2_1{}^*} \cdots \op\longrightarrow^{\pi^r_{r-1}{}^*}
 \ol\gO_r^* \longrightarrow\cdots, 
\ee
where $\pi^r_{r-1}{}^*$ are the pull-back monomorphisms. Its direct
limit $\ol\gO^*_\infty$ 
is a presheaf of graded differential
$\Bbb R$-algebras on
$J^\infty Y$. Let $\gQ^*_\infty$ be a sheaf constructed from 
$\ol\gO^*_\infty$, $\ol\gQ^*_\infty$ its canonical presheaf, 
and $\cQ^*_\infty=\G(\gQ^*_\infty)$ the structure algebra of 
sections of
the sheaf $\gQ^*_\infty$. 
In particular, $\cQ^0_\infty=C^\infty(J^\infty Y)$.
There are 
$\Bbb R$-algebra monomorphisms 
 $\ol\gO^*_\infty
\to\ol\gQ^*_\infty$ and  $\cO^*_\infty
\to\cQ^*_\infty$. 
The key point is that, since the paracompact space
$J^\infty Y$ admits a partition of unity by elements of the ring
$\cQ^0_\infty$, the sheaves of
$\cQ^0_\infty$-modules on
$J^\infty Y$ are fine and, consequently, acyclic. Therefore, the 
abstract de Rham theorem on cohomology of a sheaf resolution \cite{hir}
can be called into play in order to obtain cohomology of the graded
differential algebra $\cQ^*_\infty$. 

For short, we
agree to call elements of $\cQ^*_\infty$ the
exterior forms on
$J^\infty Y$, too.  Restricted to a
coordinate chart
$(\pi^\infty_0)^{-1}(U)$ of $J^\infty Y$, they
as like as elements of $\cO^*_\infty$ can be written in a coordinate
form, where horizontal forms 
$\{dx^\la\}$ and contact 1-forms
$\{\th^i_\La=dy^i_\La -y^i_{\la+\La}dx^\la\}$ provide local
generators of the algebra
$\cQ^*_\infty$. 
There is the canonical decomposition
\be
\cQ^*_\infty =\op\oplus_{k,s}\cQ^{k,s}_\infty, \qquad 0\leq k, \qquad
0\leq s\leq n,
\ee
of $\cQ^*_\infty$ into $\cQ^0_\infty$-modules $\cQ^{k,s}_\infty$
of $k$-contact and $s$-horizontal forms.
Accordingly, the
exterior differential on $\cQ_\infty^*$ is split
into the sum $d=d_H+d_V$ of horizontal and vertical
differentials.

Being nilpotent, the
differentials $d_V$ and $d_H$ provide the natural bicomplex
$\{\gQ^{k,m}_\infty\}$ of  the sheaf
$\gQ^*_\infty$ on $J^\infty Y$. To complete it to the
variational bicomplex, one defines the projection $\Bbb R$-module
endomorphism 
\be
&& \tau=\op\sum_{k>0} \frac1k\ol\tau\circ h_k\circ h^n, \\ 
&&\ol\tau(\f)
=(-1)^{\nm\La}\th^i\w [d_\La(\dr^\La_i\rfloor\f)], \qquad 0\leq\nm\La,
\qquad \f\in \ol\gO^{>0,n}_\infty,
\ee
of $\ol\gO^*_\infty$ such that
\be
\tau\circ d_H=0, \qquad  \tau\circ d\circ \tau -\tau\circ d=0.
\ee
Introduced on elements of the presheaf $\ol\gO^*_\infty$ 
(see, e.g., \cite{bau,book,tul}), this endomorphism is induced on the
sheaf $\gQ^*_\infty$ and its structure algebra
$\cQ^*_\infty$. Put
\be
\gE_k=\tau(\gQ^{k,n}_\infty), \qquad E_k=\tau(\cQ^{k,n}_\infty), \qquad k>0.
\ee
Since
$\tau$ is a projection operator, we have isomorphisms 
\be
\ol\gE_k=\tau(\ol\gQ^{k,n}_\infty), \qquad E_k=\G(\gE_k).
\ee
The variational operator on $\gQ^{*,n}_\infty$ is defined as the
morphism $\dl=\tau\circ d$. 
It is nilpotent, and obeys the relation 
\beq
\dl\circ\tau-\tau\circ d=0. \label{am13}
\eeq

Let $\Bbb R$ and  $\gO^*_X$ denote the constant sheaf
on
$J^\infty Y$ and the sheaf of exterior forms on $X$, respectively. The
operators $d_V$,
$d_H$,
$\tau$ and $\dl$ give the following variational bicomplex of
sheaves of differential forms on $J^\infty Y$:
\beq
\begin{array}{ccccrlcrlccrlccrlcrl}
& & & & _{d_V} & \put(0,-7){\vector(0,1){14}} & & _{d_V} &
\put(0,-7){\vector(0,1){14}} & &  & _{d_V} &
\put(0,-7){\vector(0,1){14}} & & &  _{d_V} &
\put(0,-7){\vector(0,1){14}}& & _{-\dl} & \put(0,-7){\vector(0,1){14}} \\ 
 &  & 0 & \to & &\gQ^{k,0}_\infty &\ar^{d_H} & & \gQ^{k,1}_\infty &
\ar^{d_H} &\cdots  & & \gQ^{k,m}_\infty &\ar^{d_H} &\cdots & &
\gQ^{k,n}_\infty &\ar^\tau &  & \gE_k\to  0\\  
 & &  &  & & \vdots & & & \vdots  & & & 
&\vdots  & & & &
\vdots & &   & \vdots \\ 
& & & & _{d_V} & \put(0,-7){\vector(0,1){14}} & & _{d_V} &
\put(0,-7){\vector(0,1){14}} & &  & _{d_V} &
\put(0,-7){\vector(0,1){14}} & & &  _{d_V} &
\put(0,-7){\vector(0,1){14}}& & _{-\dl} & \put(0,-7){\vector(0,1){14}} \\ 
 &  & 0 & \to & &\gQ^{1,0}_\infty &\ar^{d_H} & & \gQ^{1,1}_\infty &
\ar^{d_H} &\cdots  & & \gQ^{1,m}_\infty &\ar^{d_H} &\cdots & &
\gQ^{1,n}_\infty &\ar^\tau &  & \gE_1\to  0\\  
& & & & _{d_V} &\put(0,-7){\vector(0,1){14}} & & _{d_V} &
\put(0,-7){\vector(0,1){14}} & & &  _{d_V}
 & \put(0,-7){\vector(0,1){14}} & &  & _{d_V} & \put(0,-7){\vector(0,1){14}}
 & & _{-\dl} & \put(0,-7){\vector(0,1){14}} \\
0 & \to & \Bbb R & \to & & \gQ^0_\infty &\ar^{d_H} & & \gQ^{0,1}_\infty &
\ar^{d_H} &\cdots  & &
\gQ^{0,m}_\infty & \ar^{d_H} & \cdots & &
\gQ^{0,n}_\infty & \equiv &  & \gQ^{0,n}_\infty \\
& & & & _{\pi^{\infty*}}& \put(0,-7){\vector(0,1){14}} & & _{\pi^{\infty*}} &
\put(0,-7){\vector(0,1){14}} & & &  _{\pi^{\infty*}}
 & \put(0,-7){\vector(0,1){14}} & &  & _{\pi^{\infty*}} &
\put(0,-7){\vector(0,1){14}} & &  & \\
0 & \to & \Bbb R & \to & & \gO^0_X &\ar^d & & \gO^1_X &
\ar^d &\cdots  & &
\gO^m_X & \ar^d & \cdots & &
\gO^n_X & \ar^d & 0 &  \\
& & & & &\put(0,-5){\vector(0,1){10}} & & &
\put(0,-5){\vector(0,1){10}} & & & 
 & \put(0,-5){\vector(0,1){10}} & & &   &
\put(0,-5){\vector(0,1){10}} & &  & \\
& & & & &0 & &  & 0 & & & & 0 & & & & 0 & &  & 
\end{array}
\label{7}
\eeq
The second row and the last column of this bicomplex form the 
variational complex
\beq
0\to\Bbb R\to \gQ^0_\infty \ar^{d_H}\gQ^{0,1}_\infty\ar^{d_H}\cdots  
\op\longrightarrow^{d_H} 
\gQ^{0,n}_\infty  \op\longrightarrow^\dl \gE_1 
\op\longrightarrow^\dl 
\gE_2 \longrightarrow \cdots\, . \label{tams1}
\eeq
The corresponding variational bicomplexes and variational complexes of
graded differential algebras 
$\cQ^*_\infty$ and $\cO^*_\infty$ take place.

There are the well-known statements summarized usually as
the algebraic Poincar\'e lemma (see, e.g., \cite{olver,tul}). 

\begin{lem} \label{am12} 
If $Y$ is a contactible bundle $\Bbb R^{n+p}\to \Bbb R^n$, the
variational bicomplex of the graded differential algebra $\cO^*_\infty$
is exact.
\end{lem}

It follows that the variational bicomplex (\ref{7}) and, consequently,
the variational complex (\ref{tams1}) are exact for any smooth bundle
$Y\to X$.
Moreover, the sheaves
$\gQ^{k,m}_\infty$ in this bicomplex are fine, and so are the sheaves
$\gE_k$ in accordance with the following lemma.

\begin{lem} \label{lmp03} 
Sheaves $\gE_k$ are fine.
\end{lem}

\begin{proof}
Though the $\Bbb R$-modules $E_{k>1}$ fail to be
$\cQ^0_\infty$-modules \cite{tul}, one can use the fact that the sheaves
$\gE_{k>0}$ are projections $\tau(\gQ^{k,n}_\infty)$ of sheaves of
$\cQ^0_\infty$-modules. Let $\{U_i\}_{i\in I}$
be a 
locally finite open covering  of
$J^\infty Y$ and $\{f_i\in\cQ^0_\infty\}$ the associated partition of unity. 
For any open subset $U\subset J^\infty Y$ and any section
$\varphi$ of 
the sheaf $\gQ^{k,n}_\infty$ over $U$, let us put
$h_i(\varphi)=f_i\varphi$.
The endomorphisms $h_i$ of $\gQ^{k,n}_\infty$ yield the $\Bbb R$-module
endomorphisms 
\be
\ol h_i=\tau\circ h_i: \gE_k\ar^{\rm in} \gQ^{k,n}_\infty \ar^{h_i}
\gQ^{k,n}_\infty \ar^\tau \gE_k
\ee
of the sheaves $\gE_k$.
They possess the properties
required for $\gE_k$ to be a fine sheaf. Indeed, for each $i\in I$, ${\rm
supp}\,f_i\subset U_i$ provides a closed set  such that $\ol h_i$ is zero
outside this set, while the sum $\op\sum_{i\in I}\ol h_i$ is the identity
morphism.
\end{proof}

Thus, the columns and rows of the bicomplex (\ref{7}) as like as the
variational complex (\ref{tams1}) are sheaf resolutions, and the
abstract de Rham theorem can be applied to them. Here, we restrict our
consideration to the variational complex.

\glos{B. Cohomology of $\cQ^*_\infty$}

The variational complex (\ref{tams1}) is a resolution of the constant
sheaf $\Bbb R$ on $J^\infty Y$. Let us start from the following lemma.

\begin{lem} \label{20jpa} 
There is an
isomorphism 
\beq
H^*(J^\infty Y,\Bbb R)= H^*(Y,\Bbb R)=H^*(Y) \label{lmp80}
\eeq
between cohomology $H^*(J^\infty Y,\Bbb R)$ of $J^\infty Y$ with
coefficients in the constant sheaf $\Bbb R$, that $H^*(Y,\Bbb R)$ of $Y$, and
the de Rham cohomology $H^*(Y)$ of $Y$. 
\end{lem}

\begin{proof}
Since $Y$ is a strong deformation retract of $J^\infty Y$ \cite{ander},
the first isomorphism in (\ref{lmp80}) follows from the
Vietoris--Begle theorem
\cite{bred}, while the second
one results from the familiar de Rham theorem.
\end{proof}

Let us consider the de Rham complex of sheaves 
\beq
0\to \Bbb R\to
\gQ^0_\infty\op\longrightarrow^d\gQ^1_\infty\op\longrightarrow^d
\cdots
\label{lmp71}
 \eeq
on $J^\infty Y$ and the corresponding de Rham complex of their structure
algebras
\beq
0\to \Bbb R\to
\cQ^0_\infty\op\longrightarrow^d\cQ^1_\infty\op\longrightarrow^d
\cdots\, .
\label{5.13'}
\eeq
The complex (\ref{lmp71}) is exact due to
the Poincar\'e lemma, and is a resolution of the constant sheaf $\Bbb R$ on
$J^\infty Y$ since sheaves $\gQ^r_\infty$ are fine. Then, the abstract de
Rham theorem and Lemma
\ref{20jpa} lead to the following.

\begin{prop} \label{38jp} 
The de Rham cohomology $H^*(\cQ^*_\infty)$ 
of the graded differential algebra
$\cQ^*_\infty$  is isomorphic to that $H^*(Y)$ of the bundle $Y$.
\end{prop}

It follows that every closed form $\f\in \cQ^*_\infty$
is split into the sum
\beq
\f=\varphi +d\xi, \qquad \xi\in \cQ^*_\infty, \label{tams2} 
\eeq
where $\varphi$ is a closed form on the fiber bundle $Y$. 

Similarly, from the abstract de Rham theorem and Lemma 
\ref{20jpa},  we obtain the following. 

\begin{prop} \label{lmp05} 
There is an isomorphism
between $d_H$- and $\dl$-cohomology of the
variational complex 
\beq
0\to\Bbb R\to \cQ^0_\infty \ar^{d_H}\cQ^{0,1}_\infty\ar^{d_H}\cdots  
\op\longrightarrow^{d_H} 
\cQ^{0,n}_\infty  \op\longrightarrow^\dl E_1 
\op\longrightarrow^\dl 
E_2 \longrightarrow \cdots  \label{b317}
\eeq
and the de Rham cohomology of the fiber bundle
$Y$, namely,
\be
H^{k<n}(d_H;\cQ^*_\infty)=H^{k<n}(Y), \qquad H^{k-n}(\dl;
\cQ^*_\infty)=H^{k\geq n}(Y).
\ee
\end{prop}

This isomorphism  recovers the results of \cite{ander80,tak2}, but
notes also the following.
The relation (\ref{am13}) for $\tau$ and
the relation $h_0d=d_Hh_0$ for $h_0$ define  a homomorphism of the
de Rham complex (\ref{5.13'}) of the algebra $\cQ^*_\infty$ to its variational
complex (\ref{b317}). The corresponding homomorphism of their cohomology
groups is an isomorphism by virtue of Proposition \ref{38jp} and Proposition
\ref{lmp05}. Then, the splitting (\ref{tams2}) leads to the following
decompositions.

\begin{prop} \label{t41} 
Any $d_H$-closed form $\si\in\cQ^{0,m}$, $m< n$, is represented by a sum
\beq
\si=h_0\varphi+ d_H \xi, \qquad \xi\in \cQ^{m-1}_\infty, \label{t60}
\eeq
where $\varphi$ is a closed $m$-form on $Y$.
Any $\dl$-closed form $\si\in\cQ^{k,n}$, $k\geq 0$, is split into
\ben
&& \si=h_0\varphi + d_H\xi, \qquad k=0, \qquad \xi\in \cQ^{0,n-1}_\infty,
\label{t42a}\\ 
&& \si=\tau(\varphi) +\dl(\xi), \qquad k=1, \qquad \xi\in \cQ^{0,n}_\infty,
\label{t42b}\\
&& \si=\tau(\varphi) +\dl(\xi), \qquad k>1, \qquad \xi\in E_{k-1},
\label{t42c}
\een
where $\varphi$ is a closed $(n+k)$-form on $Y$. 
\end{prop}
 
\glos{C. Cohomology of $\cO^*_\infty$}

\begin{theo} \label{am11} 
Graded differential algebra $\cO^*_\infty$ has the same $d_H$- and
$\dl$-cohomology as $\cQ^*_\infty$.
\end{theo}

\begin{proof}
Let the common symbol $D$ stand for $d_H$ and
$\dl$. Bearing in mind decompositions
(\ref{t60}) -- (\ref{t42c}), it suffices to show that, if an
element
$\f\in
\cO^*_\infty$ is
$D$-exact in the algebra $\cQ^*_\infty$, then it is
so in the algebra
$\cO^*_\infty$. Lemma \ref{am12} states that, if
$Y$ is a contractible bundle and a $D$-exact form $\f$ on $J^\infty Y$
is of finite jet order
$[\f]$ (i.e., $\f\in\cO^*_\infty$), there exists an exterior form $\varphi\in
\cO^*_\infty$ on $J^\infty Y$ such that $\f=D\varphi$. Moreover, a glance at
the homotopy operators for $d_H$ and $\dl$ \cite{olver} shows that  the
jet order
$[\varphi]$ of $\varphi$ is bounded by an integer $N([\f])$, depending
only on the jet order of $\f$. 
Let us call this fact the finite exactness of the operator
$D$. Given an arbitrary bundle
$Y$, the finite exactness takes place on $J^\infty Y|_U$ over any domain
(i.e., a contractible open subset) $U\subset Y$.
Let us prove the following.

(i) Given a family $\{U_\al\}$ of disjoint open subsets of $Y$, let us
suppose that the finite exactness takes place on $J^\infty Y|_{U_\al}$ over
every subset $U_\al$ from this family. Then, it is true on $J^\infty Y$ over
the union
$\op\cup_\al U_\al$ of these subsets.

(ii) Suppose that the finite exactness of the operator $D$ takes place on
$J^\infty Y$ over open subsets
$U$, $V$ of $Y$ and their non-empty overlap $U\cap V$. Then, it is also true on
$J^\infty Y|_{U\cup V}$.

\noindent
{\it Proof of (i)}. Let
$\f\in\cO^*_\infty$ be a $D$-exact form on
$J^\infty Y$.
The finite exactness on $(\pi^\infty_0)^{-1}(\cup
U_\al)$ holds since $\f=D\varphi_\al$ on every $(\pi^\infty_0)^{-1}(U_\al)$
and $[\varphi_\al]< N([\f])$. 

\noindent
{\it Proof of (ii)}. Let
$\f=D\varphi\in\cO^*_\infty$ be a $D$-exact form on
$J^\infty Y$. By assumption, it can be brought into the form
$D\varphi_U$ on $(\pi^\infty_0)^{-1}(U)$ and $D\varphi_V$ on
$(\pi^\infty_0)^{-1}(V)$, where
$\varphi_U$ and $\varphi_V$ are exterior forms of bounded jet
order. Let us consider their difference $\varphi_U-\varphi_V$ on 
$(\pi^\infty_0)^{-1}(U\cap V)$. It is a $D$-exact form of bounded jet
order $[\varphi_U-\varphi_V]< N([\f])$
which, by assumption, can be written as 
$\varphi_U-\varphi_V=D\si$ where 
$\si$ is also of bounded jet order $[\si]<N(N([\f]))$. 
Lemma
\ref{am20} below shows that $\si=\si_U +\si_V$ where
$\si_U$ and
$\si_V$ are exterior forms of bounded jet order on $(\pi^\infty_0)^{-1}(U)$ and
$(\pi^\infty_0)^{-1}(V)$, respectively. Then, putting
\be
\varphi'|_U=\varphi_U-D\si_U, \qquad  
\varphi'|_V=\varphi_V+ D\si_V,
\ee
we have the form $\f$, equal to
$D\varphi'_U$ on $(\pi^\infty_0)^{-1}(U)$ and
$D\varphi'_V$ on $(\pi^\infty_0)^{-1}(V)$, respectively. Since the
difference $\varphi'_U -\varphi'_V$ on $(\pi^\infty_0)^{-1}(U\cap V)$ vanishes,
we obtain $\f=D\varphi'$ on $(\pi^\infty_0)^{-1}(U\cup V)$ where 
\be
\varphi'\op=^{\rm def}\left\{
\begin{array}{ll}
\varphi'|_U=\varphi'_U, &\\
\varphi'|_V=\varphi'_V &
\end{array}\right.
\ee
is of bounded jet order $[\varphi']<N(N([\f]))$.  

\noindent
To prove the finite
exactness of
$D$ on $J^\infty Y$, it remains to choose an appropriate cover
of $Y$. A smooth manifold $Y$ admits a countable cover $\{U_\xi\}$ by
domains $U_\xi$, $\xi\in {\bf N}$, and its refinement 
$\{U_{ij}\}$, where $j\in {\bf N}$ and $i$ runs through a finite set,
such that $U_{ij}\cap U_{ik}=\emptyset$, $j\neq k$ \cite{greub}.
Then $Y$ has a finite cover $\{U_i=\cup_j U_{ij}\}$.
Since the finite exactness of the operator $D$ takes place over any domain
$U_\xi$, it also holds over any member $U_{ij}$ of the refinement $\{U_{ij}\}$
of $\{U_\xi\}$ and, in accordance with item (i) above, over any member of
the finite cover $\{U_i\}$ of $Y$. Then by virtue of item (ii) above,
the finite 
exactness of $D$ takes place over $Y$.
\end{proof}

\begin{lem} \label{am20} 
Let $U$ and $V$ be open subsets of a bundle $Y$ and $\si\in
\gO^*_\infty$ an exterior form of bounded jet order on
$(\pi^\infty_0)^{-1}(U\cap V)\subset J^\infty Y$. Then, $\si$ is split
into  a sum $\si_U+ \si_V$ of exterior forms $\si_U$ and $\si_V$ of bounded jet
order on
$(\pi^\infty_0)^{-1}(U)$ and $(\pi^\infty_0)^{-1}(V)$, respectively. 
\end{lem} 

\begin{proof}
By taking a smooth partition of unity on $U\cup V$ subordinate to the cover
$\{U,V\}$ and passing to the function with support in $V$, one gets a
smooth real function
$f$ on
$U\cup V$ which is 0 on a neighborhood of $U-V$ and 1 on a neighborhood of
$V-U$ in $U\cup V$. Let $(\pi^\infty_0)^*f$ be the pull-back of $f$ onto
$(\pi^\infty_0)^{-1}(U\cup V)$. The exterior form $((\pi^\infty_0)^*f)\si$ is
0 on a neighborhood of $(\pi^\infty_0)^{-1}(U)$ and, therefore, can be
extended by 0 to $(\pi^\infty_0)^{-1}(U)$. Let us denote it $\si_U$.
Accordingly, the exterior form
$(1-(\pi^\infty_0)^*f)\si$ has an extension $\si_V$ by 0 to 
$(\pi^\infty_0)^{-1}(V)$. Then, $\si=\si_U +\si_V$ is a desired decomposition
because $\si_U$ and $\si_V$
are of the jet order which does not exceed that of $\si$. 
\end{proof}

\glos{D. The global inverse problem}

The expressions (\ref{t42a}) -- (\ref{t42b}) in Proposition \ref{t41} provide a
solution of the global inverse problem of the calculus of variations on fiber
bundles in the class of Lagrangians $L\in\cQ^{0,n}_\infty$ of locally finite
order \cite{ander80,tak2} (which is not so interesting for physical
applications). These expressions together with Theorem \ref{am11} give a
solution of the global inverse problem of the finite order calculus of
variations.

\begin{cor} \label{lmp112'}
(i) A finite order Lagrangian $L\in \cO^{0,n}_\infty$ is variationally trivial,
i.e.,  $\dl(L)=0$ iff 
\beq
L=h_0\varphi + d_H \xi, \qquad \xi\in \cO^{0,n-1}_\infty, \label{tams3}
\eeq
where $\varphi$ is a closed $n$-form on $Y$.
(ii) A finite order 
Euler--Lagrange-type operator satisfies the Helmholtz
condition $\dl(\cE)=0$ iff 
\be
\cE=\dl(L) + \tau(\f), \qquad L\in\cO^{0,n}_\infty, 
\ee
where $\f$ is a closed $(n+1)$-form on $Y$.
\end{cor}

Note that item (i) in Corollary \ref{lmp112'} contains the particular
result of  \cite{vin}. 

A solution of the global inverse problem of the calculus of 
variations in
the class of exterior forms of bounded jet order has been suggested in
\cite{ander80} by a computation of cohomology of a fixed
order variational sequence. However, this computation requires
rather sophisticated {\it ad hoc} technique in order to be reproduced 
(see \cite{kru90,kru98,vit} for a different
variational sequence). The theses of Corollary
\ref{lmp112'} also agree with those of \cite{ander80}, but the proof of
Theorem \ref{am11} does not give a sharp bound on the order of 
a Lagrangian.

\section{Geometry of simple graded manifolds}

The most of odd fields in quantum field theory can be described in terms of
graded manifolds. These are fermions and
odd ghosts and antifields. It should be emphasized that
graded manifolds are not supermanifolds, though every
graded manifold determines a DeWitt
$H^\infty$-supermanifold, and {\it vice versa}.  Referring the reader to
\cite{bart,kost77,stavr} for a
general theory of graded manifolds, we here focus on the most physically
relevant case of simple graded manifolds. We do not restrict a
class of graded manifolds in question, but an arbitrary graded manifold
can be brought into a certain explicit form
(see Batchelor's Theorem \ref{lmp1} below) which, of course, narrows
the class of automorphisms of a graded manifold.
n physical applications, graded manifolds
are usually given in this form from the beginning.

By a graded manifold of dimension $(n,m)$ is
meant a locally ringed space
$(Z,\gR)$ where $Z$ is an $n$-dimensional smooth manifold $Z$ and
$\gR=\gR_0\oplus\gR_1$ is a sheaf of graded commutative
algebras of rank $m$ such that \cite{bart}:
\begin{itemize}
\item there is the exact sequence of sheaves
\mar{cmp140}\beq
0\to \cR \to\gR \op\to^\si C^\infty_Z\to 0, \qquad
\cR=\gR_1+(\gR_1)^2,\label{cmp140}
\eeq
where $C^\infty_Z$ is the sheaf of smooth functions on $Z$;
\item $\cR/\cR^2$ is a locally free
$C^\infty_Z$-module of finite rank (with respect to pointwise operations),
and the sheaf $\gR$ is locally isomorphic to the
exterior algebra (or the exterior bundle) $\w_{C^\infty_Z}(\cR/\cR^2)$.
\end{itemize}
The sheaf $\gR$ is called a structure sheaf of the graded manifold
$(Z,\gR)$, while the manifold $Z$ is said to be a body of $(Z,\gR)$.
Global sections of the sheaf $\gR$ are called graded functions.
They constitute the structure module $\gR(Z)$ of the sheaf $\gR$.

A graded manifold $(Z,\gR)$, by definition, has the following local
structure.  Given a point $z\in Z$, there exists its
open neighbourhood $U$, called a splitting
domain, such that
\mar{+54}\beq
\gR(U)\cong C^\infty(U)\ot\w\Bbb R^m. \label{+54}
\eeq
It means that
the restriction $\gR\mid_U$ of the structure sheaf
$\gR$ to
$U$ is isomorphic to the sheaf $C^\infty_U\ot\w\Bbb R^m$ of sections of some
exterior bundle
$\w E^*_U= U\times \w\Bbb R^m\to U$.

The
well-known Batchelor's theorem
\cite{bart,batch1} states that such a
structure of graded manifolds is global.

\begin{theo} \label{lmp1} \mar{lmp1}
Let $(Z,\gR)$ be a graded manifold. There exists a vector bundle
$E\to Z$ with an $m$-dimensional
typical fibre $V$ such that the structure sheaf $\gR$ of
$(Z,\gR)$ is isomorphic
to the structure sheaf $\gR_E=C^\infty_Z\ot\w V^*$
of sections of the exterior bundle $\w E^*$,
whose typical fibre is the Grassmann algebra $\w V^*$.
\end{theo}

It should be emphasized that Batchelor's isomorphism in Theorem 
\ref{lmp1} fails
to be canonical. At the same time,
there are many physical models where a vector bundle $E$ is
introduced from the beginning. In this case, it suffices to consider
the structure sheaf
$\gR_E$ of the exterior bundle $\w E^*$ \cite{gaw,book00,sardijmp}.
We agree to call the pair $(Z,\gR_E)$ a simple graded
manifold. 
Its automorphisms are restricted to those, induced
by automorphisms of the vector bundle $E\to Z$. This is called the
characteristic vector bundle
of the simple graded manifold
$(Z,\gR_E)$. Accordingly, the structure
module $\gR_E(Z)=\w E^*(Z)$ of the sheaf $\gR_E$ (and of the exterior
bundle $\w E^*$)
is said to be the structure module of the simple graded manifold
$(Z,\gR_E)$. 

Given a simple graded manifold $(Z,\gR_E)$, every trivialization
chart $(U; z^A,y^a)$ of the vector bundle $E\to Z$ is a splitting
domain of $(Z,\gR_E)$. Graded functions on such a chart are
$\La$-valued function
\mar{z785}\beq
f=\op\sum_{k=0}^m \frac1{k!}f_{a_1\ldots
a_k}(z)c^{a_1}\cdots c^{a_k}, \label{z785}
\eeq
where $f_{a_1\cdots
a_k}(z)$ are smooth functions on $U$,
$\{c^a\}$ is the  fibre basis for $E^*$, and we omit the symbol of the exterior
product of elements $c$.
In particular, the sheaf epimorphism $\si$ in (\ref{cmp140}) is induced by the
body morphism of $\La$. We agree to call $\{z^A,c^a\}$ the local basis
for the graded manifold $(Z,\gR_E)$.
Transition functions $y'^a=\rho^a_b(z^A)y^b$ of bundle coordinates on $E\to Z$
induce the corresponding transformation
\mar{+6}\beq
c'^a=\rho^a_b(z^A)c^b \label{+6}
\eeq
of the associated local basis for the graded manifold $(Z,\gR_E)$ and
the according coordinate
transformation law of graded functions (\ref{z785}).

Let us note that general transformations of a graded manifold take the form
\mar{+95}\beq
c'^a=\rho^a(z^A,c^b), \label{+95}
\eeq
where
$\rho^a(z^A,c^b)$ are local graded functions. Considering only
simple graded manifolds, we actually restrict the class of graded 
manifold transformations
(\ref{+95}) to the linear ones (\ref{+6}), compatible with a given
Batchelor's isomorphism.

\begin{rem}
Although graded functions are locally represented by $\La$-valued functions
(\ref{z785}), they are not $\La$-valued functions on a manifold $Z$ 
because of the
transformation law (\ref{+6}) (or (\ref{+95})).
\end{rem}

Given a graded manifold $(Z,\gR)$, by the sheaf $\gd\gR$ of graded derivations
of $\gR$ is meant a subsheaf of endomorphisms of the structure sheaf
$\gR$ such that any section $u$ of $\gd\gR$ over an open subset $U\subset Z$
is a graded derivation of the graded  algebra
$\gR(U)$, i.e.
\mar{+832}\beq
  u(ff')=u(f)f'+(-1)^{\nw u\nw f}fu (f') \label{+832}
\eeq
for all homogeneous elements $u\in \gd\gR(U)$ and $f,f'\in \gR(U)$.
Conversely, one can show that, given open sets
$U'\subset U$, there is a surjection of the derivation modules $\gd(\gR(U))\to
\gd(\gR(U'))$ \cite{bart}.
It follows that any graded derivation of the local graded algebra
$\gR(U)$ is also a local section over $U$ of the sheaf $\gd\gR$.
  Sections of $\gd\gR$
are  called graded vector fields on the
graded manifold
$(Z,\gR)$.
The graded derivation sheaf $\gd\gR$ is a sheaf of Lie
superalgebras with respect to the bracket
\mar{+120}\beq
[u,u']=uu' + (-1)^{\nw u\nw{u'}+1}u'u. \label{+120}
\eeq

In comparison with general theory of graded manifolds,
an essential simplification is that
graded vector fields on a simple graded manifold
$(Z,\gR_E)$ can be seen as sections of a vector bundle as follows
\cite{book00,sardijmp}.

Due to the vertical splitting
$VE\cong E\times E$ (\ref{1.10}), the vertical tangent bundle
$VE$ of $E\to Z$ can be provided with the fibre bases $\{\dr/\dr
c^a\}$, which are the duals of the bases
$\{c^a\}$. These are the fibre bases for $\pr_2VE\cong E$. Then
graded vector fields on a trivialization chart $(U;z^A,y^a)$ of $E$
read
\mar{hn14}\beq
u= u^A\dr_A + u^a\frac{\dr}{\dr c^a}, \label{hn14}
\eeq
where $u^\la, u^a$ are local graded functions on $U$ \cite{bart,book00}.
In particular,
\be
\frac{\dr}{\dr c^a}\circ\frac{\dr}{\dr c^b}
=-\frac{\dr}{\dr c^b}\circ\frac{\dr}{\dr c^a}, \qquad
\dr_A\circ\frac{\dr}{\dr c^a}=\frac{\dr}{\dr c^a}\circ \dr_A.
\ee
The derivations (\ref{hn14}) act on graded functions
$f\in\gR_E(U)$ (\ref{z785}) by the rule
\mar{cmp50a}\beq
u(f_{a\ldots b}c^a\cdots c^b)=u^A\dr_A(f_{a\ldots b})c^a\cdots c^b +u^k
f_{a\ldots b}\frac{\dr}{\dr c^k}\rfloor (c^a\cdots c^b). \label{cmp50a}
\eeq
This rule implies the corresponding
coordinate transformation law
\be
u'^A =u^A, \qquad u'^a=\rho^a_ju^j +u^A\dr_A(\rho^a_j)c^j
\ee
of graded vector fields. It follows that graded vector fields (\ref{hn14})
can be represented by
sections of the vector bundle
$\cV_E\to Z$ which is locally isomorphic to the vector bundle
\be
\cV_E|_U\approx\w E^*\op\ot_Z(E\op\oplus_Z TZ)|_U,
\ee
and is characterized by an atlas of bundle coordinates 
$(z^A,z^A_{a_1\ldots a_k},v^i_{b_1\ldots b_k})$,
$k=0,\ldots,m$, possessing the transition functions
\mar{+161}\ben
&& z'^A_{i_1\ldots i_k}=\rho^{-1}{}_{i_1}^{a_1}\cdots
\rho^{-1}{}_{i_k}^{a_k} z^A_{a_1\ldots a_k}, \nonumber\\
&& v'^i_{j_1\ldots j_k}=\rho^{-1}{}_{j_1}^{b_1}\cdots
\rho^{-1}{}_{j_k}^{b_k}\left[\rho^i_jv^j_{b_1\ldots b_k}+ \frac{k!}{(k-1)!}
z^A_{b_1\ldots b_{k-1}}\dr_A\rho^i_{b_k}\right], \label{+161}
\een
which
fulfil the cocycle condition (\ref{+9}).

\begin{rem} \label{+93} \mar{+93}
One tries to construct a graded tangent bundle over a graded manifold
$(Z,\gR)$ whose sheaf of sections is the graded derivation sheaf
$\gd\gR$.
  Nevertheless, the transformation law
(\ref{+161}) shows that   the projection
\be
\cV_E\mid_U\to \pr_2VE\op\oplus_Z TZ
\ee
is not global, i.e., $\cV_E$ is not an exterior bundle. It means that the
sheaf of derivations $\gd\gR$ is not a structure sheaf of a graded manifold.
\end{rem}

There is the exact sequence
\mar{1030}\beq
0\to \w E^*\op\ot_Z E\to\cV_E\to \w E^*\op\ot_Z TZ\to 0 \label{1030}
\eeq
of vector bundles over $Z$.
Its splitting
\mar{cmp70}\beq
\wt\g:\dot z^A\dr_A \mapsto \dot z^A(\dr_A +\wt\g_A^a\frac{\dr}{\dr
c^a}) \label{cmp70}
\eeq
transforms every vector field $\tau$ on $Z$
into the graded vector field
\mar{ijmp10}\beq
\tau=\tau^A\dr_A\mapsto \nabla_\tau=\tau^A(\dr_A
+\wt\g_A^a\frac{\dr}{\dr c^a}),
\label{ijmp10}
\eeq
which is a graded derivation of the sheaf $\gR_E$ satisfying the Leibniz rule
\be
\nabla_\tau(sf)=(\tau\rfloor ds)f +s\nabla_\tau(f), \quad f\in\gR_E(Z),
\quad s\in C^\infty(Z).
\ee
Therefore, one can think of the splitting
(\ref{cmp70}) of the exact sequence (\ref{1030})
as being a graded connection
  on the simple graded manifold
$(Z,\gR_E)$ \cite{book00,sardijmp}.
In particular, this connection provides the corresponding
horizontal splitting
\be
u= u^A\dr_A + u^a\frac{\dr}{\dr c^a}=u^A(\dr_A 
+\wt\g_A^a\frac{\dr}{\dr c^a}) + (u^a-
u^A\wt\g_A^a)\frac{\dr}{\dr c^a}
\ee
of graded vector fields.

In accordance with Theorem \ref{sp11}, a graded connection (\ref{cmp70}) always
exists.

\begin{rem} \label{+94} \mar{+94}
By virtue of the isomorphism (\ref{+54}), any connection $\wt \g$ on a graded
manifold $(Z,\gR)$, restricted to a splitting domain $U$, takes the form
(\ref{cmp70}). Given two splitting domains $U$ and $U'$ of $(Z,\gR)$ with the
transition functions (\ref{+95}), the connection components $\wt\g^a_A$ obey
the transformation law
\mar{+96}\beq
\wt\g'^a_A= \wt\g^b_A\frac{\dr}{\dr c^b}\rho^a +\dr_A\rho^a. \label{+96}
\eeq
If $U$ and $U'$ are the trivialization charts of the same vector bundle $E$
in Theorem \ref{lmp1} together with the transition functions (\ref{+6}), the
transformation law (\ref{+96}) takes the form
\mar{+97}\beq
\wt\g'^a_A= \rho^a_b(z)\wt\g^b +\dr_A\rho^a_b(z)c^b. \label{+97}
\eeq
\end{rem}

\begin{rem}
It should be emphasized that the above  notion of a graded
connection differs from that
of a connection on a graded fibre bundle
$(Z,\gR)\to (X,\cB)$ in \cite{alm}. The latter
is a section of the jet graded bundle
$J^1(Z/X)\to (Z,\gR)$ of sections of the graded fibre bundle
  $(Z,\gR)\to (X,\cB)$ (see
\cite{rup} for formalism of jets of graded manifolds).
\end{rem}

\begin{ex} \label{1031} \mar{1031}
Every linear connection
\be
\g=dz^A\ot (\dr_A +\g_A{}^a{}_by^b \dr_a)
\ee
on the vector bundle $E\to Z$ yields the graded connection
\mar{cmp73}\beq
\g_S=dz^A\ot (\dr_A +\g_A{}^a{}_bc^b\frac{\dr}{\dr c^a}) \label{cmp73}
\eeq
on the simple graded manifold $(Z,\gR_E)$.
In view of Remark \ref{+94}, $\g_S$ is also a graded
connection on the graded manifold $(Z,\gR)\cong
(Z,\gR_E)$, but its linear form
(\ref{cmp73}) is not maintained under the transformation law (\ref{+96}).
\end{ex}

The curvature of the graded
connection $\nabla_\tau$ (\ref{ijmp10}) is defined by the expression:
\mar{+110}\ben
&&
R(\tau,\tau')=[\nabla_\tau,\nabla_{\tau'}]-\nabla_{[\tau,\tau']},\nonumber\\
&& R(\tau,\tau') =\tau^A\tau'^B R^a_{AB}\frac{\dr}{\dr c^a}: \gR_E\to \gR_E,
\nonumber\\
&&R^a_{AB} =\dr_A\wt\g^a_B-\dr_B\wt\g^a_A
+\wt\g^k_A\frac{\dr}{\dr c^k}\wt\g^a_B -
  \wt\g^k_B\frac{\dr}{\dr c^k} \wt\g^a_A.\label{+110}
\een
It can also be written in the form 
\mar{+111}\beq
R =\frac12 R_{AB}^a dz^A\w dz^B\ot\frac{\dr}{\dr c^a}. \label{+111}
\eeq

Let now $\cV^*_E\to  Z$ be a
vector bundle which is the pointwise $\w E^*$-dual of the
vector bundle $\cV_E\to Z$.
It is locally isomorphic to the vector bundle
\be
\cV^*_E|_U\approx \w E^*\op\ot_Z(E^*\op\oplus_Z T^*Z)|_U.
\ee
With respect to the dual bases $\{dz^A\}$ for $T^*Z$ and
$\{dc^b\}$ for $\pr_2V^*E\cong E^*$, sections of the vector bundle $\cV^*_E$
take the coordinate form
\be
\f=\f_A dz^A + \f_adc^a,
\ee
together with transition functions
\be
\f'_a=\rho^{-1}{}_a^b\f_b, \qquad \f'_A=\f_A
+\rho^{-1}{}_a^b\dr_A(\rho^a_j)\f_bc^j.
\ee
They are regarded as graded exterior one-forms on the graded manifold
$(Z,\gR_E)$.

The sheaf $\gO^1\gR_E$ of germs of sections of the vector bundle
$\cV^*_E\to Z$ is the dual of the graded derivation sheaf $\gd\gR_E$, where
the duality morphism is given by
the graded interior product
\mar{cmp65}\beq
u\rfloor \f=u^A\f_A + (-1)^{\nw{\f_a}}u^a\f_a. \label{cmp65}
\eeq
In particular, the dual of the exact sequence (\ref{1030}) is the 
exact sequence
\mar{cmp72}\beq
0\to \w E^*\op\ot_ZT^*Z\to\cV^*_E\to \w E^*\op\ot_Z E^*\to 0.
\label{cmp72}
\eeq
Any graded connection $\wt\g$ (\ref{cmp70}) yields the
splitting of the exact sequence (\ref{cmp72}), and determines
the corresponding
decomposition of graded one-forms
\be
\f=\f_A dz^A + \f_adc^a =(\f_A+\f_a\wt\g_A^a)dz^A +\f_a(dc^a
-\wt\g_A^adz^A).
\ee

Graded exterior $k$-forms $\f$ are defined as sections
of the graded exterior bundle $\op\w^k_Z\cV^*_E$ such that
\mar{+113}\beq
\f\w\si =(-1)^{\nm\f\nm\si +\nw\f\nw\si}\si\w \f, \label{+113}
\eeq
where $|.|$ denotes the form degree. For instance,
\mar{+113'}\beq
dz^A\w dc^i=-dc^i\w dz^A, \qquad dc^i\w dc^j= dc^j\w dc^i. \label{+113'}
\eeq
The graded interior product (\ref{cmp65})
is extended to higher graded exterior forms by the rule
\mar{+114}\beq
u\rfloor (\f\w\si)=(u\rfloor \f)\w \si
+(-1)^{\nm\f+\nw\f\nw{u}}\f\w(u\rfloor\si). \label{+114}
\eeq

The graded exterior differential
$d$ of graded functions is introduced by the condition
$u\rfloor df=u(f)$
for an arbitrary graded vector field $u$. It  is
extended uniquely to graded exterior forms by the rule
\mar{+114'}\beq
d(\f\w\si)= d\f\w\si +(-1)^{\nm\f}\f\w d\si, \qquad  d\circ d=0, \label{+114'}
\eeq
  and is given by the coordinate expression
\be
d\f= dz^A \w \dr_A\f +dc^a\w \frac{\dr}{\dr c^a}\f,
\ee
where the left derivatives
$\dr_\la$, $\dr/\dr c^a$ act on coefficients of graded exterior forms
by the rule
(\ref{cmp50a}), and they are graded commutative with the forms $dz^A$,
$dc^a$ \cite{gaw,kost77}.  The Lie
derivative of a graded exterior form $\f$ along a graded vector field 
$u$ is defined by
the familiar formula
\mar{+117}\beq
\bL_u\f= u\rfloor d\f + d(u\rfloor\f). \label{+117}
\eeq
It possesses the property
\be
\bL_u(\f\w\f')=\bL_u(\f)\w\f' + (-1)^{[u][\f]}\f\w\bL_u(\f').
\ee

With the graded exterior differential$d$, graded exterior forms constitute an
$\Bbb N$-, $\Bbb Z_2$-graded differential algebra
$\cO^*\cA_E$, where $\cO^*\cA_E=\cA_E=\gR_E(Z)$ denotes the structure
module of
graded functions on $Z$. It is called a graded commutative
differential algebra.  The corresponding graded de Rham complex is
\mar{+137}\beq
0\to\Bbb R\to \cA_E\ar^d \cO^1\cA_E \ar^d \cdots \cO^k\cA_E \ar^d \cdots.
\label{+137}
\eeq

Cohomology $H^q_{GR}(Z)$ of the graded de Rham
complex (\ref{+137}) is called graded de Rham cohomology
of the graded manifold $(Z,\gR_E)$.
One can compute this cohomology with the aid of the abstract de Rham theorem.
Let $\gO^k\gR_E$ denote the sheaf of germs of
graded $k$-forms on $(Z,\gR_E)$. Its structure module is
$\cO^k\cA_E=\gO^k\gR_E(Z)$. These sheaves make up the complex
\mar{1033}\beq
0\to\Bbb R\ar \gR_E \ar^d \gO^1\gR_E\ar^d\cdots \gO^k\gR_E\ar^d\cdots.
\label{1033}
\eeq
All $\gO^k\gR_E$ are sheaves of $C^\infty_Z$-modules on $Z$ and,
consequently, are fine and acyclic. Furthermore, the Poincar\'e lemma
for graded exterior forms holds \cite{bart,kost77}. It follows that
the complex (\ref{1033}) is a fine resolution of the constant
sheaf $\Bbb R$ on
the manifold $Z$.  Then, by virtue of the abstract de Rham theorem, there is
  an isomorphism
\mar{+136}\beq
H^*_{GR}(Z)=H^*(Z;\Bbb R)=H^*(Z) \label{+136}
\eeq
of the graded de Rham cohomology $H^*_{GR}(Z)$ to the
de Rham cohomology $H^*(Z)$ of the smooth manifold
$Z$ \cite{kost77}. Moreover, the cohomology isomorphism
(\ref{+136}) accompanies the cochain monomorphism
$\cO^*(Z)\to \cO^*\cA_E$
of the de Rham complex $\cO^*(Z)$ of smooth exterior forms
on $Z$ to the graded de Rham complex
(\ref{+137}). Hence, any closed graded exterior
form is split into a sum $\f=d\si +\vf$
of an exact graded exterior form $d\si\in \cO^*\cA_E$
and a closed exterior form $\vf\in \cO^*(Z)$ on $Z$.

\section{Jets of ghosts and antifields}

In field-antifield BRST theory, the antibracket is
defined by means of the variational operator.
This operator can be introduced in a rigorous algebraic way as the coboundary
operator of the variational complex of exterior forms on the infinite
jet space of physical fields, ghosts and antifields
\cite{barn,barn00,brandt,brandt01}. 
Herewith, the antibracket and the BRTS operator
are expressed in terms of jets of ghosts and physical fields.
For example, the BRST transformation of gauge
potentials $a^r_\la$ in  Yang--Mills theory reads
\be
\bs a^r_\la= C^r_\la + c^r_{pq}a^p_\la C^q,
\ee
where $C^r_\la$ are jets of ghosts $C^r$ introduced in a heuristic way.

Furthermore, the
variational complex in BRST theory on a
contractible manifold $X=\Bbb R^n$ is exact. It follows that 
the kernel of the
variational operator $\dl$ equals the image of the horizontal
differential $d_H$. Therefore, several objects in
field-antifield BRST theory on $\Bbb R^n$ are
determined modulo $d_H$-exact forms.
In particular, let us mention the iterated
cohomology $H^{k,p}(\bs|d_H)$ of the BRST bicomplex,
defined with respect to
the BRST operator
$\bs$ and the horizontal differential
$d_H$, and graded by the ghost
number $k$ and the form degree $p$.
The iterated cohomology of form degree $p=n={\rm dim}\, X$
coincides with the local BRST cohomology (i.e., the $\bs$-cohomology
modulo $d_H$).
If $X=\Bbb R^n$, an isomorphism of the
local BRST cohomology $H^{k,n}(\bs|d_H)$, $k\neq -n$, to the
cohomology
$H^{k+n}_{\rm tot}$ of the total BRST operator
$\bs +d_H$ has been proved
by constructing the descent equations \cite{brandt}.
This result has been generalized to an arbitrary connected manifold
$X$ \cite{lmp,sard01}. For this purpose, we provide a (global)
differential geometric definition of jets of odd ghosts and antifields,
and extend the variational complex to the space of these jets.

For the sake of simplicity, we consider BRST theory of
even physical fields and finitely reducible gauge
transformations. Its finite classical basis consists
of even physical fields of zero ghost number,
even and odd ghosts (including ghosts-for-ghosts) of strictly
positive ghost number, and even and odd antifields of
strictly negative ghost number. For
instance, this is the case of Yang--Mills theory.

\glos{A. Jets of odd ghosts}

There exist different geometric models of ghosts. For instance,
ghosts
in Yang--Mills theory
are often represented by the Maurer--Cartan form on the gauge
group \cite{bon}. This representation, however, is
not extended to other gauge models.
We describe all odd fields as elements of simple graded manifolds
\cite{book00,sardijmp,sard01}.

Let $Y\to X$ be the characteristic vector bundle of a simple graded
manifold $(X,\gR_Y)$. The $r$-order jet
manifold $J^rY$ of $Y$ is also a vector bundle over $X$. Let us
consider the simple graded manifold $(X,\gR_{J^rY})$, determined by the
characteristic vector bundle
$J^rY\to X$. Its local basis is $\{x^\la,c^a_\La\}$, $0\leq |\La|\leq r$,
where $\La=(\la_k,\ldots,\la_1)$ are multi-indices. It possesses
the transition functions
\mar{+471}\beq
c'^a_{\la+\La}=d_\la(\rho^a_j c^j_\La), \qquad
d_\la=\dr_\la + \op\sum_{|\La|<r}c^a_{\la+\La}\frac{\dr}{\dr c^a_\La},
\label{+471}
\eeq
where $d_\la$ is the graded total
derivative. In view
of the transition functions (\ref{+471}), one can think of $(X,\gR_{J^rY})$ as
being a graded $r$-order jet manifold of the simple graded manifold
$(X,\gR_Y)$.

Let $\cO^*\cA_{J^rY}$ be the differential algebra of graded exterior
forms on the graded jet manifold $(X,\gR_{J^rY})$.
Since $Y\to X$ is a vector bundle, the canonical fibration
$\pi^r_{r-1}:J^rY \to J^{r-1}Y$ is a linear morphism
of vector bundles over $X$ and, thereby,
yields the
corresponding morphism of graded jet manifolds
$(X,\gR_{J^rY})\to (X,\gR_{J^{r-1}Y})$ accompanied by the
pull-back  monomorphism of differential algebras
$\cO^*\cA_{J^{r-1}Y}\to \cO^*\cA_{J^rY}$. Then
we have the direct system of differential algebras
\be
\cO^*\cA_Y\ar \cO^*\cA_{J^1Y}\ar\cdots
\cO^*\cA_{J^rY}\ar^{\pi^{r+1*}_r}\cdots\,.
\ee
Its direct limit $\cO^*_\infty\cA_Y$
consists
of graded exterior forms on all graded jet manifolds $(X,\gR_{J^rY})$
modulo the pull-back identification.
It is a locally free graded
$C^\infty(X)$-algebra generated by the elements
\be
(1, c^a_\La,dx^\la, \th^a_\La=dc^a_\La -c^a_{\la +\La}dx^\la), \qquad
0\leq |\La|,
\ee
where $dx^\la$ and $\th^a_\La$ are called horizontal and contact forms,
respectively.
In particular, $\cO^0_\infty\cA_Y$ is the graded commutative ring of
graded functions on all graded jet manifolds $(X,\gR_{J^rY})$
modulo the pull-back identification.

Let us consider the sheaf
$\gQ^0_\infty\cA_Y$ of germs of graded
functions $\f\in\cO^*_\infty\cA_Y$.  It is a sheaf of graded
commutative algebras on $X$, and the pair $(X,\gQ^0_\infty\cA_Y)$
is a graded manifold. This graded manifold is the projective limit of 
the inverse system
of graded jet manifolds
\be
(X,\gR_Y)\longleftarrow (X,\gR_{J^1Y})\longleftarrow\cdots (X,\gR_{J^rY})
\longleftarrow \cdots,
\ee
and is called the graded infinite jet manifold. Then one can think
of elements of the algebra $\cO^*_\infty\cA_Y$ as being graded
exterior forms on the graded manifold $(X,\gQ^0_\infty\cA_Y)$.

There is the canonical splitting of
\be
\cO^*_\infty\cA_Y =\op\oplus_{k,s}\cO^{k,s}_\infty\cA_Y, \qquad 0\leq k, \qquad
0\leq s\leq n,
\ee
into $\cO^0_\infty\cA_Y$-modules $\cO^{k,s}_\infty\cA_Y$
of $k$-contact and $s$-horizontal graded forms.
Accordingly, the graded exterior differential $d$ on $\cO^*_\infty\cA_Y$
is split into
the sum $d=d_H+d_V$, where $d_H$ is the nilpotent graded horizontal
differential
\be
d_H(\f)=dx^\la\w d_\la(\f): \cO^{k,s}_\infty\cA_Y\to
\cO^{k,s+1}_\infty\cA_Y.
\ee

With respect to the BRST operator $\bs$, the graded exterior forms $\f\in
\cO^*_\infty\cA_Y$ are characterized by the ghost number
\be
{\rm gh}(dc^a_\La)={\rm gh}(c^a_\La)= {\rm gh}(c^a),
\ee
and one puts $\bs\circ d_H+d_H\circ\bs=0$.

\glos{B. Even physical fields and ghosts}

In order to describe odd and even elements of the classical basis of
field-antifield BRST theory  on the
same footing, we will generalize the notion of a graded manifold to graded
commutative algebras generated  both by odd and even elements
\cite{sard01}.

Let $Y=Y_0\oplus Y_1$ be the Whitney sum of vector bundles $Y_0\to
X$ and $Y_1\to X$. We regard it as
a bundle of graded vector spaces with the typical fibre $V=V_0\oplus V_1$.
Let us consider the quotient of the tensor bundle
\be
\ot Y^*=\op\oplus^\infty_{k=0} (\op\ot^k_X Y^*)
\ee
by the elements
\be
y_0y'_0 - y'_0y_0,\quad y_1y'_1 + y'_1y_1, \quad y_0y_1 -
y_1y_0
\ee
for all  $y_0,y'_0\in Y_{0x}^*$,
$y_1,y'_1\in Y_{1x}^*$, and  $x\in X$.
It is an infinite-dimensional vector bundle, further denoted by $\w Y^*$.
Global sections of $\w Y^*$ constitute a graded commutative algebra
$\cA_Y$, which is the product over $C^\infty(X)$ of the commutative algebra
$\cA_0$ of global sections of the symmetric bundle $\vee Y_0^*\to X$ and
the graded algebra $\cA_1$ of global sections of the exterior bundle $\w
Y_1^*\to X$.

Let $\gR$, $\gR_0$ and $\gR_1$ be the sheaves of germs of
sections of the vector bundles $\w Y^*$, $\vee Y_0^*$ and $\w Y_1^*$,
respectively. The pair
$(X,\gR_1)$ is a familiar simple graded manifold. Therefore, we
agree to call
$(X,\gR)$ the graded commutative manifold, determined by the characteristic
graded vector bundle $Y$. Given
a bundle coordinate chart
$(U;x^\la,y^i_0,y^a_1)$ of $Y$, the local basis for $(X,\gR)$ is
$(x^\la,c^i_0, c^a_1)$, where
$\{c^i_0\}$ and $\{c^a_1\}$ are the fibre bases for the vector bundles $Y_0^*$
and
$Y_1^*$, respectively. Then a straightforward repetition of all the above
constructions for a simple graded manifold provides us with the
differential algebra
$\cO^*_\infty\cA$ of graded commutative exterior forms on the graded 
commutative
infinite jet manifold $(X,\gQ^0_\infty\cA)$. This is a
$C^\infty(X)$-algebra generated locally by the elements
\be
(1,c^i_{0\La},c^a_{1\La}, dx^\la, \th^i_{0\La}, \th^a_{1\La}), \qquad
0\leq |\La|.
\ee
Its $C^\infty(X)$-subalgebra $\cO^*_\infty\cA_1$, generated locally by
the elements
$(1,c^i_{1\La}, dx^\la, \th^i_{1\La})$,
is exactly the differential algebra of graded exterior forms on the
graded manifold
$(X,\gQ^0\cA_1)$. The
$C^\infty(X)$-subalgebra
$\cO^*_\infty\cA_0$ of $\cO^*_\infty\cA$
generated locally by the elements
$(1,c^i_{0\La}, dx^\la, \th^i_{0\La})$,
$0\leq |\La|$, is isomorphic to the
polynomial subalgebra $P^*_\infty$ of the differential algebra 
$\cO^*_\infty$ of
exterior forms on the infinite
jet manifold $J^\infty Y_0$ of the vector bundle $Y_0\to X$
after its pull-back onto $X$ \cite{lmp,jmp01}. The algebra $\cO^*_\infty$
provides the differential calculus in classical field theory.

\glos{C. Antifields}

The jet formulation of field-antifield
BRST theory enables one to introduce antifields
on the same footing as physical fields and ghosts.
Let $\Phi^A$ be a collective symbol for physical fields and ghosts.
Let $E$ be the characteristic graded vector bundle of the graded
commutative manifold, generated by $\Phi^A$.
Treated as source
coefficients of BRST transformations, antifields $\Phi^*_A$ with the 
ghost number
\be
{\rm gh}\,\Phi^*_A=-{\rm gh}\,\Phi^A-1
\ee
are represented by
elements of the graded commutative manifold, determined by the characteristic
graded vector
bundle $\op\w^n T^*X\ot E^*$.
Then the total
characteristic graded vector bundle of a graded commutative manifold 
for a classical
basis of field-antifield BRST theory is
\be
Y=E\oplus (\op\w^n T^*X\ot E^*).
\ee

In
particular, gauge potentials $a^r_\la$ in Yang--Mills theory on a 
principal bundle
$P\to X$ are represented
by sections of the affine bundle $J^1P/G\to X$, modelled on the vector
bundle $T^*X\ot V_GP\to X$. Accordingly, the characteristic vector bundle
for their odd antifields is
\be
\op\w^n T^*X\ot TX\ot V^*_GP\to X.
\ee
As was mentioned above, the characteristic vector bundle for ghosts $C^r$
in Yang--Mills theory is the Lie algebra
bundle $V_GP\to X$.
Then the characteristic vector bundle for their even antifields is
\be
\op\w^n T^*X\ot V^*_GP\to X.
\ee
Thus, the total characteristic graded vector bundle for BRST Yang--Mills
theory is
\be
&& Y=Y_0\oplus Y_1= [(T^*X\ot V_GP)\oplus(\op\w^n T^*X\ot V^*_GP)]\oplus\\
&& \qquad[V_GP\oplus(\op\w^n T^*X\ot TX\ot V_G^*P)].
\ee

The jets $\Phi^*_{A\La}$ of antifields $\Phi^*_A$ are introduced similarly to
jets $\Phi^A_\La$ of physical fields and ghosts $\Phi^A$.

\glos{D. The variational complex in BRST theory}

The differential algebra $\cO^*_\infty\cA$ gives everything for
global formulation of Lagrangian field-antifield
BRST theory on a manifold $X$.
We restrict our consideration to the short variational complex
\mar{cmp25}\beq
0\ar \Bbb R\ar \cO^0_\infty\cA\ar^{d_H}\cO^{0,1}_\infty\cA\cdots\ar^{d_H}
  \cO^{0,n}_\infty\cA\ar^\dl \im\dl \to 0,
\label{cmp25}
\eeq
where $\dl$ is the variational operator such that $\dl\circ d_H=0$.
It is given by the expression
\be
\dl (L)=
  (-1)^{|\La|}\th^a\w d_\La (\dr^\La_a L), \qquad L\in \cO^{0,n}_\infty\cA,
\ee
with respect to a physical basis $(\zeta^a)=(\Phi^A,\Phi^*_A)$.

The variational complex (\ref{cmp25}) provides the
algebraic approach to the antibracket technique, where one can think of
elements $L$ of $\cO^{0,n}_\infty\cA$ as being Lagrangians of fields,
ghosts and antifields.  Note that, to be well-defined, a global BRST Lagrangian
should factorize through
covariant differentials of physical fields, ghosts and antifields
$D_\la\zeta^a=\zeta^a_\la-\wt\g^a_\la$, where $\wt\g$ is a connection on
the graded commutative manifold $(X,\gR)$.

In order to obtain cohomology of the variational complex (\ref{cmp25}),
let us consider the sheaf
$\gQ^*_\infty\cA$ of germs of elements
$\f\in\cO^*_\infty\cA$ and the graded differential algebra
$P^*_\infty\cA$ of global sections of this sheaf. Note that
$P^*_\infty\cA\neq \cO^*_\infty\cA$. Roughly speaking,
any element of $\cO^*_\infty\cA$ is of bounded jet order, whereas
elements of $P^*_\infty\cA$ need not be so.

We have the short variational complex of sheaves
\mar{cmp20}\beq
0\ar \Bbb R\ar \gQ^0_\infty\cA\ar^{d_H}\gQ^{0,1}_\infty\cA\cdots\ar^{d_H}
\gQ^{0,n}_\infty\cA\ar^\dl \im\dl \to 0.
\label{cmp20}
\eeq
Graded commutative exterior forms
$\f\in\cO^*_\infty\cA$ are proved to
satisfy the algebraic Poincar\'e lemma, i.e., any
closed graded commutative exterior form on the graded manifold
$(\Bbb R^n,\gQ^0_\infty\cA)$ is exact
\cite{brandt}. Consequently, the complex (\ref{cmp20}) is exact.
Since $\gQ^{0,*}_\infty\cA$ are sheaves of $C^\infty(X)$-modules on 
$X$, they are
fine and acyclic. Without inspecting the acyclicity of the sheaf
$\im\dl$, one can apply a minor modification of the abstract de Rham theorem
\cite{jmp01} to the complex
(\ref{cmp20}), and obtains that cohomology of the complex
\mar{cmp22}\beq
0\ar \Bbb R\ar
P^0_\infty\cA\ar^{d_H}P^{0,1}_\infty\cA\cdots\ar^{d_H}
  P^{0,n}_\infty\cA\ar^\dl \im\dl \to 0
\label{cmp22}
\eeq
is isomorphic to the de Rham cohomology of a manifold $X$.

Following suit of Theorem \ref{am11} and replacing
exterior forms on $J^\infty Y$ with graded commutative forms on 
$(X,\gQ^0_\infty\cA)$,
one can show that cohomology of the short variational complex (\ref{cmp25})
is isomorphic to that of the complex (\ref{cmp22}) and, consequently, to
the de Rham cohomology of $X$. Moreover, this isomorphism
is performed by the natural monomorphism of
the de Rham complex $\cO^*$ of exterior forms on $X$ to the complex 
(\ref{cmp22}).
It follows that:

(i)  every
$d_H$-closed graded form $\f\in \cO^{0,m<n}_\infty\cA$
is split into the sum
$\f=\varphi + d_H\xi$,
where $\varphi$ is a closed exterior $m$-form on $X$;

(ii) every
$\dl$-closed graded form $\f\in \cO^{0,n}_\infty\cA$ is
split into the sum
$\f=\varphi + d_H\xi$,
where $\varphi$ is an exterior $n$-form on $X$.

One should mention the important case of BRST theory where Lagrangians
are independent on coordinates $x^\la$.
Let us consider the subsheaf $\ol\gQ^*_\infty\cA$ of the sheaf
$\gQ^*_\infty\cA$ which consists of germs of $x$-independent graded
commutative exterior forms. Then we have the subcomplex
\mar{cmp27}\beq
0\ar \Bbb R\ar
\ol\gQ^0_\infty\cA\ar^{d_H}\ol\gQ^{0,1}_\infty\cA\cdots
\ar^{d_H} \ol\gQ^{0,n}_\infty\cA\ar^\dl \im\dl \to 0
\label{cmp27}
\eeq
of the complex (\ref{cmp20}) and the corresponding subcomplex
\mar{cmp28}\beq
0\ar \Bbb R\ar
\ol P^0_\infty\cA\ar^{d_H} \ol P^{0,1}_\infty\cA\cdots
\ar^{d_H} \ol P^{0,n}_\infty\cA\ar^\dl \im\dl \to 0
\label{cmp28}
\eeq
of the complex (\ref{cmp22}). Clearly,
$\ol P^{0,*}_\infty\cA\subset \cO^{0,*}_\infty\cA$, i.e., the
complex (\ref{cmp28}) is also a subcomplex of the short variational
complex (\ref{cmp25}).

The key point is that the complex of sheaves (\ref{cmp27}) fails
to be exact.
The obstruction to its exactness at the
term
$\ol\gQ^{0,k}_\infty$ consists of the germs of constant exterior $k$-forms
on $X$  \cite{barn00}.
Let us denote their sheaf
by $S^k_X$.
We have the short exact sequences of sheaves
\be
&& 0\to \im d_H \to \Ker d_H \to S^k_X \to 0, \qquad 0<k< n,\\
&& 0\to \im d_H \to \Ker \dl \to S^n_X \to 0
\ee
and the corresponding sequences of modules of their global sections
\be
&& 0\to \im d_H(X) \to \Ker d_H(X) \to S^k_X(X) \to 0, \qquad 0<k<
n,\\
&& 0\to \im d_H(X) \to \Ker \dl(X) \to S^n_X(X) \to 0.
\ee
The latter are exact because $S^{k<n}_X$ and $S^n_X$ are subsheaves 
of the sheaves
$\Ker d_H$ and $\Ker\dl$, respectively. Therefore, the $k$th cohomology
group of the  complex (\ref{cmp28})
is isomorphic to the $\Bbb R$-module
$S^k_X(X)$ of constant exterior
$k$-forms, $0<k\leq n$, on the manifold $X$.
Consequently, any $d_H$-closed graded commutative $k$-form, $0< k<
n$, and any $\dl$-closed graded commutative $n$-form $\f$, constant on $X$,
are split into the sum
$\f=\varphi + d_H\xi$ where $\varphi$ is a constant exterior form on $X$.

\end{document}